\documentclass[11pt,letterpaper,twopage,pdf]{article}
\pdfoutput=1
\usepackage{latexsym}
\usepackage[latin1]{inputenc}   
\usepackage{amsmath}
\usepackage{epsfig}
\usepackage{graphics}
\usepackage{amsfonts}
\usepackage{pifont}
\usepackage{booktabs}
\usepackage{xspace}
\usepackage{color}
\usepackage{booktabs}
\usepackage{xtab}
\usepackage{eurosym}
\usepackage{makeidx}
\usepackage{fancyhdr}

\usepackage{charter}

\newcommand{\tsc}[1]{\textsc{#1}}
\definecolor{lgrey}{rgb}{0.4,0.4,0.4}            

\newcounter{bibnumber}
\newcounter{impnumber}
\usepackage{feynmp}
\usepackage{cite}
\usepackage{xspace}
\usepackage{booktabs}
\usepackage[small, nooneline, center]{subfig}
\usepackage{amsmath,amssymb,bm}
\usepackage{longtable}

\usepackage{mciteplus}
\definecolor{lightgray}{rgb}{0.85,0.85,0.87}
\definecolor{gray}{rgb}{0.5,0.5,0.5}
\definecolor{darkgray}{rgb}{0.36,0.36,0.36}
\def\lsim{\mathrel{\rlap{\lower4pt\hbox{\hskip1pt$\sim$}}
    \raise1pt\hbox{$<$}}}                
\def\gsim{\mathrel{\rlap{\lower4pt\hbox{\hskip1pt$\sim$}}
    \raise1pt\hbox{$>$}}}                
\newcommand{\dd}[1]{\ensuremath{\mrm{d}#1}\hspace*{0.2em} }

\newcommand{\mrm}[1]{\ensuremath{\mathrm{#1}}}

\newcommand{\obs}{\ensuremath{\mathcal{O}}}

\newcommand{\pt}[1][]{\ensuremath{p_{\perp #1}}}

\newcommand{\ptmin}[1][]{\pt[\mrm{min}#1]\xspace}
\newcommand{\pdf}[1]{\ensuremath{f_{#1}}}

\newcommand{\pqcd}[2][]{\ensuremath{\sigma^{(#1)}_{#2}}}
\newcommand{\PS}[2][]{\ensuremath{\Phi_{#2}^{#1}}\xspace}
\newcommand{\dPS}[2][]{\ensuremath{\dd{\PS[#1]{#2}}}\xspace}

\newcommand{\eqRef}[1]{equation~\eqref{#1}\xspace}

\newcommand{\eqsRef}[1]{equations~\eqref{#1}\xspace}

\newcommand{\secRef}[1]{section~\ref{#1}\xspace}
\newcommand{\SecRef}[1]{Section~\ref{#1}\xspace}
\newcommand{\secsRef}[1]{sections~\ref{#1}\xspace}

\newcommand{\tabRef}[1]{table~\ref{#1}\xspace}
\newcommand{\TabRef}[1]{Table~\ref{#1}\xspace}
\newcommand{\figRef}[1]{figure~\ref{#1}\xspace}
\newcommand{\FigRef}[1]{Figure~\ref{#1}\xspace}

\newcommand{\figsRef}[1]{figures~\ref{#1}\xspace}

\newcommand{\Ar}{\tsc{Ariadne}\xspace}
\newcommand{\Co}{\tsc{Comphep}\xspace}
\newcommand{\Ca}{\tsc{Calchep}\xspace}
\newcommand{\Fw}{\tsc{Mc@nlo}\xspace}
\newcommand{\Hw}{\tsc{Herwig}\xspace}

\newcommand{\Mg}{\tsc{Madgraph}\xspace}
\newcommand{\Py}{\tsc{Pythia}\xspace}
\newcommand{\Sh}{\tsc{Sherpa}\xspace}
\newcommand{\Vc}{\tsc{Vincia}\xspace}
\newcommand{\Qgsjet}{\tsc{Qgsjet}\xspace}
\newcommand{\Dpmjet}{\tsc{Dpmjet}\xspace}
\newcommand{\Epos}{\tsc{Epos}\xspace}
\newcommand{\Sibyll}{\tsc{Sibyll}\xspace}
\newcommand{\Phojet}{\tsc{Phojet}\xspace}
\newcommand{\Pw}{\tsc{Powheg}\xspace}

\newlength{\tabcolsepsave}
\newenvironment{loopsnlegs}[1][t]{
\setlength{\tabcolsepsave}{\tabcolsep}
\setlength{\tabcolsep}{0pt}
\begin{tabular}{cc}\parbox[c]{1.1em}{\rotatebox{90}{\small $\ell$ (loops)}}&%
\begin{tabular}[#1]}{
\end{tabular}\\[-1mm]
 & \small $k$ (legs)
\end{tabular}%
\setlength{\tabcolsep}{\tabcolsepsave}
}
\newcommand{\cbox}[2]{%
\begin{minipage}[c]{1.4cm}%
\center%
{%
\parbox[c]{1.4cm}{\includegraphics*[width=1.4cm]{#1.pdf}}}%
\end{minipage}%
\hspace*{-1.4cm}%
\begin{minipage}[c]{1.4cm}
\center
#2%
\end{minipage}}
\newcommand{\cyanbox}[1]{\cbox{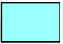}{#1}}
\newcommand{\eggbox}[1]{\cbox{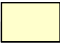}{#1}}
\newcommand{\gbox}[1]{\cbox{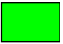}{#1}}
\newcommand{\wbox}[1]{\cbox{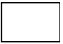}{#1}}
\newcommand{\gwbox}[1]{\cbox{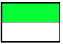}{#1}}
\newcommand{\gybox}[1]{\cbox{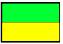}{#1}}
\newcommand{\gywbox}[1]{\cbox{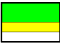}{#1}}
\newcommand{\ybox}[1]{\cbox{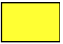}{#1}}
\newcommand{\ywbox}[1]{\cbox{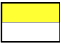}{#1}}
\newcommand{\rybox}[1]{\cbox{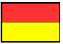}{#1}}
\newcommand{\wybox}[1]{\cbox{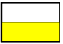}{#1}}
\newcommand{\wywbox}[1]{\cbox{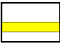}{#1}}
\newcommand{\wwybox}[1]{\cbox{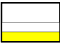}{#1}}

\makeindex

\usepackage[top=1.135in,bottom=1.17in,left=1.16in,right=1.16in]{geometry}
\begin{document}
\title{Introduction to QCD\thanks
                 {Lectures presented at TASI 2012: Theoretical Advanced
 Study Institute in Elementary Particle Physics  --- Searching for
 New Physics at Small and Large Scales. University of Colorado, Boulder,
 CO, June 4 -- 29, 2012. Based on lectures first given at ESHEP 2010
 (Raseborg, Finland); updated for AEPSHEP 2014 (Puri, India), 
 pre-SUSY 2016 (Melbourne, Australia), and for MCNET 2017 (Lund, Sweden).}}
\author{P.~Z.~Skands\thanks
                 {peter.skands@monash.edu}\\[4mm]
 Theoretical Physics, CERN, 1211 Geneva 23, Switzerland\\
 School of Physics \& Astronomy, Monash University, Clayton VIC 3800, Australia\\}
\date{Updated: July, 2017}
\maketitle
\begin{abstract}
These lectures were originally given at TASI 2012 and 
are directed at a level suitable for 
graduate students in High Energy
Physics. They are intended to give an introduction to 
the theory and phenomenology of quantum  chromodynamics (QCD),
focusing on collider physics applications. The aim is to bring the
reader to a level where informed decisions can be made 
concerning different approaches and their uncertainties. 
The material is divided into five main
areas:  1)~fundamentals, 2)~fixed-order perturbative QCD, 
3)~Monte Carlo event generators and parton showers, 4)~Matching at
Leading and Next-to-Leading Order, and 5)~Hadronization and soft
hadron-hadron physics.
\vspace*{1cm}
\begin{center}
{\bf Useful Complementary References}\\[2mm]
\begin{itemize}
\item Basic Quantum Field Theory: \cite{Peskin:1995ev}
\item Textbooks on QCD: \cite{Field:1989uq,Ellis:1991qj,Dissertori:2003pj}
\item Jets and Jet Algorithms: \cite{Salam:2010zt}
\item General-Purpose Event Generators: \cite{Buckley:2011ms}
\item The String Model: \cite{Andersson:1983ia,Andersson:1998tv}
\item Underlying Event and Multiple Parton Interactions: \cite{Sjostrand:1987su,Sjostrand:2017cdm}
\item Step-by-step PYTHIA tutorial: see ``worksheet'' available on the \href{http://home.thep.lu.se/Pythia/}{PYTHIA homepage}.
\item Monte Carlo methods and Random Numbers:
  \cite{James:1980yn,Weinzierl:2000wd}
\item The veto algorithm, trial showers, the Sudakov algorithm~\cite{Platzer:2011dq,Lonnblad:2012hz,Mrenna:2016sih}
\end{itemize}
\end{center}
\end{abstract}

\clearpage
\tableofcontents
\clearpage%

\section{Introduction \label{sec:introduction}}

When probed at very short wavelengths, QCD is essentially a theory of
free \index{Partons}`partons' --- quarks and gluons --- which only
scatter off one another through relatively small quantum corrections,
that can be systematically calculated. 
But at longer wavelengths, of order the size of the proton $\sim
1\mathrm{fm} = 10^{-15}\mathrm{m}$,  
we see strongly bound towers of hadron resonances emerge, with string-like
potentials building up if we try to separate their partonic
constituents. Due to our
inability to perform analytic calculations in 
strongly coupled field theories, QCD is therefore 
still only partially solved. Nonetheless,  all its features, across all
distance scales, are believed to be encoded in a single one-line
formula of alluring simplicity; the
\index{QCD!Lagrangian}%
Lagrangian\footnote{Throughout these notes we let it be implicit that
  ``Lagrangian'' really refers to Lagrangian density, ${\cal L}$, the
  four-dimensional space-time integral of which is the action.} of QCD.

The consequence for collider physics is that some parts of QCD can be
calculated in terms of the fundamental parameters of the Lagrangian,
whereas others must be expressed through models or functions whose effective 
parameters are not a priori calculable but which can be constrained
by fits to data. 
However, even in the absence of a
perturbative expansion, there are still several strong theorems which
hold, and which can be used to give relations between seemingly
different processes. (This is, e.g., the reason it makes sense to 
measure the partonic substructure of the proton in $ep$ collisions and
then re-use the same parametrisations for $pp$
collisions.) Thus, in the chapters 
dealing with phenomenological models we shall emphasise that the loss
of a factorised perturbative expansion is not equivalent to a total
loss of predictivity.   

An alternative approach would be to give up on calculating QCD 
and use leptons instead. Formally, this amounts to summing inclusively over
strong-interaction phenomena, when such are present. While such a
strategy might succeed in replacing what we do know about QCD by
``unity'', however, even the most adamant chromophobe would acknowledge
the following basic facts of collider physics for the next decade(s): 
1) At the LHC, the initial states are
hadrons, and hence, at
the very least, well-understood and precise parton distribution
functions (PDFs) will be required; 2) high precision will mandate
 calculations to higher orders in perturbation theory, 
which in turn will involve more QCD; 3) the requirement of lepton
\emph{isolation} makes the very definition of a lepton
 depend implicitly on QCD and 4) 
 the rate of jets that are misreconstructed as leptons in
 the experiment depends explicitly on it. 
And, 5) though many new-physics signals \emph{do} give observable
signals in the lepton sector, this is far from guaranteed, nor is it
exclusive when it occurs. 
 It would therefore be  unwise not to attempt to solve QCD to the best
 of our ability, the better to prepare ourselves for both the largest
 possible discovery reach and the highest attainable subsequent
 precision. 

Furthermore, QCD is the richest gauge theory we have so far
 encountered. Its emergent phenomena, unitarity properties, colour structure, 
 non-perturbative dynamics, quantum vs.\ classical limits, 
interplay between scale-invariant and
 scale-dependent properties, and its wide
 range of phenomenological applications, are still very much topics of
 active investigation, about which we continue to learn.  

In addition, or perhaps as a consequence, the field of QCD is
currently experiencing something of a revolution. On the perturbative
side, new methods to compute scattering amplitudes with very high
particle multiplicities are being developed, together with advanced
techniques for combining such amplitudes with all-orders resummation
frameworks. On the non-perturbative side, the wealth of data on
soft-physics processes from the LHC is
forcing us to reconsider the reliability of the standard fragmentation
models, and heavy-ion collisions are providing new insights into
the collective behavior of hadronic matter. The
study of cosmic rays impinging on the Earth's
atmosphere challenges our ability to extrapolate fragmentation models
from collider energy scales to the region of ultra-high energy cosmic
rays. And finally, dark-matter annihilation processes in space  may produce 
hadrons, whose spectra are sensitive to the modeling 
of fragmentation.

In the following, we shall focus on QCD for mainstream 
collider physics. This includes the basics of SU(3), colour factors, the running
of $\alpha_s$, factorisation, 
hard processes, infrared safety, parton showers and matching, event generators, hadronisation, and the so-called underlying event. 
While not covering everything, hopefully these topics can also serve
at least as stepping stones to more specialised
issues that have been left out, such as twistor-inspired techniques, 
heavy flavours, polarisation, or forward physics, or to topics more tangential to
other fields, such as axions, lattice QCD, or heavy-ion physics.  

\subsection{A First Hint of Colour}
\begin{figure}[t]
\begin{center}
\begin{tabular}{c}
\colorbox{gray}{\includegraphics*[scale=0.75]{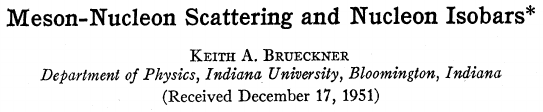}}\\[5mm]
\hspace*{2mm}\begin{minipage}{0.88\textwidth}
\small\sl  ``[...] It is concluded that the apparently anomalous features of the
scattering can be interpreted to be an indication of a resonant
meson-nucleon interaction corresponding to a nucleon isobar with spin
$\frac32$, isotopic spin $\frac32$, and with an excitation energy of
$277\,$MeV.''\\[1mm]
\end{minipage}
\end{tabular}
\caption{The title and part of the abstract of the 1951 paper
  \cite{Brueckner:1952zz} (published in 1952) in which the existence 
  of the $\Delta^{++}$ baryon was deduced, based on data from Sachs and
  Steinberger at Columbia~\cite{Chedester:1951sc}  and from Anderson,
  Fermi, Nagle, et al.~at Chicago~\cite{Fermi:1952zz}. Further studies 
  at Chicago were quickly performed
  in~\cite{Anderson:1952nw,Anderson:1952zza}. See also the memoir by
  Nagle~\cite{nagle1984delta}. 
\label{fig:Delta}}  
\end{center}
\end{figure}

Looking for new physics, as we do now at the LHC, it is instructive to 
consider the story of the discovery of colour. The first hint was
arguably the $\Delta^{++}$ \index{Baryons}baryon, discovered in 
1951~\cite{Brueckner:1952zz}. The title and part of the abstract from this
historical paper are reproduced in \figRef{fig:Delta}.
In the context of the \index{Quarks}quark model --- which first
had to be developed, successively joining together the notions of 
spin, isospin, strangeness, and 
the \index{Eightfold way}eightfold way\footnote{In physics, the ``eightfold way''
refers to the classification of the lowest-lying pseudoscalar
\index{Mesons}mesons and 
\index{SU(3)!Of Flavour}%
spin-1/2 \index{Baryons}baryons within \index{Octet}octets in SU(3)-flavour space ($u,d,s$). The
$\Delta^{++}$ is part of a spin-3/2 baryon \index{Decuplet}decuplet, a ``tenfold way'' in this
terminology.} 
--- the \index{Flavour}flavour and spin content of the $\Delta^{++}$
baryon is: 
\begin{equation}
\left\vert \Delta^{++} \right> = \left\vert
\,u_\uparrow\ u_\uparrow\ u_\uparrow \right>~,
\end{equation} 
clearly a highly symmetric configuration. However, since 
the $\Delta^{++}$ is a fermion, it must have an overall
antisymmetric wave function. In 1965, fourteen years after its
discovery, this was finally understood by the introduction of colour
\index{SU(3)}%
\index{SU(3)!Of Colour}%
as a new quantum number associated with the group SU(3)
\cite{Greenberg:1964pe,Han:1965pf}. The $\Delta^{++}$ wave function can now be made
antisymmetric by arranging its three quarks antisymmetrically 
in this new degree of freedom, 
\begin{equation}
\left\vert \Delta^{++} \right> = \epsilon^{ijk} \left\vert
\,u_{i\uparrow}\ u_{j\uparrow}\ u_{k\uparrow}\right>~,
\end{equation} 
hence solving the mystery.

More direct experimental tests of the number of colours were provided first by
measurements of the decay width of $\pi^0\to \gamma\gamma$ decays, which 
is proportional to $N_C^2$, 
and later by the famous ``R'' ratio in
$e^+e^-$ collisions ($R=\sigma(e^+e^-\to q\bar{q})/\sigma(e^+e^-\to
\mu^+\mu^-)$), which is proportional to $N_C$, see
e.g.~\cite{Dissertori:2003pj}. 
Below, in \SecRef{sec:L} we shall see how to
calculate such colour factors. 

\subsection{The Lagrangian of QCD \label{sec:L}}
\index{QCD!Lagrangian}%
Quantum Chromodynamics is based on the gauge group
\index{SU(3)}$\mrm{SU(3)}$, the 
Special Unitary group in 3 (complex) dimensions, whose elements 
are the set of unitary $3\times 3$ matrices with determinant one. 
\index{Fundamental representation}%
\index{SU(3)!Fundamental representation}%
Since there are 9 linearly independent unitary complex
matrices\footnote{A complex $N\times N$ matrix has $2N^2$ degrees of
  freedom, on which unitarity provides $N^2$ constraints.}, one of
which has determinant $-1$, there are a total of 8
independent directions in this matrix space, corresponding to eight
different generators as compared
with the single one of QED. In the context of QCD, we normally
represent this group using the 
so-called \emph{fundamental}, or \emph{defining}, representation, in
which the generators of $\mrm{SU(3)}$ appear as a set of eight traceless and
hermitean matrices, to which we return below.  
We shall refer to indices enumerating
the rows and columns of these matrices  (from 1 to 3) as
\emph{fundamental} indices, and we use the letters $i$,
$j$, $k$, \ldots, to denote them.
\index{Adjoint representation}%
\index{SU(3)!Adjoint representation}%
We refer to indices enumerating the generators (from 1 to 8),
as \emph{adjoint} 
indices\footnote{The dimension of the \emph{adjoint}, or
  \emph{vector}, representation is equal to the number of generators,
  $N^2-1=8$ for $\mrm{SU(3)}$, while the  
\index{Fundamental representation}%
\index{SU(3)!Fundamental representation}%
dimension of the fundamental representation is
  the degree of the group, $N=3$ for $\mrm{SU(3)}$.}, and we use the first
letters of the alphabet ($a$, $b$, $c$, \ldots) to denote them. 
These matrices can operate both on each other (representing
combinations of successive gauge transformations) and on a set of
$3$-vectors, the latter of 
which represent \index{Quarks}quarks in colour 
space; the quarks are \emph{triplets} under $\mrm{SU(3)}$. The matrices can be
thought of as representing gluons in colour 
space (or, more precisely, the gauge transformations carried out by
gluons), hence there are
eight different gluons; the gluons are \emph{octets} under $\mrm{SU(3)}$. 

\index{QCD!Lagrangian}%
The Lagrangian density of QCD is 
\begin{equation}
{\cal L} = \bar{\psi}_q^i(i\gamma^\mu)(D_\mu)_{ij}\psi_q^j - m_q
\bar{\psi}_q^i\psi_{qi} - \frac14 F^a_{\mu\nu}F^{a\mu\nu}~,\label{eq:L}
\end{equation}
where $\psi_q^i$ denotes a quark field with
(fundamental) colour index $i$, 
$\psi_q = ({\textcolor{red}{\psi_{qR}}},{\color{green}\psi_{qG}}, 
{\color{blue}\psi_{qB}})^T$, 
$\gamma^\mu$ is a Dirac matrix that expresses the
vector nature of the strong interaction, with $\mu$ being a Lorentz
vector index, $m_q$ allows for the
possibility of non-zero \index{Quarks}quark masses (induced by the
standard Higgs 
mechanism or similar), $F^a_{\mu\nu}$ is the gluon field strength 
tensor for a gluon\footnote{The definition of the gluon field strength
  tensor will be given below in \eqRef{eq:F}.} with (adjoint) 
colour index $a$ (i.e., $a\in[1,\ldots,8]$), 
and $D_\mu$ is the covariant derivative in QCD,
\begin{equation}
(D_{\mu})_{ij} = \delta_{ij}\partial_\mu - i g_s t_{ij}^a A_\mu^a~,\label{eq:D}
\end{equation}
\index{QCD!Coupling}
with $g_s$ the \index{alphaS@$\alpha_s$}strong coupling (related to
$\alpha_s$ by $g_s^2 = 4\pi 
\alpha_s$; we return to the strong coupling in more detail below), 
$A^a_\mu$  the gluon field with 
colour index $a$, and $t_{ij}^a$ proportional to the hermitean and
traceless \index{Gell-Mann matrices|see{SU(3)}}Gell-Mann matrices of $\mrm{SU(3)}$, 
\index{SU(3)!Generators}%
\begin{equation}
\mbox{\includegraphics*[scale=1.0]{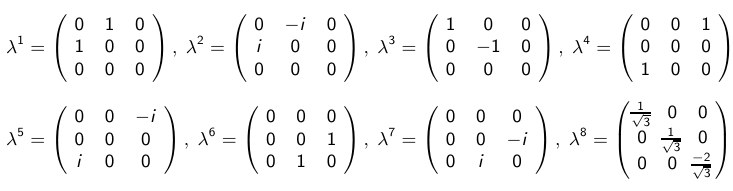}}~.
\end{equation}
These generators are just the $\mrm{SU(3)}$ analogs of the
Pauli matrices in 
$\mrm{SU(2)}$. 
By convention, the constant of proportionality is normally
taken to 
be 
\begin{equation}
t^a_{ij} = \frac12 \lambda^a_{ij}~. \label{eq:t}
\end{equation}
\index{QCD!Coupling}
This choice in turn determines the normalisation of the coupling
$g_s$, via \eqRef{eq:D}, and
fixes the values of the $\mrm{SU(3)}$ \index{Casimirs}Casimirs and structure constants, to which we return below. 

\begin{figure}[t]
\begin{center}
\begin{minipage}[h]{4.6cm}
\begin{center}
$A^1_\mu$\\
\includegraphics*[scale=0.75]{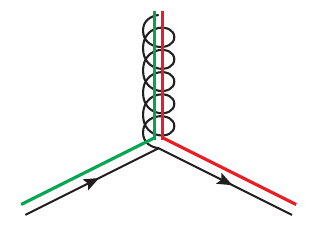}\\[-3mm]
$\psi_{q\textcolor{green}{G}}$\hfill$\psi_{q\textcolor{red}{R}}$
\end{center}
\end{minipage}~~~
\parbox{0.4\textwidth}{
$
\begin{array}{ccccc}
\propto & - \frac{i}{2} g_s & \bar{\psi}_{q\color{red}R}  & \lambda^{1} & \psi_{q\color{green}G} 
\\[2mm]
= & -\frac{i}{2}g_s & \left(\begin{array}{ccc} \textcolor{red}{1} & \color{green} 0 &
  \color{blue} 0 
\end{array}\right) & 
\left(\begin{array}{ccc}
0 & 1 & 0  \\
1 & 0 & 0 \\
0 & 0 & 0
\end{array}\right) & 
 \left(\begin{array}{c}
\textcolor{red}{0} \\
\color{green}1 \\
\color{blue}0
\end{array}\right) \end{array}
$}
\caption{Illustration of a 
\index{Quarks}\index{Gluons}$qqg$ vertex in QCD, before
  summing/averaging over colours: a gluon in a state represented by $\lambda^1$
  interacts with quarks in the states $\psi_{qR}$ and
  $\psi_{qG}$. \label{fig:qg}}
\end{center}
\end{figure}

An example of the colour flow for a
quark-gluon interaction in colour 
space is given in \figRef{fig:qg}.
Normally, of course, we sum over all the colour indices, so this
example merely gives a pictorial representation of what one particular
(non-zero) term in the colour sum looks like.

\subsection{Colour Factors}
\index{QCD!Colour factors}
\index{Colour factors}%
\index{Colour-space indices|see{Colour connections}}%
\index{Matrix elements}%
Typically, we do not measure colour in the final state ---
instead we average over all possible incoming colours and sum over all
possible outgoing ones, wherefore QCD scattering amplitudes (squared) in
practice always contain sums over quark fields contracted with
\index{SU(3)!Generators}Gell-Mann matrices. These contractions in turn
produce traces  
which yield the \index{Colour factors}\emph{colour factors} that are associated to each QCD
process, and which basically count the number of ``paths through
colour space'' that the process at hand can take\footnote{The
  convention choice represented by \eqRef{eq:t} introduces a
  ``spurious'' factor of 2 for each power of the coupling $\alpha_s$. 
Although one could in principle absorb that factor into a redefinition
of the coupling, effectively redefining the normalisation of ``unit
colour charge'', the standard definition of $\alpha_s$ is now so
entrenched that alternative choices would be counter-productive, at
least in the context of a pedagogical review.}.

A very simple example of a colour factor is given by the decay process $Z\to
q\bar{q}$. This vertex contains a simple $\delta_{ij}$ in colour
space; the outgoing quark and antiquark must have identical 
(anti-)col\-ours. Squaring the corresponding matrix element and summing over
final-state colours yields a colour factor of
\begin{equation}
e^+e^-\to Z \to q\bar{q}~~~:~~~\sum_{\mrm{colours}}|M|^2 \propto
\delta_{ij}\delta_{ji} = \mrm{Tr}\{\delta\} = N_C = 3~,
\end{equation}
since $i$ and $j$ are quark (i.e., 3-dimensional
fundamental) indices. This factor corresponds directly to the 3 different
``paths through colour space'' that the process at hand can take; the
produced quarks can be red, green, or blue. 

A next-to-simplest example is given by $q\bar{q}\to
\gamma^*/Z\to\ell^+\ell^-$ (usually referred to as the
\index{Drell-Yan}Drell-Yan 
process~\cite{Drell:1970wh}),  
which is just a crossing of the previous one. By crossing
symmetry, the squared matrix element, including the colour factor, is
exactly the same as before, but since the quarks are here incoming, we
must \emph{average} rather than sum over their colours, leading to
\begin{equation}
q\bar{q}\to Z\to e^+e^-~~~:~~~\frac{1}{9}\sum_{\mrm{colours}}|M|^2 \propto \frac19\delta_{ij}\delta_{ji} = \frac19 \mrm{Tr}\{\delta\} = \frac13~,
\end{equation}
where the colour factor now expresses a \emph{suppression} which can
be interpreted as due to the fact that only quarks of matching colours
are able to collide and produce a $Z$ boson. The chance that a quark
and an antiquark picked at random from the colliding hadrons have 
matching colours is $1/N_C$. 
\begin{figure}[t]
\end{figure}

\begin{figure}[t]
\centering\vspace*{-8mm}
\begin{tabular}{ccc}
\rotatebox{360}{\includegraphics*[scale=0.93]{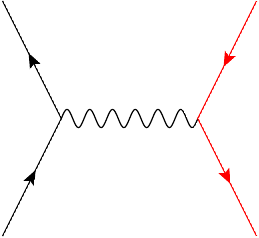}} \ \ 
& \ \ \includegraphics*[scale=0.93,angle=180,origin=c]{ee2qq}
\ \ & \ \ \includegraphics*[scale=0.9,angle=297,origin=c]{ee2qq}\\
Hadronic $Z$ decay & \index{Drell-Yan}Drell-Yan & \index{DIS}DIS \\[1mm]
$e^-e^+ \to \gamma^*/Z^0 \to q\bar{q}$ &
$q\bar{q} \to \gamma^*/Z^0 \to \ell^+\ell^-$ &
$\ell \bar{q} \stackrel{\gamma^*/Z^*}{\to} \ell \bar{q}$
\\[2mm] 
$\propto N_C$ & $\propto 1/N_C$ & $\propto 1$
\end{tabular}
\caption{Illustration of the three crossings of the interaction of a
  lepton current (black) with a \index{Quarks}quark current (red) 
  via an intermediate photon or
  $Z$ boson, with corresponding colour factors. \label{fig:Zcrossings}}
\end{figure}

Similarly, $\ell q \to
\ell q$ via $t$-channel photon exchange (usually called Deep
Inelastic Scattering --- \index{DIS}\index{Deep inelastic scattering|see{DIS}}DIS --- with ``deep'' referring to a 
large virtuality of the exchanged photon), constitutes yet another
crossing of the same basic process, 
see \figRef{fig:Zcrossings}. \index{Colour factors}The colour factor in this case 
comes out as unity. 

To illustrate what happens when we insert (and sum over)
quark-gluon
vertices, such as the one depicted in \figRef{fig:qg}, we take
the process $Z\to3\,$jets. \index{Colour factors}The colour factor for
this process can be 
computed as follows, with the accompanying illustration showing a
corresponding diagram (squared) with explicit colour-space indices on
each vertex:\\
\index{Colour connections}
\begin{equation}
\mbox{
\begin{tabular}{cc}
\parbox{5.2cm}{
$Z \to qg\bar{q}$~~~:~~~\\
\[
\begin{array}{rcl}
\displaystyle\sum_{\mrm{colours}}|M|^2 & \propto & \displaystyle
\delta_{ij}t_{jk}^a t_{k\ell
    }^a\delta_{\ell i} \\
& = & \displaystyle
\mrm{Tr}\{t^at^a\}\\[4mm] & = & \displaystyle
  \frac12\mrm{Tr}\{\delta\} = 4~,
\end{array}
\]}
&
\parbox{8.5cm}{\includegraphics*[scale=0.6]{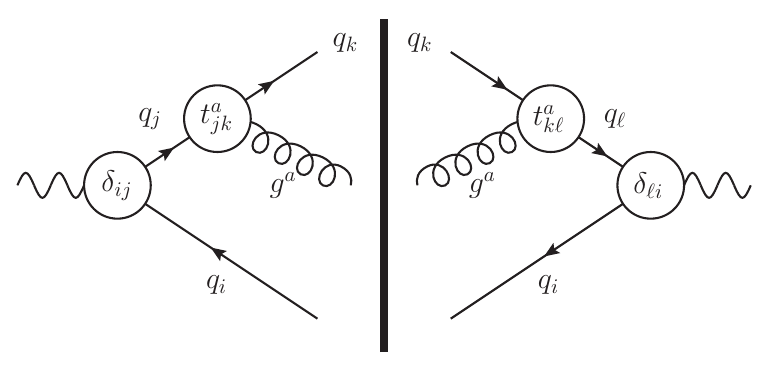}
}
\end{tabular}}
\end{equation}
where the last $\mrm{Tr}\{\delta\} = 8$, since the trace runs over
the 8-dimensional adjoint indices. If we
want to ``count the paths through colour space'', we should leave out
the factor $\frac12$ which comes from the normalisation convention for
the $t$ matrices, \eqRef{eq:t}, hence this process can take 8
different paths through colour space, one for each gluon basis state.

The tedious task of taking traces over $t$
matrices can be greatly alleviated by use of the relations given in
\TabRef{tab:lambda}.  
\index{Traces in SU(3)|see{SU(3)}}%
\index{SU(3)!Trace relations}%
\index{QCD!Trace relations|see{SU(3)}}%
\begin{table}
\begin{center}
\scalebox{1.04}{\begin{tabular}{ccc}
\toprule
\index{SU(3)!Trace relations}Trace Relation & Indices & Occurs in Diagram Squared
\\
\midrule
$\mrm{Tr}\{t^at^b\} = T_R\, \delta^{ab}$ & $a,b\in[1,\ldots,8]$
& \parbox[c]{4cm}{\includegraphics*[scale=0.5]{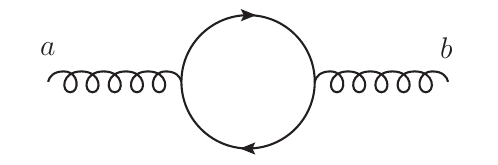}}\\
$\sum_a t^a_{ij}t^a_{jk} = C_F\, \delta_{ik}$ &%
\parbox[c]{3cm}{\begin{center}
$a\in[1,\ldots,8]$\\
$i,j,k\in[1,\ldots,3]$\end{center}}
& \parbox[c]{4cm}{\includegraphics*[scale=0.5]{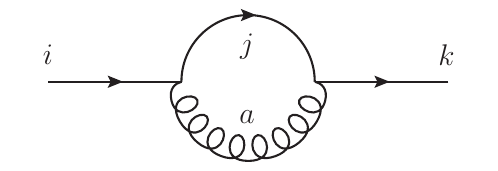}}\\
$\sum_{c,d} f^{acd} f^{bcd} = C_A\, \delta^{ab}$ & $a,b,c,d\in[1,\ldots,8]$
& \parbox[c]{4cm}{\includegraphics*[scale=0.5]{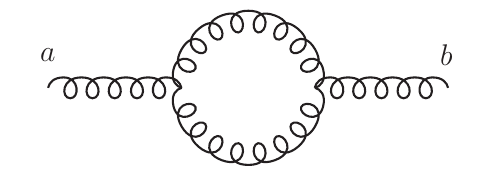}}\\
$ t^a_{ij}t^a_{k\ell} = T_R \left(\delta_{jk}\delta_{i\ell}
- \frac{1}{N_C}\delta_{ij}\delta_{k\ell}\right)$ & $i,j,k,\ell\in[1,\ldots,3]$
& \parbox[c]{4cm}{\includegraphics*[scale=0.5]{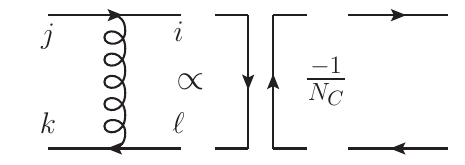}}\hspace*{-0.2cm}(Fierz)\\
\bottomrule
\end{tabular}}
\caption{Trace relations for $t$ matrices (convention-independent). 
 More relations
  can be found in \cite[Section 1.2]{Ellis:1991qj} and in 
  \cite[Appendix A.3]{Peskin:1995ev}.
\label{tab:lambda}}
\end{center}
\end{table}
In the standard normalisation convention for the \index{SU(3)}$\mrm{SU(3)}$ generators,
\eqRef{eq:t}, the \index{Casimirs}Casimirs of $\mrm{SU(3)}$ appearing in
\TabRef{tab:lambda} are\footnote{See, e.g., \cite[Appendix
    A.3]{Peskin:1995ev} for how to obtain the Casimirs in other
  normalisation conventions. As an example, choosing $t^a_{ij} = \lambda_{ij}^a/\sqrt{2}$ would yield $T_R=1$, $C_F=T_R(N_C^2-1)/N_C=8/3$, $C_A=3$.} 
\index{Casimirs}\index{TR@$T_R$}\index{CA@$C_A$}\index{CF@$C_F$}
\begin{equation}
T_R = \frac12 \hspace*{2cm} C_F = \frac43 \hspace*{2cm} C_A = N_C = 3~.
\end{equation}
In addition, the gluon self-coupling on the third line in
\TabRef{tab:lambda} involves factors of $f^{abc}$. These
\index{QCD!Structure constants|see{SU(3)}}%
are called the \index{SU(3)!Structure constants}\emph{structure constants} of QCD and they enter via 
the non-Abelian term in the \index{Gluons}gluon field strength tensor appearing in
\eqRef{eq:L}, 
\begin{equation}
F^a_{\mu\nu} = \underbrace{\partial_\mu A_\nu^a - \partial_\nu
  A^a_\mu}_{\mathrm{Abelian}} +
\underbrace{ g_s f^{abc} A_\mu^b A_\nu^c}_{\mathrm{non-Abelian}}~. \label{eq:F}
\end{equation}

\noindent\begin{minipage}[t]{0.46\textwidth}
The structure constants of $\mrm{SU(3)}$ are listed in the table to the
right. They define the \emph{adjoint}, or \emph{vector}, representation of $\mrm{SU(3)}$
and are related to the fundamental-representation generators via the
commutator relations
\begin{equation}
t^at^b - t^bt^a = [t^a,t^b] = i f^{abc} t_c~,
\end{equation} 
or equivalently,
\begin{equation}
if^{abc}~=~2\mrm{Tr}\{t^c[t^a,t^b]\}~.
\end{equation}
Thus, it is a matter of choice whether one prefers to express colour
space on a basis of fundamental-representation $t$ matrices, or via
the structure constants $f$, and one can go back and forth between the
two.
\end{minipage}%
\hfill%
\colorbox{darkgray}{%
\colorbox{lightgray}{%
\begin{minipage}[t]{0.46\textwidth}
\vspace*{3mm}\begin{center}
\textbf{Structure Constants of SU(3)}
\begin{equation}
f_{123} = 1
\end{equation}
\begin{equation}
f_{147} = f_{246} = f_{257} = f_{345} = \frac12
\end{equation}
\begin{equation}
f_{156} = f_{367} = -\frac12
\end{equation}
\begin{equation}
f_{458} = f_{678} = \frac{\sqrt{3}}{2}
\end{equation}
Antisymmetric in all indices\\[3mm]
All other $f_{abc}=0$\vspace*{3mm}\\
\end{center}
\end{minipage}%
}}\vskip1mm

\begin{figure}[t]
\begin{center}
\begin{minipage}[h]{4.6cm}
\begin{center}
$A_\nu^4(k_2)$\\
\includegraphics*[scale=0.75]{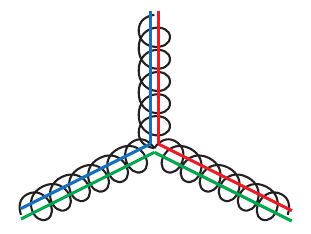}\\[-3mm]
$A^6_\rho(k_1)$\hfill$A_\mu^2(k_3)$
\end{center}
\end{minipage}~~~
\parbox{0.35\textwidth}{
$
\begin{array}{cccc}
\propto & - g_s \ f^{246} \!\! & \!\! [ (k_3 - k_2)^\rho g^{\mu\nu}  \\ 
& & +(k_2 - k_1)^\mu g^{\nu\rho} \\ 
& &+(k_1 - k_3)^\nu g^{\rho\mu}]
\end{array}
$}\vspace*{1mm}
\caption{Illustration of a \index{Gluons}$ggg$ vertex in QCD, before
  summing/averaging over colours: interaction between gluons in the 
  states $\lambda^2$, $\lambda^4$, and $\lambda^6$ is represented by
  the structure constant $f^{246}$. 
\label{fig:gg}}
\end{center}
\end{figure}
 Expanding the $F_{\mu\nu}F^{\mu\nu}$ term of the
Lagrangian using \eqRef{eq:F}, we see that there is a 3-gluon and a
4-gluon vertex that involve $f^{abc}$, the latter of which has two
powers of $f$ and two powers of the coupling. 

Finally, the last line of \TabRef{tab:lambda} is not really a trace
relation but instead a useful so-called Fierz transformation, which
expresses products of $t$ matrices in terms of Kronecker $\delta$ functions. 
It is often used, for instance, in shower Monte Carlo
applications, to assist in mapping between colour flows in $N_C = 3$,
in which cross sections and splitting probabilities are calculated, 
and those in $N_C\to\infty$ (``leading colour''), used to represent colour flow in
the MC ``event record''.

A \index{Gluons}gluon self-interaction vertex is
illustrated in \figRef{fig:gg}, to be compared with the quark-gluon
one in \figRef{fig:qg}. We remind the reader that gauge boson
self-interactions are a hallmark of non-Abelian theories and that their
presence leads to some of the main differences between QED and
QCD. One should also keep in mind 
that the \index{Colour factors}colour factor for the vertex in \figRef{fig:gg}, \index{CA@$C_A$}$C_A$, 
is roughly twice as large as that for a quark, \index{CF@$C_F$}$C_F$.

\subsection{The Strong Coupling \label{sec:coupling}}
\index{QCD!Coupling}
\index{Jets}
\index{alphaS@$\alpha_s$}To first approximation, QCD is 
\index{QCD!Scale invariance}\emph{scale invariant}. That is, if one
``zooms in'' on a QCD jet, one will find a repeated self-similar 
pattern of jets within jets within jets, reminiscent of
fractals. 
In the context of QCD, this property was originally 
called \index{Lightcone scaling|see{QCD Scale invariance}}light-cone scaling, or 
\index{Bjorken scaling|see{QCD Scale invariance}}Bj{\o}rken scaling. 
This type of scaling is closely related to the class of
angle-preserving symmetries, called \index{Conformal
invariance}\emph{conformal} symmetries. In physics 
today, the terms ``conformal'' and ``scale invariant'' are used 
interchangeably\footnote{Strictly speaking, conformal symmetry is more
restrictive than just scale invariance, but examples of
scale-invariant field theories that are not conformal are rare.}.
Conformal invariance is a mathematical property of several
QCD-``like'' theories which are now being studied (such as $N=4$
supersymmetric relatives of QCD). It is also 
related to the physics of so-called ``unparticles'', though that is a
relation that goes beyond the scope of these lectures.

Regardless of the labelling, 
if the  \index{alphaS@$\alpha_s$}strong coupling did not run (we shall
return to the running 
of the coupling below), Bj{\o}rken scaling would be absolutely true. QCD
would be a theory with a fixed coupling, the same at all scales. 
This simplified picture already captures some of the most important
properties of QCD, as we shall discuss presently.  

\index{QCD!Scale invariance}%
In the limit of exact Bj{\o}rken scaling --- QCD at fixed coupling
--- properties of high-energy interactions are determined 
only by \emph{dimensionless} kinematic quantities, such as scattering
angles (pseudorapidities) and ratios of energy
scales\footnote{Originally, the observed approximate agreement with
this was used as a powerful argument
for pointlike substructure in hadrons; since measurements at different
energies are sensitive to different resolution scales, independence of the absolute
energy scale is indicative of the absence of other fundamental
scales in the problem and hence of pointlike constituents.}.
For applications of QCD to high-energy collider physics, an important
consequence of Bj{\o}rken scaling is thus that the rate of 
\index{Parton showers}%
\index{Bremsstrahlung|see{Parton showers}}
bremsstrahlung
jets, with a given transverse momentum, scales in direct proportion to
the hardness 
of the fundamental partonic scattering process they are produced in
association with. This agrees well with our intuition about accelerated
charges; the harder you ``kick'' them, the harder the radiation they
produce.  

For instance, in the limit of exact scaling, a
measurement of the rate of 10-GeV jets produced in association with an
ordinary $Z$ 
boson could be used as a direct prediction of the rate of 100-GeV jets
that would be 
produced in association with a 900-GeV $Z'$ boson, and so 
forth. Our intuition about how many bremsstrahlung jets a given type of
process is likely to have should therefore be governed first and
foremost by the \emph{ratios} of scales that appear in that particular
process, as has been  highlighted in a number of studies focusing on
the mass and $p_\perp$ scales appearing, e.g., in
Beyond-the-Standard-Model (BSM) 
physics processes
\cite{Plehn:2005cq,Alwall:2008qv,Papaefstathiou:2009hp,Krohn:2011zp}. 
\index{QCD!Scale invariance}Bj{\o}rken scaling 
\index{Scale invariance|see{QCD}}
is also fundamental to the understanding of jet substructure in QCD, see, e.g.,
\cite{Vermilion:2011nm,Altheimer:2012mn}.  

\index{alphaS@$\alpha_s$!Running coupling}%
On top of the underlying scaling behavior, the running coupling will
introduce a dependence on the absolute scale, implying more radiation
at low scales than at high ones. The running is logarithmic with
\index{alphaS@$\alpha_s$!beta function}%
energy, and is governed by the so-called \emph{beta function}, 
\index{alphaS@$\alpha_s$}
\begin{equation}
Q^2 \frac{\partial \alpha_s}{\partial Q^2} = \frac{\partial
  \alpha_s}{\partial \ln Q^2} =
\beta(\alpha_s)~, \label{eq:running}
\end{equation}
where the function driving the energy dependence, the \index{Beta function}{beta
  function}, is defined as
\begin{equation}
\beta(\alpha_s) = -\alpha_s^2(b_0 +
b_1\alpha_s + b_2\alpha_s^2 + \ldots)~,\label{eq:beta}
\end{equation}
with LO (1-loop) and NLO (2-loop) coefficients
\begin{eqnarray}
b_0 & = & \frac{11C_A - 4 T_R n_f}{12\pi}~,\\[3mm]
b_1 & = & \frac{17C_A^2 - 10 T_R C_A n_f - 6 T_R C_F n_f}{24\pi^2} ~=~
\frac{153-19 n_f}{24\pi^2}~.\label{eq:b}
\end{eqnarray}
In the $b_0$ coefficient, the first term is due to
\index{Gluons!Contribution to beta function}gluon loops while the
second is due to \index{Quarks!Contribution to beta function}quark
ones. Similarly, the first 
term of the $b_1$ coefficient arises from double gluon loops,
while the second and third represent mixed quark-gluon ones. 
At higher loop orders, the $b_i$ coefficients depend explicitly on the
renormalisation scheme that is used. A brief discussion can be found in the
PDG review on QCD~\cite{pdg2012}, with more elaborate ones
contained in \cite{Dissertori:2003pj,Ellis:1991qj}. 
Note that, if there are additional coloured particles beyond the
Standard-Model ones, loops involving those particles enter
 at energy scales above the masses of the
new particles, thus modifying the  \index{alphaS@$\alpha_s$}running of the coupling at high scales. 
This is discussed, e.g., for supersymmetric models in
\cite{Martin:1997ns}. For the running of other SM couplings, see
e.g.,~\cite{Langacker:2010zza}. 

\begin{figure}[t]
\centering
\includegraphics*[scale=0.45]{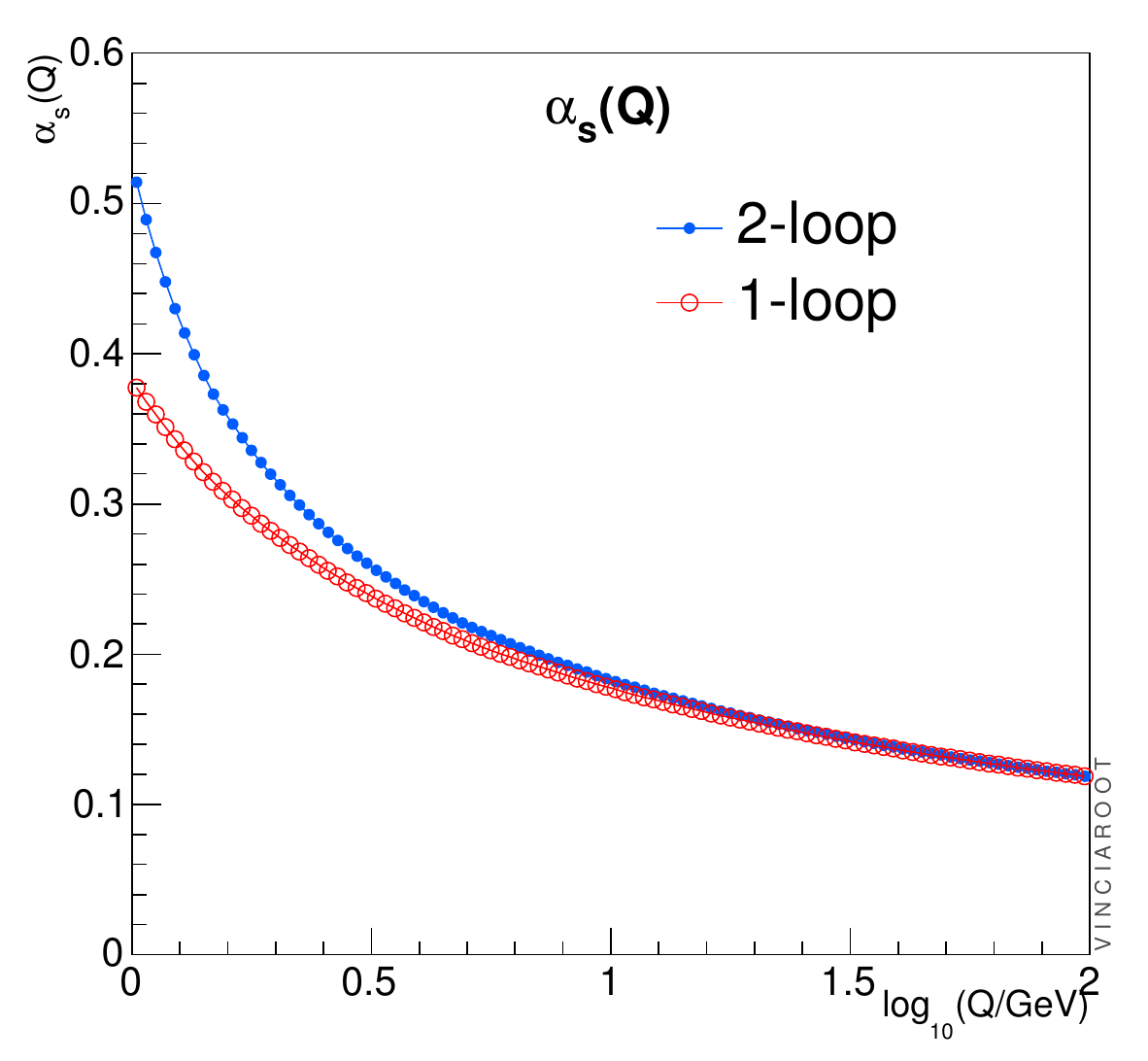}
\caption{Illustration of the running of
 $\alpha_s$ at 1- (open 
  circles) and 2-loop
  order (filled circles), 
starting from the same value of $\alpha_s(M_Z)=0.12$. 
\label{fig:asRun}}
\end{figure}

\index{alphaS@$\alpha_s$!Running coupling}%
Numerically, the value of the  \index{alphaS@$\alpha_s$}strong coupling is usually specified by
giving its value at the specific 
reference scale $Q^2=M^2_Z$, from which we can obtain its
value at any other scale by solving \eqRef{eq:running}, 
\begin{equation}
\alpha_s(Q^2) = \alpha_s(M_Z^2) \frac{1}{1+b_0
  \alpha_s(M_Z^2)\ln\frac{Q^2}{M_Z^2} + {\cal O}(\alpha_s^2)}~,
\label{eq:alphaq2}
\end{equation}
with relations including the ${\cal O}(\alpha_s^2)$ terms 
available, e.g., in \cite{Ellis:1991qj}. 
Relations between scales 
not involving $M_Z^2$ can obviously be obtained by just replacing $M_Z^2$
by some other scale $Q'^2$ everywhere in \eqRef{eq:alphaq2}. A
comparison of running at one- and two-loop order, in both cases starting from
$\alpha_s(M_Z)=0.12$, is given in \figRef{fig:asRun}.
As is evident from the figure, the 2-loop running is somewhat faster
than the 1-loop one.

\index{alphaS@$\alpha_s$!Running coupling}%
As an application, let us prove that the 
logarithmic running of the coupling implies that an intrinsically 
multi-scale problem can be converted to a single-scale one, up to
corrections suppressed by two powers of $\alpha_s$, 
by taking the geometric mean of the scales involved. This follows from
expanding an arbitrary product of individual  \index{alphaS@$\alpha_s$}$\alpha_s$ factors around an
arbitrary scale $\mu$, using \eqRef{eq:alphaq2}, 
\begin{eqnarray}
\alpha_s(\mu_1)\alpha_s(\mu_2)\cdots\alpha_s(\mu_n) & = &
\prod_{i=1}^{n} \alpha_s(\mu) \left(1 +
b_0\,\alpha_s\ln\left(\frac{\mu^2}{\mu_i^2}\right) + {\cal O}(\alpha_s^2)\right)
\nonumber\\[2mm]
& = & \alpha_s^n(\mu) \left(1 + b_0\, \alpha_s \ln \left(
 \frac{\mu^{2n}}{\mu_1^2\mu_2^2\cdots\mu_n^2}\right) +  {\cal
   O}(\alpha_s^2) \right)~,
\end{eqnarray}
whereby the specific single-scale choice $\mu^n =
\mu_1\mu_2\cdots\mu_n$ (the geometric mean) can
be seen to push the difference between the two sides of the equation one order higher
than would be the case for any other combination of scales\footnote{In
  a fixed-order calculation, the individual scales $\mu_i$,
would correspond, e.g., to the $n$ hardest scales appearing in an infrared
safe sequential clustering algorithm applied to the given momentum
configuration.}. 

The appearance of the number of \index{Flavour}flavours, $n_f$, in $b_0$ implies that the
slope of the running depends on the number of contributing
\index{Flavour}flavours. Since full QCD is best approximated by $n_f=3$
below the charm threshold, by $n_f=4$ and $5$ from there to the $b$
and $t$ thresholds, respectively, and then by $n_f=6$ at scales
higher than $m_t$, it is therefore important to be aware that 
the running changes slope across quark \index{Flavour}flavour
thresholds. Likewise, it would change across the threshold for any coloured
new-physics particles that might exist, with a magnitude depending on
the particles' colour and spin quantum numbers.

\index{alphaS@$\alpha_s$!Running coupling}%
\index{alphaS@$\alpha_s$}
The negative overall sign of \eqRef{eq:beta}, combined with the fact
that $b_0 > 0$ (for $n_f \le 16$), leads to the famous
result\footnote{
Perhaps the highest pinnacle of fame for \eqRef{eq:beta} was reached
when the sign of it featured in an episode of the TV series ``Big Bang
Theory''.} 
that the QCD coupling effectively \emph{decreases} with
 energy, called \index{Asymptotic freedom}asymptotic 
freedom, for the discovery of which the \index{Nobel prize}Nobel prize in physics was
awarded to D.~Gross, H.~Politzer, and F.~Wilczek in 2004. An extract
of the prize announcement runs as follows:
\begin{center}
\begin{minipage}{0.84\textwidth}
\sl  What this year's Laureates discovered was something that, at
first sight, seemed completely contradictory. The interpretation of
their mathematical result was that the closer the quarks are to each
other, the \emph{weaker} is the ``colour charge''. When the quarks are
really close to each other, the force is so weak that they behave
almost as free particles\footnote{More correctly, it is the coupling
  rather than the  
  force which becomes weak as the distance decreases. 
  The $1/r^2$ Coulomb singularity of the force is only dampened, not removed, 
  by the diminishing coupling.}. 
This phenomenon is called ``asymptotic
freedom''. The converse is true when the quarks move apart: the force
becomes stronger when the distance increases\footnote{More correctly,
 it is the potential which grows, linearly, while the force becomes
 constant.}. 
\end{minipage}
\end{center}

\begin{figure}[t]
\begin{center}\hspace*{-0.25cm}
\parbox[c]{3.1cm}{\includegraphics*[scale=0.65]{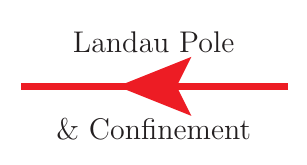}}
\parbox[c]{8cm}{\includegraphics*[scale=0.5]{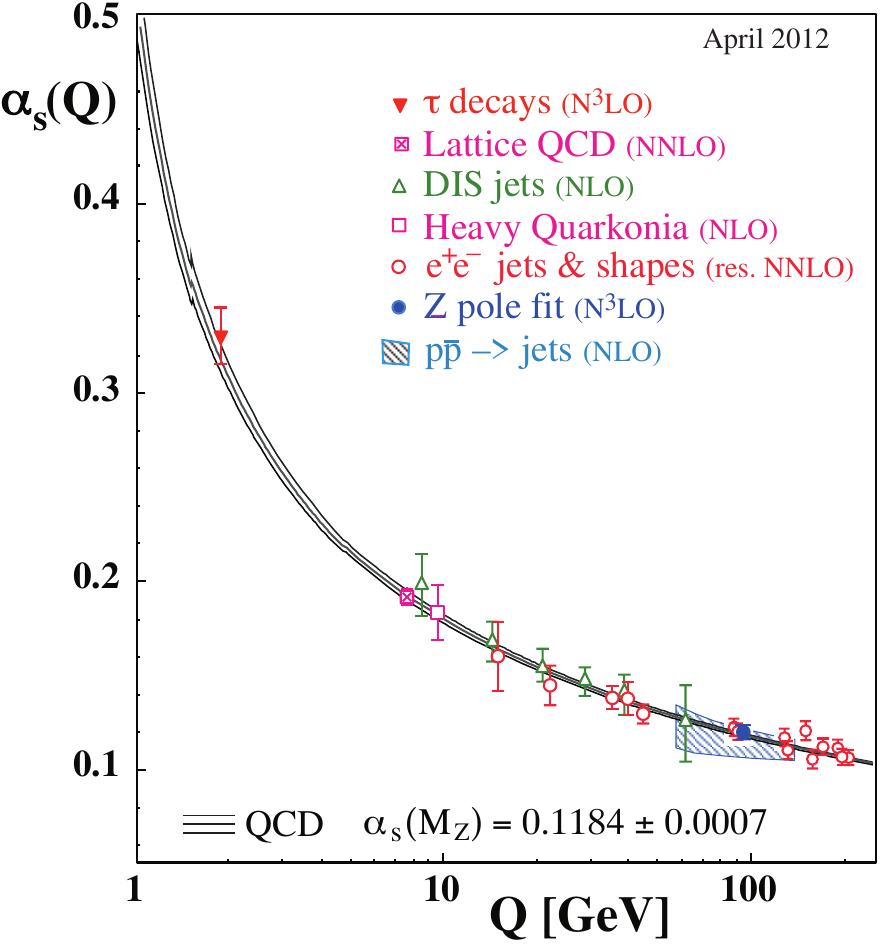}}\hspace*{-1mm}
\parbox[c]{3.1cm}{\includegraphics*[scale=0.65]{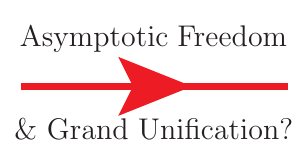}}
\caption{Illustration of the running of $\alpha_s$ in a theoretical
  calculation (band) and in physical processes at
  different characteristic scales, from
  \cite{pdg2012,Bethke:2012jm}. The little kinks at $Q=m_{c}$ and
  $Q=m_b$ are
  caused by discontinuities in the running across the flavour
  thresholds.\label{fig:alphas}}  
\end{center}           
\end{figure}

\index{Running coupling|see{alphaS@$\alpha_s$}}%
\index{alphaS@$\alpha_s$!Running coupling}%
Among the consequences of \index{Asymptotic freedom}asymptotic freedom is that perturbation
theory becomes better behaved at higher absolute energies, due to the
effectively decreasing coupling. Perturbative calculations for our
900-GeV $Z'$ boson from before should therefore be slightly faster
converging than equivalent calculations for the 90-GeV one. 
Furthermore, since the running of  \index{alphaS@$\alpha_s$}$\alpha_s$ explicitly
breaks Bj{\o}rken scaling, we also expect to see small changes in jet
shapes and in jet production ratios as we vary the energy. For
instance, since high-$p_\perp$ jets
start out with a smaller effective coupling, their intrinsic shape
(irrespective of boost effects) is
somewhat narrower than for low-$p_\perp$ jets, an issue which can be
important for jet calibration. Our current understanding of the
running of the QCD coupling is summarised by the plot in
\figRef{fig:alphas}, taken from a recent comprehensive review by S.\ Bethke
\cite{pdg2012,Bethke:2012jm}. A complementary up-to-date overview of
$\alpha_s$ determinations can be found in~\cite{d'Enterria:2015toz}. 

\index{alphaS@$\alpha_s$!Running coupling}%
As a final remark on \index{Asymptotic freedom}asymptotic freedom, note
that the decreasing 
value of the  \index{alphaS@$\alpha_s$}strong coupling with energy must eventually cause it to
become comparable to the electromagnetic and weak ones, at some energy
scale. Beyond that point, which may lie at energies of order
$10^{15}-10^{17}\,$GeV (though it may be lower if as yet undiscovered
particles generate large corrections to the running), 
we do not know  what the further evolution of the combined theory will 
actually look like, or whether it will continue to exhibit
\index{Asymptotic freedom}asymptotic
freedom. 

\index{alphaS@$\alpha_s$}%
\index{alphaS@$\alpha_s$!Running coupling}%
\index{alphaS@$\alpha_s$!LambdaQCD@$\Lambda_{\mathrm{QCD}}$}%
Now consider what happens when we run the coupling in the other
direction, towards smaller energies. 
Taken at face value, the numerical value of the coupling diverges
rapidly at scales below 1 GeV, as illustrated by the curves
disappearing off the left-hand edge of the plot in
\figRef{fig:alphas}. To make this divergence
explicit, one can rewrite
\eqRef{eq:alphaq2} in the following form, 
 \index{alphaS@$\alpha_s$}
\begin{equation}
\alpha_s(Q^2) = \frac{1}{b_0 \ln \frac{Q^2}{\Lambda^2}}~,\label{eq:alphasLam}
\end{equation}
where 
\begin{equation}
\Lambda \sim 200\, \mbox{MeV}
\end{equation}
\index{alphaS@$\alpha_s$!LambdaQCD@$\Lambda_{\mathrm{QCD}}$}%
\index{alphaS@$\alpha_s$!Landau Pole|see{$\Lambda_{\mathrm{QCD}}$}}%
\index{LambdaQCD@$\Lambda_{\mathrm{QCD}}$|see{alphaS@$\alpha_s$}}%
specifies the energy scale at which the perturbative coupling would nominally become
infinite, called the Landau pole. (Note, however, that this only
parametrises the purely \emph{perturbative} result, which is not
reliable at \index{Strong coupling}strong coupling, so \eqRef{eq:alphasLam} should 
not be taken to imply that the physical behavior of full QCD should
exhibit a divergence for $Q\to \Lambda$.) 

\index{alphaS@$\alpha_s$}%
\index{alphaS@$\alpha_s$!Running coupling}%
\index{alphaS@$\alpha_s$!LambdaQCD@$\Lambda_{\mathrm{QCD}}$}%
Finally, one should be aware that there is a multitude of different
ways of defining both $\Lambda$ and $\alpha_s(M_Z)$. At the very
least, the numerical value one obtains depends both on the
renormalisation scheme used (with the dimensional-regularisation-based
``modified minimal subtraction'' scheme, $\overline{\mbox{MS}}$, being the
most common one) and on the perturbative order of the calculations 
used to extract them. As a rule of thumb, fits to experimental data typically yield 
smaller values for $\alpha_s(M_Z)$ the higher the order of the
calculation used to extract it (see, e.g.,
\cite{Bethke:2009jm,Dissertori:2009ik,Bethke:2012jm,pdg2012}), with  $
\alpha_s(M_Z)\vert_{\mrm{LO}} \gsim \alpha_s(M_Z)\vert_{\mrm{NLO}}
\gsim \alpha_s(M_Z)\vert_{\mrm{NNLO}}$. 
Further, since the number of \index{Flavour}flavours changes the slope
of the running, the location of the Landau pole for fixed
$\alpha_s(M_Z)$ depends explicitly on the number of \index{Flavour}flavours used in
the running. Thus each value of $n_f$ is associated with its own
value of $\Lambda$, with the following matching relations across
thresholds guaranteeing continuity of the coupling at one loop,
\index{LambdaQCD@$\Lambda_{\mathrm{QCD}}$|see{$\alpha_s$}}
\index{alphaS@$\alpha_s$!LambdaQCD@$\Lambda_{\mathrm{QCD}}$}%
\begin{eqnarray}
n_f = 5 \leftrightarrow 6 ~~~:~~~~~~\Lambda_6 = \Lambda_5
  \left(\frac{\Lambda_5}{m_t}\right)^{\frac{2}{21}} & & 
\Lambda_5 = \Lambda_6
  \left(\frac{m_t}{\Lambda_6}\right)^{\frac{2}{23}} ~, \\[2mm]
n_f = 4 \leftrightarrow 5 ~~~:~~~~~~\Lambda_5 = \Lambda_4
  \left(\frac{\Lambda_4}{m_b}\right)^{\frac{2}{23}} & & 
\Lambda_4 = \Lambda_5
  \left(\frac{m_b}{\Lambda_5}\right)^{\frac{2}{25}} ~, \\[2mm]
n_f = 3 \leftrightarrow 4 ~~~:~~~~~~\Lambda_4 = \Lambda_3 
  \left(\frac{\Lambda_3}{m_c}\right)^{\frac{2}{25}} & &
\Lambda_3 = \Lambda_4 
  \left(\frac{m_c}{\Lambda_4}\right)^{\frac{2}{27}} ~.
\end{eqnarray}

\index{alphaS@$\alpha_s$}%
\index{alphaS@$\alpha_s$!Running coupling}%
It is sometimes stated that QCD only has a single free
parameter, the  \index{alphaS@$\alpha_s$}strong coupling. 
However, even in the perturbative
region, the beta function depends explicitly on the number of
quark \index{Flavour}flavours, as we have seen, and thereby also on the quark
masses. Furthermore, in the non-perturbative region around or below
$\Lambda_{\mrm{QCD}}$, the value of the 
perturbative coupling, as obtained, e.g., from \eqRef{eq:alphasLam},
gives little or no insight into the behavior of the full theory. 
Instead, universal functions (such as parton densities, form factors,
fragmentation functions, etc), effective theories (such as the
Operator Product Expansion, Chiral Perturbation Theory, or Heavy Quark
Effective Theory), or phenomenological models (such as Regge Theory or
the String and Cluster Hadronisation Models) must be used, which in
turn depend on additional non-perturbative parameters whose relation to, e.g.,
$\alpha_s(M_Z)$, is not a priori known. 

\index{Lattice QCD}
For some of these questions,
such as hadron masses, lattice QCD can furnish important
additional insight, but for multi-scale and/or time-evolution
problems, the applicability of lattice methods is still severely
restricted; the lattice formulation of QCD requires 
  a Wick rotation to
  Euclidean space. The time-coordinate can then be treated on an
  equal footing with the other dimensions, but intrinsically
  Minkowskian problems, such as the time evolution of a system, are
   inaccessible. The limited size of current lattices
  also severely constrain the scale hierarchies that it is possible to
  ``fit'' between the lattice spacing and the lattice size. 

\index{Landau pole|see{$\alpha_s$}}%
\index{QCD!Landau Pole|see{$\alpha_s$}}%
\index{Renormalisation|see{$\alpha_s$}}%
\index{QCD!Renormalisation|see{$\alpha_s$}}%

\subsection{Colour States}
\index{Coherence}%
A final example of the application of the underlying $\mrm{SU(3)}$ group
theory to QCD is given by considering which colour states we can
obtain by combinations of quarks and gluons. The simplest example of
this is the combination of a quark and antiquark. We can form a total
of nine different colour-anticolour combinations, which fall into two
irreducible representations of $\mrm{SU(3)}$:
\begin{equation}
3 \otimes \overline{3} = 8 \oplus 1~.\label{eq:33bar}
\end{equation}
The singlet corresponds to the symmetric wave function 
$\frac{1}{\sqrt{3}}\left(\left|R\bar{R}\right>+\left|G\bar{G}\right>+\left|B\bar{B}\right>\right)$, 
which is invariant under $\mrm{SU(3)}$ transformations (the definition of a
singlet). The other eight linearly independent 
combinations (which can be represented by one for each Gell-Mann
matrix, with the singlet corresponding to the identity matrix) transform
into each other under $\mrm{SU(3)}$. Thus, although we sometimes talk about
colour-singlet states as 
being made up, e.g., of ``red-antired'', that is not quite precise
language. The actual state $\left|R\bar{R}\right>$ is \emph{not} a
pure colour singlet.  Although it does
have a non-zero \emph{projection} onto the singlet wave function
above, it also has non-zero projections onto the two members of
the octet that correspond to the diagonal Gell-Mann
matrices. Intuitively, one can also easily realise this by noting that
an $\mrm{SU(3)}$ rotation of $\left|R\bar{R}\right>$ would in general turn it into a
different state, say $\left|B\bar{B}\right>$, whereas a true colour singlet
would be invariant. 
Finally, we can also realise from \eqRef{eq:33bar} that a random
(colour-uncorrelated) quark-antiquark pair has a $1/N^2=1/9$ 
chance to be in an overall colour-singlet state; otherwise it is in
an octet. 

Similarly, there are also nine possible quark-quark (or
antiquark-antiquark) combinations, six of which are symmetric
under interchange of the two quarks and three of which are antisymmetric:
\index{Sextet}%
\begin{equation}
6 ~=~ \left(\begin{array}{c}
\left|RR\right>\\
\left|GG\right>\\
\left|BB\right>\\
\frac{1}{\sqrt{2}}\left(\left|RG\right> + \left|GR\right>\right)\\
\frac{1}{\sqrt{2}}\left(\left|GB\right> + \left|BG\right>\right)\\
\frac{1}{\sqrt{2}}\left(\left|BR\right> + \left|RB\right>\right)
\end{array}\right)
~~~~~~~~~
\bar{3} = \left(\begin{array}{c}
\frac{1}{\sqrt{2}}\left(\left|RG\right> - \left|GR\right>\right)\\
\frac{1}{\sqrt{2}}\left(\left|GB\right> - \left|BG\right>\right)\\
\frac{1}{\sqrt{2}}\left(\left|BR\right> - \left|RB\right>\right)
\end{array}\right)~.
\end{equation}
The members of the sextet transform into (linear combinations of) 
each other under $\mrm{SU(3)}$ transformations, and similarly for the
members of the antitriplet, hence neither of these can be reduced
further. The breakdown into
irreducible $\mrm{SU(3)}$ multiplets is therefore
\begin{equation}
3 \otimes 3 = 6 \oplus \overline{3}~.
\end{equation}
Thus, an uncorrelated pair of quarks has a $1/3$ chance to add to an overall
anti-triplet state (corresponding to coherent
superpositions like ``red + green = antiblue''\footnote{In the context of
  hadronisation models, 
  this coherent superposition of two quarks in an overall antitriplet
  state is sometimes called a
  \index{Diquarks}``diquark'' (at low $m_{qq}$)
  \index{String junctions}or a ``string junction'' (at high $m_{qq}$), see
  \secRef{sec:stringModel}; it corresponds to the antisymmatric ``red
  + green = antiblue'' combination needed to create a baryon
  wavefunction. }); otherwise it is in an overall 
sextet state. 

Note that the emphasis on
the quark-(anti)quark pair being \emph{uncorrelated} is important;
production processes that correlate the produced partons, like $Z\to q\bar{q}$ or $g\to q\bar{q}$, will
project out specific components (here the singlet and octet,
respectively). 
Note also that, if the quark
and (anti)quark are on opposite sides of the universe (i.e., living in
two different hadrons), the QCD \emph{dynamics} will not care what
overall colour state they 
are in, so for the formation of multi-partonic states in QCD, obviously the
spatial part of the wave functions (causality at the very least) 
will also play a role. Here, we are considering \emph{only} the colour part
of the wave functions. 
Some additional examples are 
\begin{eqnarray}
8\otimes 8 & = & 27 \oplus 10 \oplus \overline{10} \oplus 8 \oplus 8
\oplus 1 ~,\\ 
3 \otimes 8 & = & 15 \oplus 6 \oplus 3~,\\
3 \otimes 6 & = & 10 \oplus 8~,\\
3\otimes3\otimes3 & = & (6 \oplus \overline{3}) \otimes 3 = 10 \oplus 8
\oplus 8 \oplus 1 ~.
\end{eqnarray}
Physically, the 27 in the first line corresponds to a completely
incoherent addition of the colour charges of two gluons;
\index{Decuplet}the decuplets are slightly more coherent (with a lower
total colour charge), the octets
yet more, and the singlet corresponds to the combination of two gluons
that have precisely equal and opposite colour charges, so that their
total charge is zero. 
Further extensions and generalisations of these combination rules can
\index{Young tableaux}be obtained, e.g., using the method of Young
tableaux~\cite{young1901,youngSagan}.

\clearpage
\section{Hard Processes\label{sec:pQCD}}
\index{Matrix elements}%
Our main tool for solving QCD at high energy scales, 
$Q\gg \Lambda_{\mrm{QCD}}$, is
perturbative quantum field theory, the starting point for which is
Matrix Elements (MEs) which can be calculated systematically at fixed
orders (FO) in the strong coupling $\alpha_s$. 
At least at lowest order (LO), the procedure is
standard textbook material \cite{Peskin:1995ev} and it has also by now been
highly automated, by the advent of tools like 
\Mg~\cite{Alwall:2011uj}, 
\Ca~\cite{Pukhov:2004ca} / 
\Co~\cite{Boos:2004kh}, and several others
\cite{Kanaki:2000ey,Krauss:2001iv,Moretti:2001zz,Kilian:2007gr,Cafarella:2007pc,Bahr:2008pv,Gleisberg:2008fv}.  
Here, we  require only that the reader has a
basic familiarity with the methods involved from graduate-level particle
physics courses based, e.g., on \cite{Peskin:1995ev,Dissertori:2003pj}. 
Our main concern are the uses to which these calculations 
are put, their limitations, and ways to improve on the results obtained
with them.

\begin{figure}[t]
\center
\begin{tabular}{ccc}
\begin{minipage}[c]{5.5cm}
$p_1$\hfill $p_3$\\[-1mm]
\includegraphics*[scale=0.5]{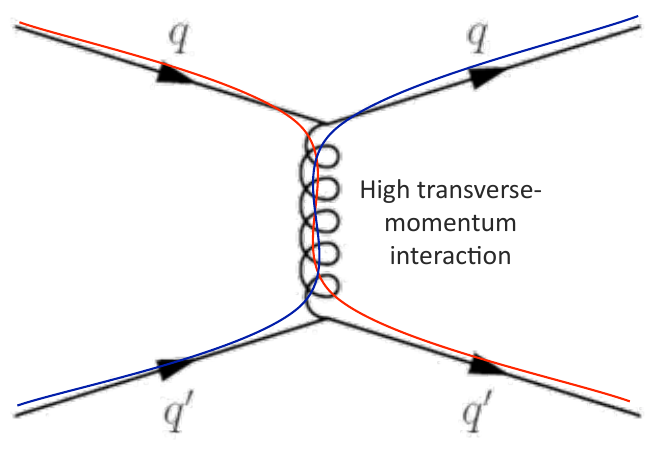}\\[-4.5mm]
$p_2$\hfill $p_4$
\end{minipage}
& ~~~&
\begin{minipage}[c]{7cm}
\includegraphics*[scale=0.3]{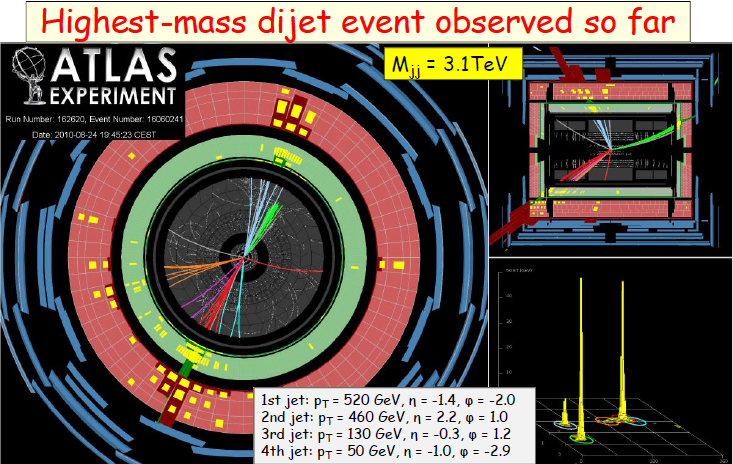}
\end{minipage}
\end{tabular}
\caption{{\sl Left:} Rutherford scattering of quarks in QCD,
  exemplifying the type 
  of process that dominates the short-distance interaction cross section at
  hadron colliders. {\sl Right:} an example of what such a reaction
  looks like in a detector, in this case the ATLAS experiment.
\label{fig:rutherford} }
\end{figure}
For illustration, take one of the
most commonly occurring processes in hadron collisions:
Rutherford scattering of two quarks via a $t$-channel gluon exchange
--- \figRef{fig:rutherford} --- which at leading order 
has the differential cross section
\begin{equation}
qq'\to qq' ~~~:~~~\frac{\dd{\sigma}}{\dd{\hat{t}}} =
\frac{\pi}{\hat{s}^2}\, \frac{4}{9}\,
\alpha_s^2\, \frac{\hat{s}^2 + \hat{u}^2}{\hat{t}^2}~,
\end{equation}
with the $2\to 2$ Mandelstam variables (``hatted'' to emphasize that
they refer to a partonic $2\to 2$ scattering rather than the full
$pp\to\mbox{jets}$
process)
\begin{eqnarray}
\hat{s} & = & (p_1+p_2)^2 ~,\\[1.5mm]
\hat{t} & = &(p_3-p_1)^2 = -\hat{s}\frac{(1-\cos\hat\theta)}{2} ~,\\
\hat{u} & = &(p_4-p_1)^2 = -\hat{s}\frac{(1+\cos\hat\theta)}{2}~.
\end{eqnarray}

Reality, however, is more complicated; the picture on the right-hand pane of
\figRef{fig:rutherford} shows a real dijet event, as recorded by the
ATLAS experiment.  
The complications to be addressed when going from left to right in
\figRef{fig:rutherford} are: 
firstly, additional jets, a.k.a.\ real-emission corrections, 
which can significantly change the topology of the final state, potentially
shifting jets in or out of an experimentally defined acceptance
region. Secondly, loop factors, a.k.a.\ virtual corrections, change
the number of 
 available quantum paths through phase space, and hence modify 
the normalization of the cross section (total \emph{and}
differential). And finally, additional corrections are generated by
confinement and by the so-called  underlying event. These corrections 
must be taken into account to complete our understanding of QCD and
connect the short-distance physics with macroscopic experiments.  
Apart from the perturbative expansion itself, the most powerful tool
we have to organize this vast calculation, is factorization. 

\index{Factorization}%
\subsection{Factorization \label{sec:factorization}}
\begin{figure}[t]
\begin{center}
\includegraphics*[scale=0.55]{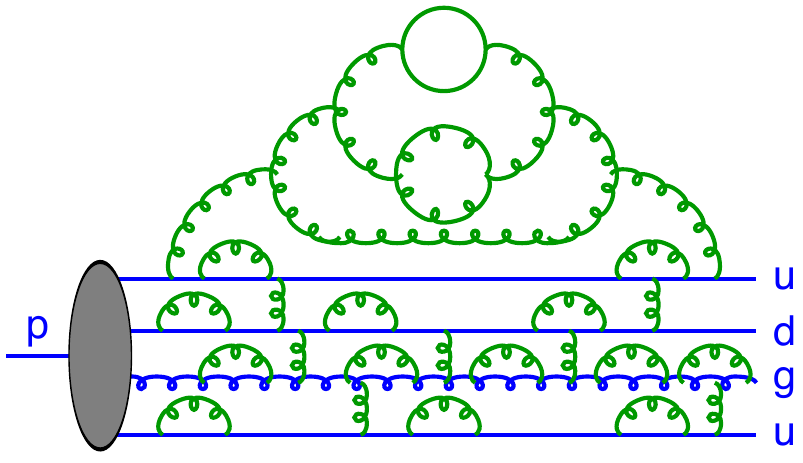}
\caption{Illustration
  of partonic fluctuations inside a proton beam  (from
 \cite{Sjostrand:2006su}). 
\label{fig:pdfs}}
\end{center}
\end{figure}
In high-energy scattering problems involving hadrons in the initial
state, we immediately face the complication that hadrons are composite,
with a time-dependent structure illustrated in
\figRef{fig:pdfs}; there are partons  within clouds of further partons,
constantly being emitted and absorbed. Thus, before we can use
perturbatively calculated partonic scattering matrix elements, we must
first address the partonic structure of the colliding hadron(s). 

For the hadron to remain intact, the fluctuations inside it must involve
momentum transfers smaller than the confinement scale. Indeed,
high-virtuality fluctuations are suppressed by powers of
\begin{equation}
\frac{\alpha_s\, \Lambda^2}{|k|^2}~,
\end{equation}
with $\Lambda$ the confinement scale ($\sim$ 200 MeV,
 see \secRef{sec:coupling}) and $|k|$ the virtuality of the
fluctuation. Thus, most fluctuations occur over timescales 
$\sim 1/\Lambda$. 

\index{DIS}%
A hard perturbative probe, on the other hand, such as the exchanged
photon in DIS (\figRef{fig:Zcrossings}), interacts over a much shorter
timescale $1/Q \ll 1/\Lambda$, during which the partonic fluctuations in
the struck hadron appear almost frozen. The hard probe effectively takes
an instantaneous snapshot of the hadron structure, at a characteristic
resolution given by $\sim 1/Q$.

This is formalized by the \emph{factorization
theorem}~\cite{Collins:1981uw} (see also the TASI lectures by
George Sterman~\cite{Sterman:1995fz}), which expresses the independence of
long-wavelength (soft) structure on the nature of the hard
(short-distance) process. Originally formulated for DIS, factorization
allows us to write the cross section for lepton-hadron scattering as a
convolution of a non-perturbative but universal (i.e.,
process-independent) parton density function (PDF) and a
perturbatively calculable partonic scattering cross section. Denoting
the fraction of the hadron momentum carried by parton $i$ by $x_i$,
\begin{equation}
\vec{p}_i \ = \ x_i\, \vec{p}_h~, \label{eq:x}
\end{equation}
we may write the lepton-hadron cross section on factorized form (see,
e.g., \cite{Brock:1993sz,Dissertori:2003pj}),
\begin{equation}
\sigma_{\ell h} = \sum_i \int_0^1 \dd{x_i}\int\dPS{f}
 \pdf{i/h}(x_i,\mu_F^2)\, \frac{\dd{\hat{\sigma}_{\ell i\to 
 f}(x_i,\PS{f},\mu_F^2)}}{\dd{x_i}\dPS{f}}~, \label{eq:sigmaDis}
\end{equation}
with $i$ an index running over all possible parton
types\footnote{Typically, only quarks and gluons are included in this
sum, but also photons and even leptons can in principle be
included. Similarly, parton density functions are normally used to
describe hadrons, but can also be defined, e.g., to describe the cloud
of virtual photons (and fermion pairs) surrounding an electron.} in the
incoming hadron and $f$ enumerating all possible (partonic) final
states, with Lorentz-invariant phase space, \PS{f}. 

\index{PDFs}%
The \emph{parton density functions} (PDFs), $f_{i/h}$, parametrize the
distribution of partons inside the target hadron. They are not a priori  
calculable and must be constrained by fits to data. This is discussed in
\secRef{sec:pdfs}. 

The \emph{partonic cross section}, $\mathrm{d}{\hat{\sigma}}$, 
knows nothing of the target hadron apart from the fact that it contained
the struck parton. It is calculable within perturbation theory, as will
be discussed in \secRef{sec:fixed-order}.  

\index{Factorization!Factorization scale}%
The dividing line between
the two is drawn at an arbitrary (``user-defined'') scale $\mu_F$,
called the \emph{factorization scale}. 
There is some arbitrariness involved in choosing a value for
$\mu_F$. Some heuristic arguments to guide in the choice of factorization
scale are the following. On the
long-distance side, the PDFs include a (re)summation of fluctuations
inside fluctuations up to virtualities of order $\mu_F$. 
It would therefore not make much sense to take $\mu_F$ 
significantly larger than the scales characterizing resolved particles
on the short-distance side of the calculation (i.e., the particles
appearing explicitly in $\PS{f}$); 
otherwise the PDFs would be including sums over fluctuations that happen
on timescales shorter than those probed by the physical
process. Similarly, $\mu_F$ should also not be taken much lower than the
scale(s) appearing in the hard process. 
\index{Uncertainties!Factorization}
For matrix elements characterized by a single well-defined scale,
such as the $Q^2$ scale in DIS or the
invariant-mass scale $\hat{s}$ in Drell-Yan production
($q\bar{q}\to Z/\gamma^*\to \ell^+\ell^-$), such arguments essentially
fix the preferred scale choice to $\mu_F=Q$ or $\mu_F=\sqrt{\hat{s}}$,
respectively, which may then be varied by a factor
of 2 (or larger) around the nominal value in order to estimate
uncertainties. For multi-scale problems, however, such as $pp\to
Z/W+n\,$jets, there are several a priori equally good choices
available, from the lowest to the highest QCD scales that can be
constructed from the final-state momenta, usually with several
dissenting groups of theorists arguing over which particular choice is
best. Suggesting that one might simply \emph{measure} the scale
would not really be an improvement, as the factorization
scale is fundamentally unphysical and therefore
unobservable (similarly to gauge or convention choices). 
\index{NLO}%
One plausible strategy is to
look at higher-order (NLO or NNLO) calculations, in which correction
terms appear that cancel the dependence on the scale choice, stabilizing the final
result. From such comparisons, a
``most stable'' initial scale choice can in principle be determined,
which then furnishes a 
reasonable starting point, but we  emphasize that the
question \emph{is} intrinsically ambiguous, and no golden recipe
is likely to magically give all the right answers. The best we can do is
to vary the value of $\mu_F$ not only by an overall factor, but also by
exploring different possible forms for its 
functional dependence on the momenta appearing in
\PS{f}. A complementary useful discussion of the pros and cons of different
factorization scale choices can be found in the TASI lectures by
Tilman Plehn~\cite{Plehn:2008zs}. 

\index{NLO}%
Secondly, and more technically, at NLO and beyond one also has to settle on a
\emph{factorization scheme} in which to do the calculations. 
For all practical
purposes, students focusing on LHC physics are only likely to
encounter one such scheme, the modified minimal subtraction
($\overline{\mrm{MS}}$) one already mentioned in the discussion of the
definition of the strong coupling in \SecRef{sec:coupling}. At the
level of these lectures, we shall therefore not elaborate further on this choice
here.   

We note that factorization 
can also be applied multiple times, to break up a complicated calculation
into simpler  pieces that can be treated as approximately independent.
This will be very useful when dealing with successive emissions in a
parton shower, \secRef{sec:parton-showers}, or
when factoring off  decays of long-lived particles from a hard
production process, \secRef{sec:decays}. 

We round off the discussion of factorization by mentioning a few caveats
the reader should be aware of. (See \cite{Sterman:1995fz} for a more
technical treatment.)

\index{Factorization!Caveats}%
\index{Uncertainties!Factorization}%
\index{Twist}%
Firstly, the proof only applies to the first term in an operator
product expansion in ``twist'' = mass dimension - spin. Since
operators with higher mass dimensions are suppressed by the hard scale
to some power, this leading twist approximation becomes exact in the
limit $Q \to \infty$, while at finite $Q$ it neglects corrections of
order 
\begin{equation}
\mbox{Higher Twist :
 }\frac{[\ln(Q^2/\Lambda^2)]^{m<2n}}{Q^{2n}}~~~(n=2~\mbox{for DIS})~.
\end{equation}
In \secRef{sec:soft}, we shall discuss some corrections that go beyond
this approximation, in the context of multiple parton-parton
interactions.

\index{DIS}
\index{Drell-Yan}
\index{Inclusive cross sections}
Secondly, the proof only really applies to inclusive cross sections in
DIS~\cite{Collins:1981uw}  
and in Drell-Yan~\cite{Collins:1984kg}. 
For all other hadron-initiated processes, 
factorization is an ansatz. For a general hadron-hadron process, we
write the assumed factorizable cross section as: 
\begin{equation}
\dd{\sigma_{h_1h_2}} \ = \ \sum_{i,j}\int_0^1\dd{x_i}\int_0^1\dd{x_j}\sum_f\int
 \dPS{f} \pdf{i/h_1}(x_i,\mu_F^2)\,\pdf{j/h_2}(x_j,\mu_F^2)\,
 \frac{\dd{\hat{\sigma}_{ij\to f}}}{\dd{x_i}\dd{x_j}\dd{\Phi_f}}
~. \label{eq:factorization}
\end{equation}
Note that, if $\dd{\hat{\sigma}}$ is divergent (as, e.g., Rutherford
scattering is) then the integral over $\dPS{f}$ must be regulated, 
e.g.\ by imposing some explicit minimal transverse-momentum cut 
and/or other phase-space restrictions.

\subsection{Parton Densities \label{sec:pdfs}}
\index{PDFs}%
\index{Partons}%
\index{Parton Distributions|see{PDFs}}%

The parton density function, $\pdf{i/h}(x_, \mu_F^2)$, represents the effective
density of partons of type/flavor $i$, as a function of the momentum
fraction\footnote{Recall: the $x$ fraction is defined in \eqRef{eq:x}.}
$x_i$, when a hadron 
of type $h$ is probed at the factorization scale $\mu_F$. 
The PDFs are
non-perturbative functions which are not a priori calculable, but a
perturbative differential equation governing their evolution with
$\mu_F$ can be 
obtained by requiring that physical scattering cross sections, such as
the one for DIS in \eqRef{eq:sigmaDis}, be
independent of $\mu_F$ to the calculated
orders~\cite{Altarelli:1977zs}. 
\index{Altarelli-Parisi|see{DGLAP}}
\index{DGLAP kernels}%
\index{DGLAP equation}%
The resulting \emph{renormalization group equation} (RGE) is called the
DGLAP\footnote{DGLAP: 
Dokshitzer-Gribov-Lipatov-Altarelli-Parisi~\cite{Gribov:1972ri,Altarelli:1977zs,Dokshitzer:1977sg}.} equation 
and can be used to ``run'' the PDFs from one perturbative
resolution scale to another (its evolution kernels are the same as those
used in parton showers, to which we return in
\secRef{sec:parton-showers}). 

This means that we only need to determine 
the form of the PDF as a function of $x$ a single (arbitrary) scale,
$\mu_0$. We can then get its form at any other scale $\mu_F$ by simple
RGE evolution.
In the context of PDF fits (constraining
the form of the PDF functions by fitting cross sections to experimental
data, e.g., from DIS~\cite{Mason:2007zz,CooperSarkar:2012tx},
Drell-Yan~\cite{Alekhin:2006zm,deOliveira:2012ji}, and
$pp\to\mbox{jets}$~\cite{Alekhin:2011sk}), 
the reference scale
$\mu_0$ is usually taken to be relatively low, of order one or a 
few GeV.

\begin{figure}[t]
\centering
\includegraphics[scale=0.42]{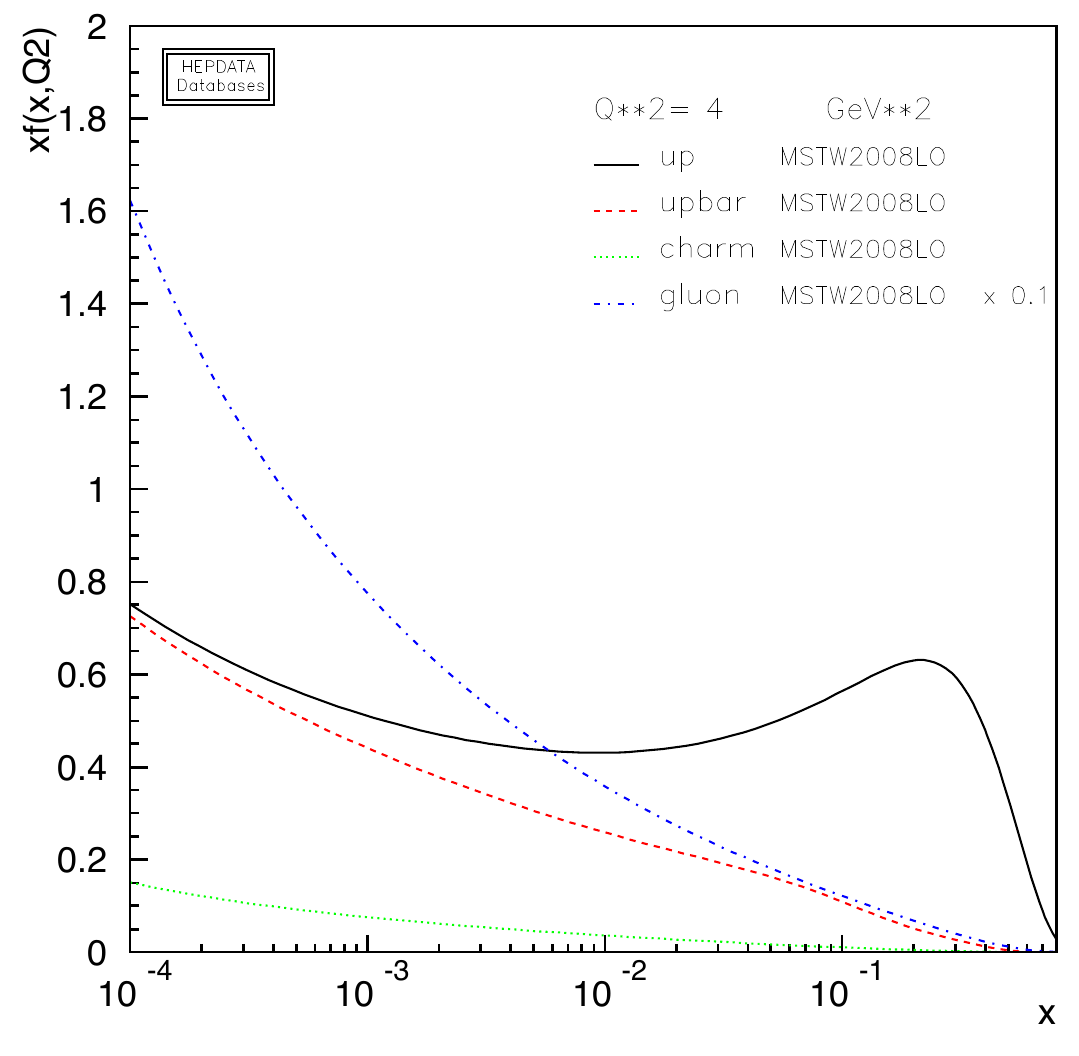} \ 
\includegraphics[scale=0.42]{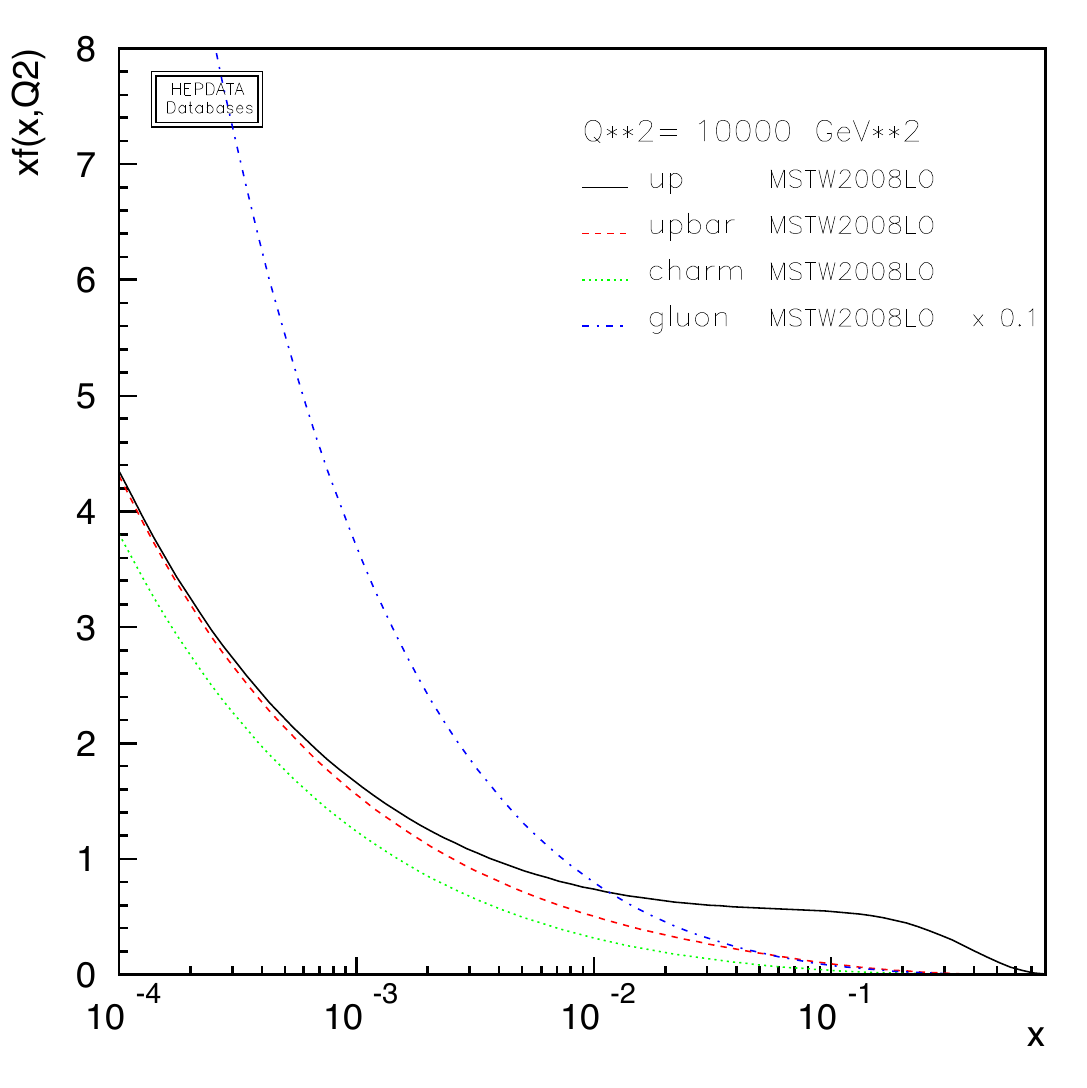}
\caption{Illustration of the change of the $u$ (black), $\bar{u}$ (red, dashed),
 $c$ (green, dotted), and $g$ (blue, dot-dashed) distributions, from 
$Q=\mu_F= 2\,\mathrm{GeV}$ (left) to $Q = \mu_F = 100\,\mathrm{GeV}$ (right). 
Note that a factor 0.1 has been
 applied to the gluon distribution. Plots made using the HEPDATA online
 tool~\cite{Buckley:2010jn}.\label{fig:pdfEvol}}
\end{figure}
\index{Quarks!PDFs}%
\index{Gluons!PDFs}%
The behavior 
of the PDFs as we evolve $\mu_F$ from a low scale, 2 GeV, to a high one, 100
GeV, is illustrated in \figRef{fig:pdfEvol}, 
for the MSTW\footnote{MSTW: Martin-Stirling-Thorne-Watt.} 2008
LO\footnote{The ``LO'' means 
that the fit was performed using LO matrix elements in the cross section
formulae.} PDF set~\cite{Martin:2009iq}. At low
$Q=\mu_F = 2\,\mathrm{GeV}$ (left), the proton structure is dominated by a few hard
quarks (a ``valence bump'' is clearly visible around $x \sim 0.2$), 
while at
higher scales $Q = 100\,\mathrm{GeV}$ (right) we predominantly resolve fluctuations
within fluctuations, yielding increasingly large gluon- and sea-quark
distributions with rather small $x$ values, while the valence quarks play
a progressively smaller role.

\index{Uncertainties!PDF sets}%
We note that different collaborations, like CTEQ, MSTW, NNPDF, etc., use  different 
ans\"atze for the form of $f(x,\mu_0^2)$. They may also include
different data in the fits, and/or treat or weight the data
differently. Thus, results from different groups may not always be
mutually compatible. 
\begin{figure}[t]
\centering
\includegraphics*[scale=0.45]{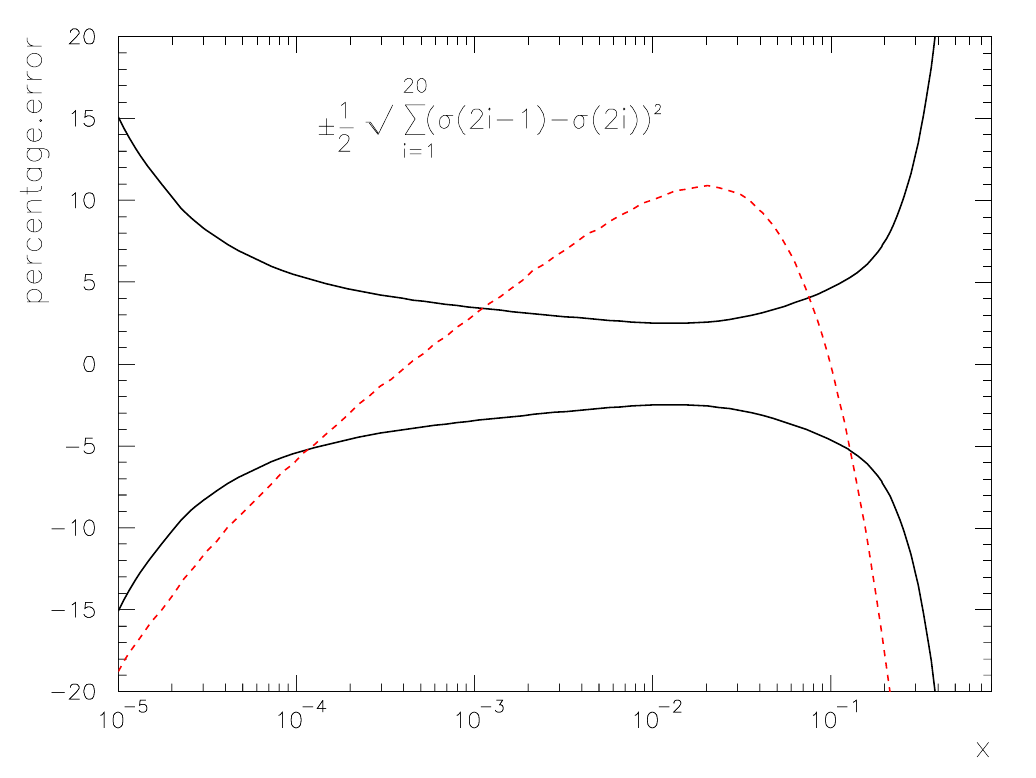}
\caption{Illustration of the difference between the MSTW 2008 and CTEQ6 LO
 gluon PDFs at $\mu_F = 10\,\mathrm{GeV}$. 
All curves are normalized to the central MSTW 2008
 prediction. The black solid lines show the 90\% CL MSTW variations,
 while the dashed red line shows the CTEQ6L1 distribution.\label{fig:pdfUnc}}
\end{figure}
An example is given in \figRef{fig:pdfUnc}, which shows the difference
 between the CTEQ6L1 gluon PDF~\cite{Pumplin:2002vw}~(red dashed) and
 the MSTW 2008 LO 
 one~\cite{Martin:2009iq}, normalized to MSTW  (which would thus
 be a flat line at zero),  at $\mu_F=10\,\mathrm{GeV}$. 
The $y$ axis shows the relative difference between the sets, in per
 cent. Also shown are 
 the 90\% CL contours computed from the uncertainty variations included in the 
MSTW 2008 LO set (black). Using only the MSTW uncertainty band, 
 one would arrive at an estimated $\sim 5\%$ uncertainty over most of
 the $x$ range, while including the CTEQ6L1 set would increase that to
 $\sim 10\%$. At NLO, this discrepancy is reduced, but not removed.
 A significant effort is currently being undertaken within the PDF community 
to agree on common, and more comprehensive, ways of defining PDF uncertainty
bands~\cite{Alekhin:2011sk,Watt:2012tq}. This is complicated due to the
 different ways of defining $f(x,\mu_0^2)$ and due to 
the experimental data sets not always
being fully compatible with one another. For the
time being, it is recommended to try at least sets from two different
groups, for a comprehensive uncertainty estimate. 

\index{Structure functions}%
Occasionally, the words \emph{structure functions} and \emph{parton
densities} 
are
used interchangeably. However, there is an important distinction between
the two, which we find often in (quantum) physics: the former is a
physical observable used to parametrize the DIS cross sections (see
e.g.~\cite{Dissertori:2003pj}), while the latter is a ``fundamental''
quantity extracted from it. In particular, since the parton densities
are not, themselves, physically observable, they can only be defined
within a specific factorization scheme, order by order in perturbation
theory. The only exception is at leading order, at which they have the
simple physical interpretation of parton number densities. When going to
higher orders, we tend to keep the simple intuitive picture from LO in
mind, but one should be aware that the fundamental relationship between
PDFs and measured quantities is now more complicated (due to the
interplay between the PDFs and the real and virtual corrections to the
LO cross section), and that the parton densities no longer have a clear
probabilistic interpretation starting from NLO. 

\begin{figure}[t]
\centering
\includegraphics[scale=0.42]{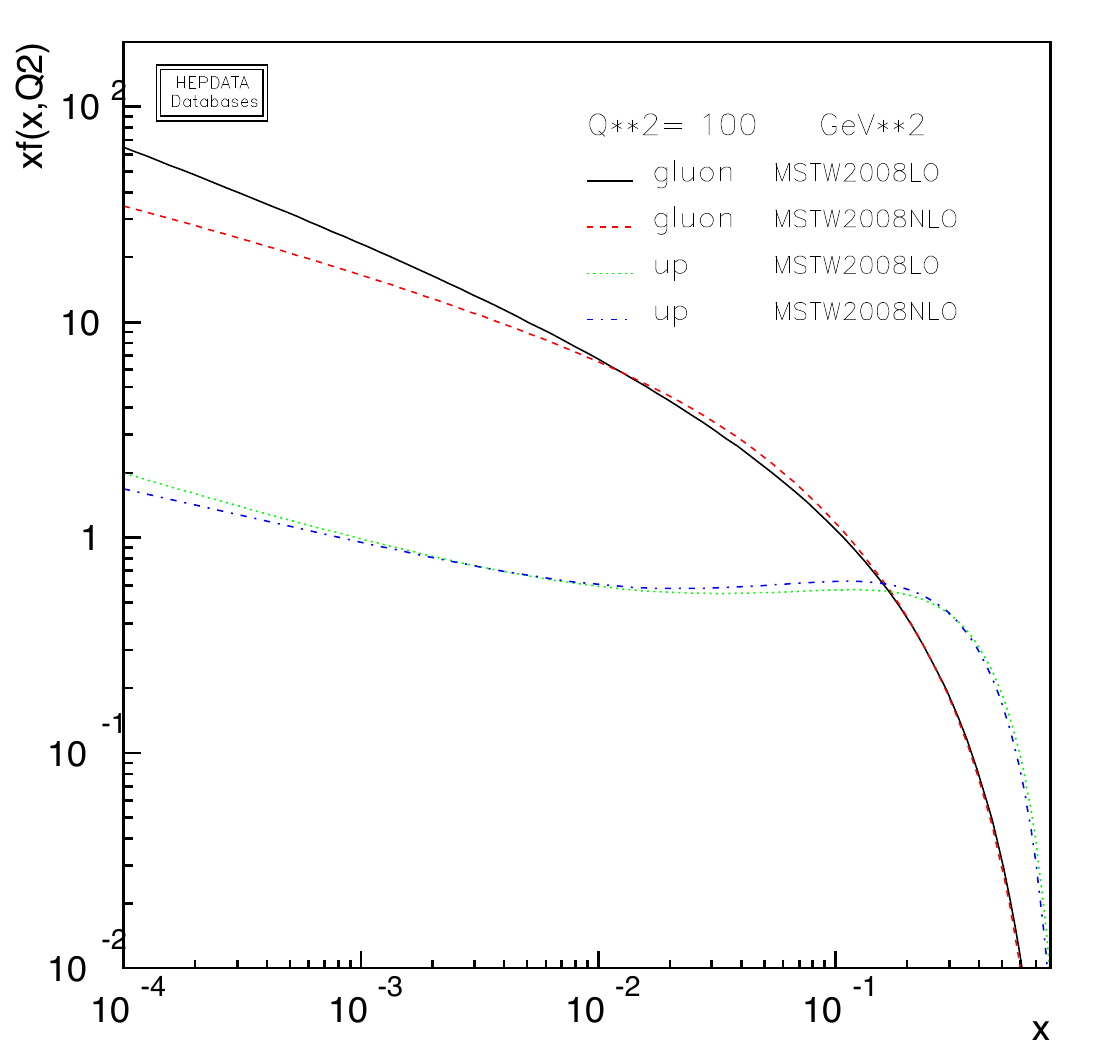}
\caption{Illustration of the change between PDF fits using LO and NLO
 matrix elements: the $g$ distribution at LO (black) and NLO (red,
 dashed), and the $u$ distribution at LO (green, dotted) and NLO (blue,
 dot-dashed), for the MSTW 2008 PDF sets~\cite{Martin:2009iq}, 
at $Q = \mu_F = 10\,\mathrm{GeV}$. 
Plots made using the HEPDATA online 
 tool~\cite{Buckley:2010jn}.\label{fig:pdforders}}
\end{figure}
The reader should also be aware that there is some ambiguity whether 
NLO PDFs should be used for 
 LO calculations. In principle, the higher-order PDFs are better
  constrained and the difference between, e.g., an NLO and an LO set
  should formally be beyond LO precision, so that one might be tempted
  to simply use the highest-order available PDFs for any calculation. 
  However, higher-order terms can sometimes be absorbed, at least partially,
  into effective lower-order  coefficients. In the
  context of PDFs, the fit parameters of lower-order PDFs will 
  attempt to compensate for missing higher-order
  contributions in the matrix 
  elements. To the extent those higher-order contributions are
  \emph{universal}, this is both desirable and
  self-consistent. This leads to some typical qualitative differences
  between LO and NLO PDFs, illustrated in \figRef{fig:pdforders}: 
  NLO PDFs tend to be smaller at low $x$ and
      slightly larger at high $x$, than LO ones. Thus, it is quite
      possible that using an NLO PDF in conjunction with LO matrix
      elements can give a worse agreement with data than LO PDFs do. 

\index{Monte Carlo!Event generators}%
\index{Parton showers}%
\index{Parton showers!Initial-state evolution}%
Finally, another oft-raised question concerns which PDF sets to use for the
  parton-shower evolution in Monte Carlo
  generators. Importantly, the equations driving the 
initial-state showers in Monte Carlo models are only sensitive to 
\emph{ratios} of PDFs~\cite{Bengtsson:1986gz}. Since the shower
  evolution typically 
  only has leading-logarithmic (LL) precision, it should be
  theoretically consistent to use any (LO or better) 
PDF set to drive the
  evolution. However, similarly to above, there will be subleading
  differences between different choices, and one is justified in
  worrying about the level of physical effects that could be generated. 
Unfortunately, there is 
  currently no way to ensure 100\% self-consistency. 
  Since PDF fits are not done with MC codes, but
  instead use analytical resummation models (see, e.g., the TASI
  lectures by Sterman~\cite{Sterman:1995fz}), which are not identical to
  their MC counterparts, the PDF fits are essentially ``tuned'' to a
  slightly different resummation than that incorporated in a given MC
  model. In practice, not much is known about the size and impact
  of this ambiguity~\cite{Gieseke:2004tc}. Known differences include: 
  the size of phase space (purely collinear massless PDF evolution
  vs.\ the finite-transverse-momentum massive  MC phase space),  
  the treatment of momentum conservation and recoil effects,
  additional higher-order effects explicitly or implicitly included in
  the MC evolution, choice of renormalization scheme
  and scale, and, for those MC algorithms that do not rely on
  collinear (DGLAP, see \cite{Dissertori:2003pj}) 
  splitting kernels (e.g., the various kinds of dipole
  evolution algorithms, see \cite{Bern:2008ef}), differences in the
  effective factorization 
  scheme. 

As a baseline, we recommend simply using whatever PDF set the given MC
  model was originally tuned with, since this should de facto  (by
  fitting the available 
  data) reabsorb as much of the inconsistency as
  possible. Furthermore, it should be emphasized that underlying-event
  and minimum-bias models 
  based on multi-parton interactions
  (see \secRef{sec:soft-processes}) usually make the explicit
  assumption 
that 
  the PDFs can be interpreted as physical number densities even 
  down to very low $Q$ and $x$, a property which is generally 
  only true for LO PDFs. It must therefore be strongly discouraged to use 
  (N)NLO PDF sets in this context. 

\subsection{Fixed-Order QCD \label{sec:fixed-order}}
\index{Matrix elements}%

Consider the production of an arbitrary final state, $F$ (e.g., a
Higgs boson, a $t\bar{t}$ pair, etc). 
Schematically, we may  express the (perturbative) all-orders
differential cross section for an observable \obs, 
in the following way:
\begin{equation}
\left.\frac{\dd{\sigma_F}}{\dd{\obs}}\right\vert_{\textcolor{blue}{\mrm{ME}}}
= \underbrace{\sum_{k=0}^\infty\int \dPS{F+k} }_{\Sigma~\mbox{legs}} 
 \Big\vert \underbrace{\sum_{\ell=0}^{\infty} {\cal
   M}_{F+k}^{(\ell)}}_{\Sigma~\mbox{loops}} \Big\vert^2 \,
\delta\left(\obs-\obs(\PS{F+k})\right)~,\label{eq:fixed-order}
\end{equation}
where, for compactness, we have suppressed all PDF and luminosity
normalization factors. 
${\cal M}_{F+k}^{(\ell)}$ is the amplitude for producing $F$ in
association with $k$ additional final-state partons, ``legs'', 
and with $\ell$ additional loops. The sums start at $k=0$ and $\ell=0$,
corresponding to the Leading Order for producing $F$, while higher terms
represent real and virtual corrections, respectively. 

The purpose of the $\delta$ function is
to project out hypersurfaces of constant value of \obs\ 
in the full $\dPS{F+k}$ phase space, with $\obs(\PS{F+k})$ a function that
defines $\obs$ evaluated on each specific momentum configuration, 
$\PS{F+k}$. (Without the 
$\delta$ function, the formula would give the total integrated
cross section, instead of the cross section differentially in $\obs$.) 

We recover the various fixed-order truncations of perturbative QCD
\index{pQCD} (pQCD) 
by limiting the
nested sums in \eqRef{eq:fixed-order} to include only specific values
of $k+\ell$. Thus, 
\index{LO}%
\index{NLO}%
\index{NNLO}%
\index{Jets}
\begin{center}
\begin{tabular}{lcp{9.5cm}}
$k=0$, $\ell=0$ &$\implies$& Leading Order  (usually tree-level) 
for $F$ production\\[2mm]
$k=n$, $\ell=0$ &$\implies$& Leading Order for $F+n\,$jets\\[2mm]
$k+\ell\le n$,  &$\implies$& N$^n$LO for $F$ {\small (includes N$^{n-1}$LO for
  $F+1\,$jet, N$^{n-2}$LO for $F+2\,$jets, and so on up to LO for $F+n\,$jets)}~.\\
\end{tabular}
\end{center}

\index{Collinear limit}%
\index{Soft limit}%
For $k\ge1$, we are  not considering inclusive
$F$ production; instead, we are considering the process
$F+k$ jets. If we simply  
integrate over all momenta, as implied by the integration over \dPS{F+k}
in \eqRef{eq:fixed-order}, we would be including configurations in
which one or more of the $k$ partons are collinear or soft. Such
configurations are infrared divergent in QCD and hence must be
\index{Infrared divergences}
regulated. Since we talk about \emph{collinear} and \emph{soft}
divergences (the origins of which will be discussed in more detail
in \secsRef{sec:subtraction} and \ref{sec:parton-showers}), 
cuts on \emph{angles} and \emph{energies} and/or cuts on
combinations, like \emph{transverse momenta}, can be used
to cut away the problematic regions of phase space. 

Recall, however, that pQCD is approximately scale invariant. This
implies that 
\index{QCD!Scale invariance}%
a regularization cut on a dimensionful quantity, like energy or
transverse momentum, should be formulated as a \emph{ratio} of scales,
rather than as an absolute number. 
\index{LO}%
For example, a jet with $p_\perp = 50\,\mathrm{GeV}$ 
would be considered hard and well-separated if produced in association with
an ordinary $Z$ boson (with hard scale $M_Z=91.2\,\mathrm{GeV}$), 
while the same jet would be considered soft if produced
in association with a 900-GeV $Z'$ boson
(see \cite{Plehn:2005cq,Alwall:2008qv} for more explicit
examples).

\index{NLO!Infrared divergences}%
\index{Infrared divergences}%
The essence of the point is that, 
if the regularization scale is taken too low, logarithmic
enhancements of the type 
\begin{equation}
\alpha_s^n\ln^{m\le 2n}\left(\frac{Q_F^2}{Q_k^2}\right)
\label{eq:logs}
\end{equation}
will generate progressively larger corrections, order by order, which
will spoil any fixed-order truncation of the perturbative
series. Here, $Q_F$ is the hard scale associated with the process
under consideration, while $Q_k$ is the scale associated with an
additional parton, $k$. 

\index{LO}%
A good rule of thumb
is that if $\sigma_{k+1} \approx \sigma_{k}$ (at whatever order you
are calculating), then the perturbative series is 
converging too slowly for a fixed-order truncation
of it to be reliable. For fixed-order perturbation theory to be
applicable, you must place your cuts on the hard process such that 
$\sigma_{k+1} \ll \sigma_{k}$. In the
discussion of parton showers in
\SecRef{sec:parton-showers}, we shall see how the region of
applicability of perturbation theory can be extended.  

\index{NLO}%
\index{Unitarity}%
\index{KLN theorem}%
\index{NLO!Infrared divergences}%
\index{Infrared divergences}%
The virtual amplitudes, for $\ell \ge 1$, are divergent
for any point in phase space. However, as encapsulated by the famous KLN theorem
\cite{Kinoshita:1962ur,Lee:1964is}, unitarity (which essentially
expresses probability conservation) puts a powerful
constraint on the IR divergences\footnote{The loop integrals also
  exhibit UV divergences, but these are dealt with by
  renormalization.}, 
forcing them to cancel exactly 
against those coming from the unresolved real emissions 
that we had to cut out above, order by order, 
making the complete answer for fixed
$k+\ell = n$ finite\footnote{Formally, the KLN theorem states that the
sum over degenerate quantum states is finite. In context of
fixed-order perturbation theory, this is exemplified by states with infinitely
collinear and/or soft radiation being degenerate with the
corresponding states with loop corrections; they cannot be told apart
by any physical observable.}
Nonetheless, since this cancellation happens
between contributions that formally live in different phase spaces, 
a main aspect of loop-level higher-order calculations is how to
arrange for this cancellation in practice, either analytically or 
numerically, with many different methods currently on the market. We
shall discuss the idea behind subtraction approaches 
in \secRef{sec:subtraction}. 

\begin{figure}[t]\vspace*{-0.15cm}
\begin{center}
\scalebox{0.90}{
\begin{tabular}{l}
\large\bf F @ LO\\[2mm]
\begin{loopsnlegs}[c]{p{0.25cm}|ccccc}
 \small 2&~\wbox{\pqcd[2]{0}} & \wbox{\pqcd[2]{1}} & \ldots \\[2mm]
 \small 1&~\wbox{\pqcd[1]{0}} & \wbox{\pqcd[1]{1}}  
   & \wbox{\pqcd[1]{2}} & \ldots \\[2mm]
 \small 0&~\gbox{\pqcd[0]{0}} & \wbox{\pqcd[0]{1}} 
   & \wbox{\pqcd[0]{2}} 
   & \wbox{\pqcd[0]{3}} & \ldots \\
\hline
& \small 0 & \small 1 & \small 2 & \small 3 & \ldots
 \end{loopsnlegs}\end{tabular}\hspace*{-0.7cm}
\raisebox{0.9cm}{\begin{minipage}{1.9cm}
\center\includegraphics*[scale=0.23]{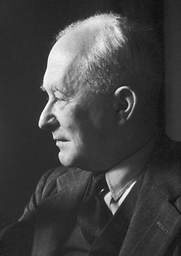}
\tiny Max Born, 1882-1970\\
\index{Nobel prize}Nobel 1954
\end{minipage}}%
\begin{tabular}{l}
\large\bf F + 2 @ LO\\[2mm]
\begin{loopsnlegs}[c]{p{0.25cm}|ccccc}
 \small 2&~\wbox{\pqcd[2]{0}} & \wbox{\pqcd[2]{1}} & \ldots
&  \multicolumn{2}{l}{\tiny 
\colorbox{red}{\parbox[b]{1.75cm}{\raggedright\textcolor{white}{LO for $F+2$\\
      $\to \infty$ for $F+1$\\
      $\to \infty$ for $F+0$\\
}}
}} \\[2mm]
 \small 1&~\wbox{\pqcd[1]{0}} & \wbox{\pqcd[1]{1}}  
   & \wbox{\pqcd[1]{2}} & \ldots \\[2mm]
 \small 0&~\wbox{\pqcd[0]{0}} & \wbox{\pqcd[0]{1}} 
   & \gwbox{\pqcd[0]{2}} 
   & \wbox{\pqcd[0]{3}} & \ldots \\
\hline
& \small 0 & \small 1 & \small 2 & \small 3 & \ldots
 \end{loopsnlegs}
\end{tabular}}
\caption{Coefficients of the perturbative series covered by LO calculations. 
{\sl Left:} $F$ production at lowest order. {\sl Right:} $F+2\,$jets at LO, with the
  half-shaded box illustrating the restriction to the region of phase
  space with exactly 2 resolved jets.
  The total power of $\alpha_s$ for each coefficient is $n = k+\ell$. 
(Photo of Max Born from \url{nobelprize.org}).
\label{fig:loopsnlegs}}
\end{center}
\end{figure}

\index{Matrix elements}
A convenient way of illustrating the terms of the perturbative series that
a given matrix-element-based calculation includes is given in
\figRef{fig:loopsnlegs}. 
In the left-hand pane, the shaded box corresponds to 
the lowest-order ``Born-level'' matrix element squared. This coefficient 
is non-singular and hence can be integrated over all of phase space,
which we illustrate by letting the shaded area fill all of the
relevant box. 
A different kind of leading-order calculation is illustrated in
the right-hand pane of \figRef{fig:loopsnlegs}, where the shaded box
corresponds to the lowest-order matrix element squared for
$F+2\,$jets. 
\index{Soft limit}\index{Collinear limit}
This coefficient diverges in the part of phase space
where one or both of the jets are unresolved (i.e., soft or collinear), 
  and hence integrations can only cover the hard part of
  phase space, which we reflect by only shading the upper half of
  the relevant box. 

\begin{figure}[t]
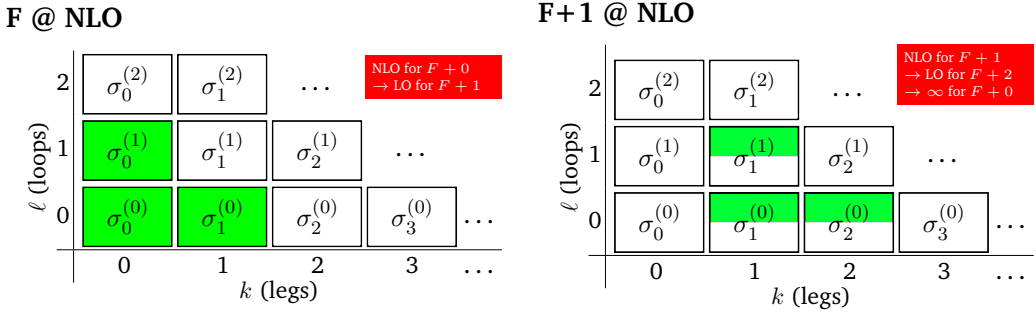

\begin{center}%
\scalebox{0.90}{
\begin{tabular}{l}
\large\bf F @ NLO\\[2mm]
\begin{loopsnlegs}[c]{p{0.25cm}|ccccc}
 \small 2&~\wbox{\pqcd[2]{0}} & \wbox{\pqcd[2]{1}} & \ldots &
 \multicolumn{2}{l}{\tiny 
\colorbox{red}{\parbox[b]{1.75cm}{\raggedright\textcolor{white}{NLO for
      $F+0$\\
$\to$ LO
      for $F+1$}}
}}
\\[2mm]
 \small 1&~\gbox{\pqcd[1]{0}} & \wbox{\pqcd[1]{1}}  
   & \wbox{\pqcd[1]{2}} & \ldots \\[2mm]
 \small 0&~\gbox{\pqcd[0]{0}} & \gbox{\pqcd[0]{1}} 
   & \wbox{\pqcd[0]{2}} &\wbox{\pqcd[0]{3}} & \ldots \\
\hline
& \small 0 & \small 1 & \small 2 & \small 3 & \ldots
 \end{loopsnlegs}
\end{tabular}\hspace*{-1mm}
\begin{tabular}{l}
\large\bf F+1 @ NLO\\[2mm]
\begin{loopsnlegs}[c]{p{0.25cm}|ccccc}
 \small 2&~\wbox{\pqcd[2]{0}} & \wbox{\pqcd[2]{1}} & \ldots &
 \multicolumn{2}{l}{\tiny 
\colorbox{red}{\parbox[b]{1.75cm}{\raggedright\textcolor{white}{NLO for
      $F+1$\\
     $\to$ LO
      for $F+2$ \\$\to \infty$ for $F + 0$}}
}}
\\[2mm]
 \small 1&~\wbox{\pqcd[1]{0}} & \gwbox{\pqcd[1]{1}}  
   & \wbox{\pqcd[1]{2}} & \ldots \\[2mm]
 \small 0&~\wbox{\pqcd[0]{0}} & \gwbox{\pqcd[0]{1}} 
   & \gwbox{\pqcd[0]{2}} 
   & \wbox{\pqcd[0]{3}} & \ldots \\
\hline
& \small 0 & \small 1 & \small 2 & \small 3 & \ldots
 \end{loopsnlegs}
\end{tabular}}
\caption{Coefficients of the perturbative series covered by NLO calculations. 
{\sl Left:} 
   $F$ production at NLO. {\sl Right:} $F+1\,$jet at NLO, with
  half-shaded boxes illustrating the restriction to the region of phase
  space with exactly 1 resolved jet.
  The total power of $\alpha_s$ for each coefficient is $n = k+\ell$.
\label{fig:loopsnlegs2}}
\end{center}
\end{figure}

\index{NLO}
\FigRef{fig:loopsnlegs2} illustrates the inclusion of NLO
virtual corrections. 
To prevent confusion, first a point on notation: by 
$\sigma_0^{(1)}$, we intend
\begin{equation}
\sigma_0^{(1)} = \int \dPS{0} \, 2\mrm{Re}[{\cal M}_{0}^{(1)} {\cal M}_0^{(0)*}]~,
\end{equation}
which is of order $\alpha_s$ relative to the Born
level. 
\index{NLO}
\index{NNLO}
Compare, e.g., with the expansion of
\eqRef{eq:fixed-order} to order $k+\ell = 1$. In particular, $\sigma_0^{(1)}$
should \emph{not} be confused with the integral over the 1-loop matrix
element squared (which would be of relative order $\alpha_s^2$ and
hence forms part of the NNLO coefficient $\sigma_0^{(2)}$). Returning
to \figRef{fig:loopsnlegs2}, the unitary cancellations between real
and virtual singularities imply that we can now extend the
integration of the real correction in the left-hand pane over all of
phase space, while retaining a finite total cross section,
\index{NLO}
\begin{equation}
\begin{array}{rclll}
\displaystyle\sigma^\mrm{NLO}_0 & = 
 &\displaystyle\int \dPS{0} |{\cal M}^{(0)}_0|^2 
 & \displaystyle + \int \dPS{1}|{\cal M}^{(0)}_1|^2
 & \displaystyle + \int \dPS{0} \,2\mrm{Re}[{\cal M}_{0}^{(1)} {\cal M}_0^{(0)*}] 
\\[6mm]
 & = &\displaystyle \sigma^{(0)}_0
 & \displaystyle +\ \sigma^{(0)}_{1}
 & \displaystyle +\ \sigma^{(1)}_{0} 
~,\end{array} \label{eq:sigmaNLO}
\end{equation}
with $\sigma_0^{(0)}$ the finite Born-level cross section, 
and the positive divergence caused by integrating the second term over
all of phase space  is canceled by a negative one coming from the
integration over loop momenta in 
the third term. One method
for arranging the cancellation of singularities  --- subtraction --- is 
discussed in \secRef{sec:subtraction}. 

However,  if our starting point for the NLO
calculation is a process which already has a non-zero number of hard
jets, we must 
continue to impose that at least that number of jets must still be resolved in the
final-state integrations,
\begin{equation}
\begin{array}{rclll}
\displaystyle\sigma^\mrm{NLO}_1(\ptmin) & = 
 & \displaystyle
   \int_{\pt>\ptmin}\hspace*{-11mm} \dPS{1}\, |{\cal M}^{(0)}_1|^2
 & \displaystyle 
   + \int_{\pt[1]>\ptmin}\hspace*{-11mm} \dPS{2}\,
   |{\cal M}^{(0)}_2|^2
 &\displaystyle 
   + \int_{\pt>\ptmin}\hspace*{-11mm} \dPS{1} \, 
   2\mrm{Re}[{\cal M}_{1}^{(1)} {\cal M}_1^{(0)*}] 
\\[6mm]
 & = &\displaystyle \sigma^{(0)}_1(\pt>\ptmin)
 & \displaystyle +\ \sigma^{(0)}_{2}(\pt[1]>\ptmin)
 & \displaystyle +\ \sigma^{(1)}_{1}(\pt>\ptmin)
~,\end{array}\label{eq:sigmanlo1}
\end{equation}
where the restriction to at least one jet having 
$\pt>\ptmin$ has been illustrated in the right-hand 
pane of \figRef{fig:loopsnlegs2} by shading only the upper part of the
relevant boxes. In the second term in \eqRef{eq:sigmanlo1}, 
the notation $\pt[1]$ is used to
\index{Jets}
denote that the integral runs over the phase space in which at least
one ``jet'' (which may consist of one or two partons) must be resolved
with respect to $\ptmin$. Here, therefore, an explicit dependence on
the algorithm used to define ``a jet'' enters for the first time. This
is discussed in more detail in the 2009 ESHEP lectures by Salam
\cite{Salam:2010zt}. 
\index{Jets}%
\index{Jets!Definitions|see{\cite{Salam:2010zt}}}
\index{Jets!Algorithms|see{\cite{Salam:2010zt}}}

\begin{figure}[t]
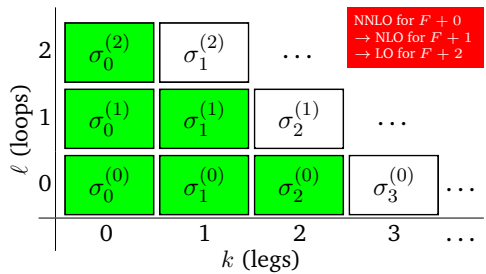

\scalebox{0.90}{\large\bf F @ NNLO}
\begin{center}
\scalebox{0.90}{
\begin{loopsnlegs}[c]{p{0.25cm}|ccccc}
 \small 2&~\gbox{\pqcd[2]{0}} & \wbox{\pqcd[2]{1}} & \ldots &
 \multicolumn{2}{l}{\tiny 
\colorbox{red}{\parbox[b]{1.75cm}{\raggedright\textcolor{white}{NNLO for
      $F+0$\\
     $\to$ NLO
      for $F+1$ \\$\to$ LO for $F + 2$}}
}}
\\[2mm]
 \small 1&~\gbox{\pqcd[1]{0}} & \gbox{\pqcd[1]{1}}  
   & \wbox{\pqcd[1]{2}} & \ldots \\[2mm]
 \small 0&~\gbox{\pqcd[0]{0}} & \gbox{\pqcd[0]{1}} 
   & \gbox{\pqcd[0]{2}} &\wbox{\pqcd[0]{3}} & \ldots \\
\hline
& \small 0 & \small 1 & \small 2 & \small 3 & \ldots
 \end{loopsnlegs}}
\caption{Coefficients of the perturbative series covered by an NNLO
  calculation. 
  The total power of $\alpha_s$ for each coefficient is $n =
  k+\ell$. Green  shading represents the full perturbative 
  coefficient at the respective $k$ and $\ell$.
\label{fig:loopsnlegs3}}
\end{center}
\end{figure}

\index{NNLO}%
To extend the integration to cover also the case of 2 unresolved jets,
we must combine the left- and right-hand parts of
\figRef{fig:loopsnlegs2} and add the new coefficient
\begin{equation}
\sigma_0^{(2)} = |{\cal M}_0^{(1)}|^2 + 2\mrm{Re}[ {\cal
    M}_0^{(2)}{\cal M}_0^{(0)*} ]~,
\end{equation}
 as illustrated by the diagram in \figRef{fig:loopsnlegs3}. 

\index{NLO}
\index{Subtraction}
\index{Unitarity}
\index{KLN theorem}
\subsection{The Subtraction Idea \label{sec:subtraction}}
According to the KLN theorem, the IR singularities coming from
integrating over collinear and soft real-emission configurations
should cancel, order by order, by those coming from the IR divergent
loop integrals. This implies that we should be able to rewrite e.g.\
the NLO cross section, \eqRef{eq:sigmaNLO}, as 
\begin{eqnarray}
\displaystyle\sigma^\mrm{NLO} 
 & = &\displaystyle \sigma^\mathrm{Born}
\ + \ \mathrm{Finite}\left\{\int \dPS{F+1} |{\cal M}^{(0)}_{F+1}|^2\right\}
\ + \ 
\mathrm{Finite}\left\{\int \dPS{F} \, 2\mrm{Re}[{\cal M}_{F}^{(1)} {\cal M}_{F}^{(0)*}]\right\} 
\displaystyle ~,
\end{eqnarray}
with the second and third terms having had their common (but
opposite-sign) singularities canceled out and some explicitly finite
quantities remaining. 

The first step towards this goal is to classify all IR singularities
that could appear in the amplitudes. We know that the IR limits are
universal, so they can be classified using a set of
process-independent functions that only has to be worked out once and
for all. A widely used such set of functions are the 
\index{Catani-Seymour dipoles}
\emph{Catani-Seymour} (CS) dipole ones~\cite{Catani:1996vz,Catani:1996jh},
a method which by now has even been partially automated~\cite{Nagy:2003tz,Frederix:2008hu}. 
Here,  
\index{Antennae}
we shall instead use a formalism based
on \emph{antennae}~\cite{Kosower:1997zr,Kosower:2003bh,GehrmannDeRidder:2005cm}.    
\index{NLO}The distinction between the two is basically that one antenna is made
up of two dipole ``ends'', hence the antenna formalism tends to
generate somewhat fewer terms. At NLO, however, there is no fundamental
incompatibility --- the antennae we use here can always be
partitioned into two dipole ends, if so desired. (Note: 
\index{NNLO}%
\index{Sector decomposition}%
only the antenna method has been 
successfully generalized to 
NNLO~\cite{GehrmannDeRidder:2007hr,Weinzierl:2008iv}. Other NNLO 
techniques, not covered here, are \emph{sector decomposition}, 
see \cite{Heinrich:2008si,Boughezal:2011jf}, and the generic formalism 
for hadroproduction of colorless states presented in~\cite{Catani:2007vq}.)

\index{NLO}
\index{Subtraction}
At NLO, the idea with subtraction is thus to rewrite the NLO cross section
by adding and subtracting a simple function, $\dd{\sigma_S}$, 
that encapsulates all the
IR limits, 
\index{Kinoshita-Lee-Nauenberg|see{KLN theorem}}
\index{KLN theorem}
\begin{eqnarray}
\sigma^\mrm{NLO} & = & 
 \sigma^\mathrm{Born} 
 \ + \ \int \dPS{F+1} \underbrace{\left(|{\cal M}^{(0)}_{F+1}|^2
{\color{red}{ \ - \ \dd{\sigma_S^\mrm{NLO}}}}\right)}_{\mbox{Finite by
Universality}} \nonumber \\[2mm]
&& \ + \ \underbrace{\int \dPS{F} \, 2\mrm{Re}[{\cal M}_{F}^{(1)} {\cal
M}_F^{(0)*}] \ {\color{red}+} \int \dPS{F+1} {\color{red}\dd{\sigma_S^\mrm{NLO}}}
}_{\mbox{Finite by KLN}}~.\label{eq:subtraction}
\end{eqnarray}
The task now is to construct a suitable form for $\dd{\sigma_S}$. A
main requirement is that it 
should be sufficiently simple that the integral in the last term can
be done analytically, in dimensional regularization, 
so that the IR poles it generates can be canceled against those
from the loop term.

\index{Soft limit}
\index{Eikonal}
To build a set of universal terms that parametrize the IR singularities of
any amplitude, we start from the observation that gauge theory
amplitudes factorize in the \emph{soft limit}, as follows:
\begin{eqnarray}
\left\vert{\cal M}_{F+1} (\ldots, i,j,k, \ldots )\right\vert^2
  & \stackrel{j_g\to 0}{\to} & g_s^2\ N_C \left( \frac{2 s_{ik}}{s_{ij}
  s_{jk}} - \frac{2m_i^2}{s_{ij}^2} - \frac{2 m_k^2}{s_{jk}^2}\right) 
\left\vert{\cal M}_{F} (\ldots, i,k, \ldots )\right\vert^2
~,\label{eq:eikonal}
\end{eqnarray}
\index{Color connections}%
where parton $j$ is a soft gluon, partons 
$i$, $j$, and $k$ form a chain of color-space index contractions 
(we say they are \emph{color-connected}), 
$g_s$ is the strong coupling, 
and the terms in parenthesis are 
\index{Eikonal}%
\index{Soft eikonal|see{Eikonal}}%
\index{Color factors}%
called the \emph{soft eikonal factor}. We here show it including 
mass corrections, which appear if $i$ and $k$ have non-zero rest
masses, with the invariants 
$s_{ab}$ then defined as 
\begin{equation}
s_{ab} \equiv 2p_a\cdot p_b = (p_a+p_b)^2 - m_a^2 - m_b^2~.
\end{equation}
The color factor, $N_C$, is valid for the leading-color
contribution, regardless of whether the $i$ and $k$ partons are quarks
or gluons. At subleading color, an additional soft-eikonal
factor identical to the one above but with a color factor 
proportional to $-1/N_C$ arises for each $q\bar{q}$ pair
combination. 
\index{CA@$C_A$}
\index{CF@$C_F$}
This, e.g., modifies the effective color factor for $q\bar{q}\to
qg\bar{q}$ from $N_C$ to $N_C(1-1/N_C) = 2C_F$, in agreement with the
color factor for quarks being $C_F$ rather than $C_A$.

\index{Collinear limit}%
\index{DGLAP kernels}%
Similarly, amplitudes also factorize in the \emph{collinear limit}
(partons $i$ and $j$ parallel, so $s_{ij}\to 0$), in which
the eikonal factor above is replaced by the 
famous Dokshitzer-Gribov-Lipatov-Altarelli-Parisi (DGLAP)
splitting kernels~\cite{Gribov:1972ri,Altarelli:1977zs,Dokshitzer:1977sg}, 
which 
were already mentioned in \secRef{sec:pdfs}, in 
the context of PDF evolution. 
They are also the basis of conventional 
parton-shower models, to which we return in \secRef{sec:parton-showers}. 

Essentially,
what antenna functions, CS dipoles, and the like, all do, is to 
combine the soft (eikonal) and
collinear (Altarelli-Parisi) limits into one universal set of functions
that achieve the correct limiting behavior for \emph{both} 
\index{Antennae}
soft and collinear radiation. 
To give an explicit example, the \emph{antenna function} for gluon
emission from a color-connected $q\bar{q}$ pair can be derived from
the matrix elements squared for the process $Z^0 \to q\bar{q} \to
qg\bar{q}$~\cite{Gustafson:1987rq}, 
\index{Eikonal}
\begin{equation}
\frac{\left\vert{\cal M}(Z^0\to q_i g_j \bar{q}_k)\right\vert^2}
{\left\vert{\cal M}(Z^0\to q_I \bar{q}_K)\right\vert^2}
\ = \ g_s^2 \,
2C_F \, \left[ \underbrace{\frac{2s_{ik}}{s_{ij}s_{jk}}}_{\mrm{eikonal}}
+ \frac{1}{s_{IK}}\underbrace{\left(
\frac{s_{jk}}{s_{ij}}
+ \frac{s_{ij}}{s_{jk}}
\right)}_\mrm{collinear} \right]~,\label{eq:subZ}
\end{equation}
where we have neglected mass corrections
(see~\cite{GehrmannDeRidder:2009fz,GehrmannDeRidder:2011dm} for
massive expressions) 
and we recognize the universal eikonal soft factor from \eqRef{eq:eikonal}
in the first term. The two additional terms are less singular, and are
required to obtain the correct collinear (Altarelli-Parisi) 
limits as $s_{ij}\to 0$ or $s_{jk} \to 0$. 

However, since the singularity structure is universal, we could
equally well have used the process
$H^0 \to q\bar{q} \to qg\bar{q}$ to derive the antenna function.
Our antenna function would then have come out as~\cite{GehrmannDeRidder:2011dm},
\index{Eikonal}
\begin{equation}
\frac{\left\vert{\cal M}(H^0\to q_i g_j \bar{q}_k)\right\vert^2}
{\left\vert{\cal M}(H^0\to q_I \bar{q}_K)\right\vert^2}
\ = \ g_s^2 \,
2C_F \, \left[ \underbrace{\frac{2s_{ik}}{s_{ij}s_{jk}}}_{\mrm{eikonal}}
+ \frac{1}{s_{IK}}\underbrace{\left(
\frac{s_{jk}}{s_{ij}}
+ \frac{s_{ij}}{s_{jk}}
\right)}_\mrm{collinear}
+ \underbrace{\frac{2}{s_{IK}}}_{\mrm{finite}}\right]~,
\label{eq:subH}
\end{equation}
where the additional term $2/s_{IK}$ is non-singular (``finite'') over
all of phase space. Thus, we here see an explicit example that the
singularities are process-independent 
while the non-singular terms are process-dependent. 
Since we add and subtract
the same term in \eqRef{eq:subtraction}, the final answer does not depend
on the choice of finite terms. 
\index{Subtraction schemes}
\index{Antennae}
\index{Antenna functions|see{Antennae}}
We say that they correspond to different \emph{subtraction
schemes}. One standard antenna subtraction scheme, which uses the
antenna function defined in \eqRef{eq:subZ} rather than the one
in \eqRef{eq:subH}, is the Gehrmann-Gehrmann-de Ridder-Glover (GGG)
one, given in \cite{GehrmannDeRidder:2005cm}. 

If there is more than one color antenna in the Born-level process, the
form of $\dd{\sigma_S}$ is obtained as a sum over terms, each of which
captures one specific soft limit and either all or ``half'' of a
collinear one, depending on the specific scheme and the type of
parton,
\index{Subtraction}
\begin{equation}
\dd{\sigma_S} \ = \ \sum_{j} A_{IK\to ijk} \, 
\left\vert {\cal M}_F(\ldots,I,K,\ldots)\right\vert^2~,
\end{equation}
with the sum running over all  singular 
$3\to 2$ ``clusterings'' of the $(F+1)$-parton state to $F$
partons. An analysis of the different ways of partitioning the collinear
singularity of gluons among neighboring antenna is beyond the scope of
this introduction, but useful discussions can be found
in~\cite{Larkoski:2009ah,Giele:2011cb,LopezVillarejo:2011ap}. 

\index{Jets!Infrared safety}%
\index{Infrared safety}%
\index{pQCD}%
\subsection{Infrared Safety \label{sec:IRsafe}}
A further requirement for being able to perform calculations within
perturbative QCD is that the observable be \emph{infrared safe}. 
Note: by ``infrared'',
we here mean any  limit 
that involves a low scale (i.e., any non-UV limit), 
without regard to  whether it is collinear or soft.

The property of infrared safety defines a special class of observables which
have \emph{minimal sensitivity} 
to long-distance physics, and which can be consistently computed in
pQCD. 
An observable is infrared safe if:
\begin{enumerate}
\item \index{Soft limit}{\sl(Safety against soft radiation):} Adding any number of infinitely soft particles should not change
  the value of the observable. 
\item \index{Collinear limit}{\sl(Safety against collinear radiation):} Splitting an existing particle up into two comoving particles, with
arbitrary fractions $z$ and $1-z$, respectively, of 
the original momentum, should not change the value of the
observable. 
\end{enumerate}
If both of these conditions are satisfied, any long-distance non-perturbative
corrections will be suppressed by the ratio of the 
long-distance scale to the short-distance one to some
(observable-dependent) power, typically
\index{Infrared safety}%
\index{Jets!Infrared safety}
\begin{equation}
\mbox{IR Safe Observables: IR corrections~~~$\propto$~~~} \frac{Q_{\mathrm{IR}}^2}{Q_{\mathrm{UV}}^2} 
\end{equation}
where $Q_\mathrm{UV}$ denotes
a generic hard scale in the problem, and $Q_\mrm{IR} \sim
\Lambda_\mrm{QCD} \sim \mathcal{O}(\mrm{1\ GeV})$. 

Due to this \emph{power suppression}, 
IR safe observables are not so sensitive to our lack
of ability to solve the strongly coupled IR physics, unless of course
we go to processes for which the relevant hard scale, $Q_{\mrm{UV}}$, is small
(such as minimum-bias, soft jets, or small-scale jet substructure). 
Even when a high scale is present, however, as in resonance decays, jet
fragmentation, or underlying-event-type studies, infrared safety only
guarantees us that infrared corrections are small, not that they are zero. 
Thus, ultimately, we run into a precision barrier even for IR safe observables, 
which only a reliable understanding of the
long-distance physics itself can address. 

To constrain models of long-distance physics, one needs 
infrared \emph{sensitive} observables.
Hence it is not always the case that infrared safe
observables are preferable --- the purpose decides the tool.
Instead of the suppressed corrections above, the perturbative
prediction for such observables contains logarithms of the type
already encountered in \eqRef{eq:logs},
\index{Infrared safety}%
\index{Jets!Infrared safety}%
\begin{equation}
\mbox{IR Sensitive Observables: IR Corrections~~~$\propto$~~~} 
\alpha_s^n\log^{m\le
2n}\left(\frac{Q_{\mathrm{UV}}^2}{Q_{\mathrm{IR}}^2}\right)~,
\end{equation}
 which grow
increasingly large as $Q_{\mathrm{IR}}/Q_{\mathrm{UV}}\to 0$.   
As an example, consider such a fundamental quantity as particle 
multiplicities (= number of particles);
in the absence of nontrivial infrared
effects, the number of partons tends logarithmically to infinity 
as the IR cutoff is lowered. Similarly, the distinction between
a charged and a neutral pion only occurs in the very last phase of
hadronization, and hence observables that only include charged tracks,
for instance, are always IR sensitive\footnote{This remains true in principle 
even if the tracks are clustered into jets, although the energy
clustered in this way does provide a lower bound on $Q_{\mrm{UV}}$ in
the given event, since ``charged + neutral $>$ charged-only''.}.

\index{Event shapes}%
Two important categories of infrared safe observables that are widely used are
\emph{event shapes} and \emph{jet algorithms}. Jet algorithms are
perhaps nowhere as pedagogically described as in the 2009 ESHEP lectures by 
Salam \cite[Chapter 5]{Salam:2010zt}. Event shapes in the context of
hadron colliders have not yet been as widely explored, but the basic
phenomenology is introduced also by Salam and collaborators in
\cite{Banfi:2010xy}, with first measurements reported by CMS and ATLAS 
\cite{Khachatryan:2011dx,Aad:2012} and a proposal to use them also for the
characterization of soft-QCD (``minimum-bias'') events put forth
in \cite{Wraight:2011ej}. 

\index{Infrared safety}%
\index{Hadronization}%
Let us here merely emphasize that the real reason to prefer infrared safe
jet algorithms over unsafe ones 
is not that they  necessarily give very different
or ``better'' answers in the experiment --- experiments are infrared safe by
definition, and the difference between infrared safe and unsafe
algorithms may not even be visible when running the
algorithm on experimental data --- but that it is only possible to
compute perturbative QCD predictions for the infrared safe ones. Any
measurement performed with an infrared unsafe algorithm can only be
compared to calculations that include a detailed hadronization model. This
both limits the number of calculations that can be compared to and
also adds an a priori unknown sensitivity to the details of the 
hadronization description, details which one would rather investigate and
constrain separately, in the framework of more dedicated fragmentation
studies. 

\index{Anti-kT algorithm|see{Jets}}%
\index{Jets!Anti-kT algorithm}%
For LHC phenomenology, the preferred IR safe algorithm for jet
reconstruction is currently the \emph{anti-kT}
one~\cite{Cacciari:2008gp}, with size parameter
$R$ varying between 0.4 and 0.7, though larger sizes can be
motivated in certain contexts, e.g., to look for highly energetic jets
and/or the boosted decay products of high-mass
objects~\cite{Abdesselam:2010pt,Altheimer:2012mn}.   
\index{Underlying event}%
This algorithm generates circular-looking
jets, so subtracting off the energy believed to be associated with the
\emph{underlying event} (UE, see \secRef{sec:soft-processes}) 
is particularly simple. 

\index{Parton showers}%
\index{Jets!kT algorithm}%
\index{Jets!Cambridge/Aachen algorithm}%
For jet substructure, typically either the
``kT'' or ``Cambridge/Aachen'' algorithms are used,
see e.g.\ \cite{Abdesselam:2010pt,Altheimer:2012mn}. 
The clustering measures used in these algorithms more closely mimic
the singularity structure of QCD bremsstrahlung and they are therefore
particularly well suited to ``unravel'' a tree of QCD branchings~\cite{Salam:2010zt}, 
such as a parton shower generates. The Cambridge/Aachen algorithm may
also be used to characterize the underlying 
event, see \cite{Cacciari:2009dp}.

\clearpage
\index{Event generators|see{Monte Carlo}}
\index{Monte Carlo!Event generators}
\section{Monte Carlo Event Generators \label{sec:MC}}

In this section, we discuss the physics of Monte Carlo event generators and
their mathematical foundations, at an introductory level. We shall attempt
to convey the main ideas as clearly as possible without burying them
in an avalanche of technical details. References to
more detailed discussions are included where applicable.
We assume the reader is already familiar with the contents of the
preceding section on hard processes. 

The task of a Monte Carlo event generator is to calculate
everything that happens in a high-energy collision, from the hard
short-distance physics to the long wavelengths of hadronisation and
hadron decays. Obviously, this requires some compromises to be
made. General-purpose generators like \Hw~\cite{Corcella:2000bw,Bahr:2008pv,Bellm:2015jjp},
\Py~\cite{Sjostrand:2006za,Sjostrand:2014zea}, and \Sh~\cite{Gleisberg:2008ta}, 
start from low-order (LO or NLO) descriptions of the perturbative hard physics
and then attempt to include the ``most significant'' corrections, such
as higher-order matrix-element corrections and parton showers,
resonance decays and finite-width effects, 
underlying event, beam remnants, hadronisation, and hadron
decays. Each of them had slightly different origins, which carries
through to the emphasis placed on various physics aspects today:
\begin{itemize}
\item \Py. \index{PYTHIA} 
Successor to \textsc{Jetset} (begun in 1978). Originated in
hadronisation studies. Main feature: the Lund string fragmentation model. 
\item \Hw. \index{HERWIG}
\index{Angular ordering}
Successor to \textsc{Earwig} (begun in 1984). Originated in
perturbative coherence studies. Main feature:
angular-ordered parton showers.
\item \Sh. \index{SHERPA}
Begun in 2000. Originated in studies of the matching of 
hard-emission matrix elements with parton showers. Main feature: 
CKKW matching. 
\end{itemize}
There is also a large number of more specialised generators, mainly
for hard processes within and beyond the SM, 
a few that offer alternative shower
models, and ones specializing in soft-inclusive and/or heavy-ion
physics.

An important aspect of contemporary generators is the
ability to combine specialised ones with general-purpose ones, via interfaces. 
The most common interface
between partonic hard-process and parton-shower generators is  the
\index{Les Houches Event Files|see{LHEF}}%
\index{LHEF}%
Les Houches Event File (LHEF) standard, defined
in~\cite{Boos:2001cv,Alwall:2006yp} and ``spoken'' by most modern
generator tools. For interfaces to experimental analysis
packages (like \textsc{Rivet}~\cite{Buckley:2010ar}) 
and detector simulations
(like \textsc{Geant}~\cite{Agostinelli:2002hh}), 
typically the HepMC standard is used~\cite{Dobbs:2001ck}. 

Hard processes were the topic
of \secRef{sec:pQCD}. In this section, we shall focus
mainly on parton showers, with some brief comments on resonance
decays at the end. 
\SecRef{sec:matching} then concerns the matching of matrix elements 
and parton showers. Finally, models of hadronisation and the
underlying event  are the topic of \secRef{sec:soft}.

Several of the discussions below rely on material from the section on Monte
Carlo Event Generators in the PDG Review of Particle Physics~\cite{pdg2012} and
on the more comprehensive  
\index{MCnet review}%
review by the {\sl MCnet} collaboration 
in \cite{Buckley:2011ms}. The latter also contains brief
descriptions of the  
physics implementations of each of the main general-purpose event
generators on the 
market, together with a guide on how to use (and not use) them 
in various connections, and a collection of comparisons to important
experimental 
distributions. We highly recommend readers to obtain a copy of that
review, as it is the most comprehensive and up-to-date review of event
generators currently available. Another useful and pedagogical review
on event generators is contained in the 2006 ESHEP lectures
by Torbj\"orn Sj\"ostrand \cite{Sjostrand:2006su}, with a more recent update in
\cite{Sjostrand:2009ad}. 

\index{Monte Carlo!Integration}
\subsection{The Monte Carlo Method}
A ubiquitous problem in fundamental physics is the following:
given a source located some distance from a detector, predict the
number of counts that should be observed within the solid angle spanned
by the detector (or within a bin of its phase-space acceptance), as a
function of the properties of the source, the intervening medium,
and the efficiency of the detector. Essentially, the task is to
compute integrals of the form
\begin{equation}
N_\mrm{Count}(\Delta\Omega) \ = \ \int_{\Delta \Omega} \dd{\Omega} \frac{\dd{\sigma}}{\dd{\Omega}}~,
\end{equation}
with $\dd{\sigma}$ a differential cross section for the process of
interest. 

In particle physics, phase space has three dimensions per final-state 
particle (minus four for overall
four-momentum-conservation). 
Thus, for problems with more than a few outgoing particles, the
dimensionality of phase space increases rapidly. 
At LEP, for instance, the total multiplicity of neutral +
charged hadrons (before weak decays) was typically $\sim$ 30 particles, 
for about 86 dimensions. 

\index{Convergence}%
\index{Numerical integration}%
\index{Monte Carlo!Integration}%
The standard 1D numerical-integration methods 
give very slow convergence rates for
higher-dimensional problems. For illustration, a table of convergence
rates in 1 and $d$ dimensions is given in \tabRef{tab:convergence},
comparing the Trapezoidal (2-point) rule and Simpson's (3-point) rule
to random-number-based Monte Carlo. 
\begin{table}[t]
\centering
\index{Uncertainties!Monte Carlo statistics}%
\index{Monte Carlo!Statistics}%
\index{Monte Carlo!Uncertainties}%
\begin{tabular}{c|cc|c}
\toprule
\mbox{Relative uncertainty with $n$ points} &
1-Dim & $d$-Dim & $n_{\mrm{eval}}$/point \\
\midrule
\mbox{Trapezoidal Rule}  & $1/n^2$ & $1/n^{2/d}$ & $2^d$\\
\mbox{Simpson's Rule}  & $1/n^4$ & $1/n^{4/d}$ & $3^d$\\
\midrule
\mbox{Monte Carlo} & $1/\sqrt{n}$ & $1/\sqrt{n}$& $1$ \\
\bottomrule
\end{tabular}
\caption{Relative uncertainty after $n$ evaluations, in $1$ and $d$
dimensions, for two traditional
numerical integration methods and stochastic Monte Carlo. The 
last column shows the number of function evaluations that are
required per point, in $d$ dimensions. 
\label{tab:convergence}}
\end{table}
In 1D, the $1/n^2$ convergence rate of the 
Trapezoidal rule is much faster than the stochastic $1/\sqrt{n}$ of
random-number Monte Carlo, and
Simpson's rule converges even faster. However, as we go to 
$d$ dimensions, 
the convergence rate of the 
$n$-point rules all degrade with $d$ (while the number of function
evaluations required for each ``point'' simultaneously increases). The
MC convergence rate, on the other hand, remains the simple stochastic 
$1/\sqrt{n}$, independent of $d$, and each point still only requires
one function evaluation. 
\index{Monte Carlo!Integration}%
These are some of
the main reasons that MC is the preferred numerical integration
technique for high-dimensional problems. In addition, the random
phase-space vectors it generates can be re-used in many ways, for
instance as input to iterative solutions, to compute many different
observables simultaneously, and/or to hand ``events'' to propagation
and detector-simulation codes. 

Therefore, virtually
all numerical cross section calculations are based on Monte Carlo
techniques in one form or another, the simplest being the 
\index{Monte Carlo!RAMBO}%
\index{RAMBO}%
\textsc{Rambo} algorithm \cite{Kleiss:1985gy} which can be expressed
in about half a page of code and generates a flat scan over $n$-body
phase space\footnote{Strictly speaking, \textsc{Rambo} is only truly
  uniform for massless particles. Its massive variant makes up for
  phase-space biases by returning weighted momentum configurations.}. 

\index{Infrared divergences}%
However, due to 
the infrared singularities in perturbative QCD, and due to the
presence of short-lived resonances, the functions to be
integrated, $|{\cal{M}}_{F+k}|^2$, can be highly non-uniform,
especially for large $k$.
This implies that we will have to be clever in the way
we sample phase space if we want the integration to converge in any
\index{RAMBO}%
reasonable amount of time --- simple algorithms like \textsc{Rambo} 
quickly become inefficient for $k$ greater than a few. 
To address this bottleneck, 
the simplest step up
from \textsc{Rambo} is to introduce generic (i.e., automated) 
importance-sampling methods, such as offered by the  
\index{Monte Carlo!VEGAS}%
\index{VEGAS|see{Monte Carlo}}%
\textsc{Vegas} algorithm \cite{Lepage:1977sw,Lepage:1980dq}. This is still the
dominant basic technique, although most modern codes do employ several 
additional refinements, such as several different copies of \textsc{Vegas}
running in parallel (multi-channel integration), to further optimise
the sampling.  
Alternatively, a few algorithms incorporate the singularity structure of QCD
explicitly in their phase-space sampling, either by directly generating momenta
distributed according to the leading-order QCD singularities, in a
sort of ``QCD-preweighted'' analog of \textsc{Rambo}, called
\index{SARGE}%
\index{Markov chains}%
\textsc{Sarge} \cite{Draggiotis:2000gm}, or by using all-orders
Markovian parton showers to generate them 
\index{VINCIA}
(\textsc{Vincia} \cite{Giele:2011cb,LopezVillarejo:2011ap,Fischer:2016vfv}).

\begin{figure}[t]
\centering
\begin{minipage}{0.45\textwidth}
\includegraphics*[scale=0.5]{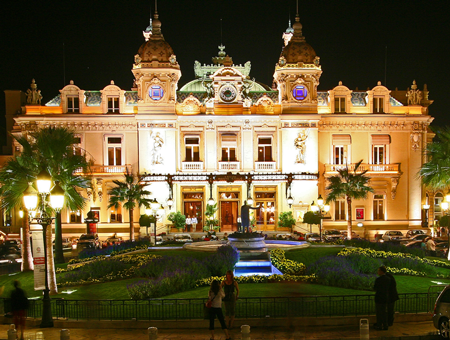}
\end{minipage}
 \hfill  
\index{Convergence}%
\index{Monte Carlo!Convergence}%
\begin{minipage}[c]{0.46\textwidth}\it
``This risk, that convergence is only given with a certain probability,
is inherent in Monte Carlo calculations and is the reason why this
technique was named after the world's most famous gambling
casino. Indeed, the name is doubly appropriate because the style of
gambling in the Monte Carlo casino, not to be confused with the noisy
and tasteless gambling houses of Las Vegas and Reno, is serious and
sophisticated.''
\flushright \color{gray}
F. James, ``Monte Carlo theory and practice'',\\ 
Rept.~Prog.~Phys.~43~(1980)~1145 
\end{minipage}
\caption{{\sl Left:} The casino in Monaco. {\sl Right:} 
\index{Monte Carlo!Monaco}%
extract from \cite{James:1980yn} 
concerning the nature of Monte Carlo techniques.
\label{fig:montecarlo}}
\end{figure}

\index{Convergence}
\index{Monte Carlo!Convergence}
\index{Monte Carlo!Integration}
The price of using random numbers is that we must generalise our
notion of convergence. 
In calculus, we say that a sequence $\{A\}$ \emph{converges} to $B$ if
an $n$ exists for which the difference $|A_{i>n} -B|<\epsilon$ for any
$\epsilon > 0$. In random-number-based techniques, we cannot
completely rule out the possibility of very pathological sequences of
``dice rolls'' leading to large deviations from the true goal, 
hence we are restricted to
say that $\{A\}$ converges to $B$ if an $n$ exists for which \emph{the
probability} for $|A_{i>n} - B| < \epsilon$, for any $\epsilon > 0$,
is greater than $P$, for any $P\in[0,1]$~\cite{James:1980yn}. 
This risk, that convergence
is only given with a certain probability, is the reason why Monte
Carlo techniques were named after the famous casino in Monaco, 
illustrated in \figRef{fig:montecarlo}. 

\index{Parton showers}
\subsection{Theoretical Basis of Parton Showers\label{sec:parton-showers}}

\index{pQCD}
In \secRef{sec:pQCD}, we noted two 
conditions that had to be valid for fixed-order truncations of the
perturbative series to be valid: firstly, the strong coupling
$\alpha_s$ must be small for perturbation theory to be valid at
all. This restricts us to the region in which all scales $Q_i\gg
\Lambda_\mrm{QCD}$. We shall maintain this restriction in this
section, i.e., we are still considering \emph{perturbative QCD}. 
Secondly, however, in order to be allowed to \emph{truncate} the
perturbative 
series, we had to require  $\sigma_{k+1} \ll \sigma_k$, i.e., the
corrections at successive orders must become successively smaller, which --- due
to the enhancements from soft/collinear singular (conformal) dynamics
--- effectively restricted us to consider only the phase-space 
region in which all jets
are ``hard and well-separated'', equivalent to requiring all $Q_i/Q_j \approx
1$. 
\index{Jets}
In this section, we shall see how to lift this restriction,
extending the applicability of perturbation theory into regions that
include scale 
hierarchies, $Q_i \gg Q_j \gg \Lambda_\mrm{QCD}$, such as occur for
soft jets, jet substructure, etc.

In fact, the simultaneous restriction to all resolved scales being
larger than $\Lambda_{\mrm{QCD}}$ \emph{and} no large hierarchies is
extremely severe, if taken at face value. 
Since we collide and observe  hadrons ($\to$ low
scales) while
simultaneously wishing to study short-distance physics processes ($\to$ high
scales), it would appear trivial to conclude that fixed-order pQCD is not 
applicable to collider physics at all. So why do we still use it?

\index{Factorisation}
The answer lies in the fact that we actually never truly perform a fixed-order
calculation in QCD. Let us repeat the factorised formula for the cross
section, \eqRef{eq:factorisation}, now inserting also a function, $D$,
to represent the fragmentation of the final-state partons into
observable hadrons, 
\index{PDFs}
\index{Fragmentation functions}
\index{Factorisation!Factorisation scale}
\begin{equation}
\frac{\dd{\sigma}}{\dd{\obs}} = 
\sum_{i,j}\int_{0}^{1}\dd{x_i}\dd{x_j}\sum_f\int\!\dPS{f}\,
\pdf{i/h_1}(x_i,\mu^2_F)\pdf{j/h_2}(x_j,\mu^2_F)\,\frac{\dd{\hat{\sigma}_{ij\to
      f}}}{\dd{\hat\obs}}D_f(\hat{\obs}\to\obs,\mu^2_F)~,
\label{eq:factorisation2}
\end{equation}
with $\hat{\cal O}$ denoting the observable evaluated on the partonic
final state, and ${\cal O}$ the observable evaluated on the
hadronic final state, after fragmentation. 
Although the partonic cross section, $\dd{\hat{\sigma}_{ij\to f}}$, 
does represent a fixed-order
calculation, the parton densities, $f_{i/h_1}$ and
$f_{j/h_2}$,  
include so-called resummations of perturbative corrections \emph{to
  all orders} from the initial scale of
order the mass of the proton, up to the factorisation scale, $\mu_F$
(see \secRef{sec:pdfs} and/or the TASI lectures by
Sterman~\cite{Sterman:1995fz}).  
Note that the oft-stated mantra that the PDFs are purely 
non-perturbative functions is therefore misleading.
 True, they are defined as essentially non-perturbative functions at
 some very low 
scale, $\mu_0$ $\sim$ a few GeV, but, 
if $\mu_F$ is taken large, they necessarily incorporate a
significant amount of perturbative physics as well.
On the ``fixed-order side'', all we have left to ensure in 
$\dd{\sigma_{ij\to f}}$ is then
that there are no large hierarchies remaining between $\mu_F$ and the
QCD scales appearing in \PS{f}. 
Likewise, in the final
state, the fragmentation functions, $D_f$, include infinite-order
resummations of perturbative corrections all the way \emph{from}
$\mu_F$ down to some low scale, with
similar caveats concerning mantras about their non-perturbative 
nature as for the PDFs. 

\subsubsection{Step One: Infinite Legs}
The infinite-order resummations that are 
included in objects such as the PDFs and FFs in
\eqRef{eq:factorisation2} (and in their parton-shower equivalents)
rely on some very simple and powerful
properties of gauge field theories that were already touched on
in \secRef{sec:pQCD}. In particular, we saw
in \secRef{sec:subtraction}
that we can represent all the infrared (IR) limits of any NLO
amplitude with a set of simple universal functions, based solely
on knowing which partons are color-connected (i.e., have color-space
index contractions) with one another. 

\begin{figure}[t]
\centering
\includegraphics*[scale=0.5]{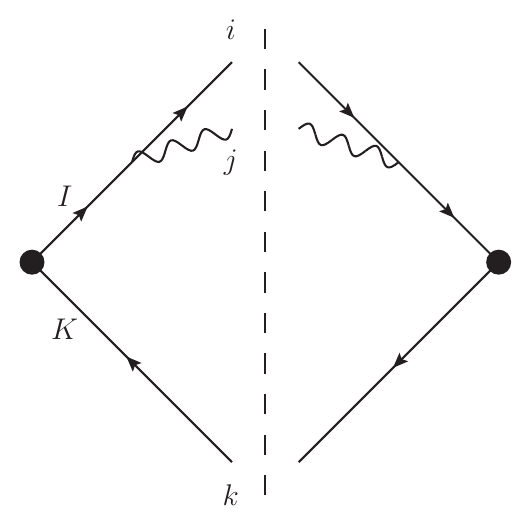}\hspace*{1cm}
\includegraphics*[scale=0.5]{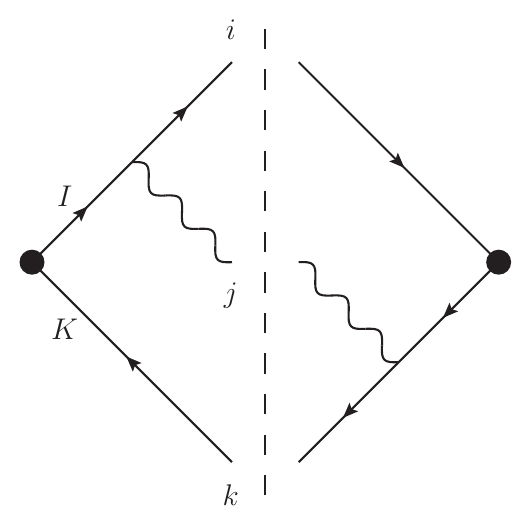}
\caption{Diagrams (squared) giving rise to collinear ({\sl left}) and soft ({\sl
right}) singularities. \index{Collinear limit} \index{Soft limit}
 \label{fig:SoftColl}}
\end{figure}
\index{Collinear limit}
The diagrams in \figRef{fig:SoftColl} show the basic origin of the
universal IR singularities of gauge theory amplitudes. On the left is
shown a diagram (squared) in which an emission with small $s_{ij}$ 
interferes with itself. In the collinear limit, $s_{ij} \to 0$, the
propagator of the parent parton, $I$, goes on shell; the 
singularity of the associated propagator factor is the
origin of the $1/s_{ij}$ collinear singularities. 
\index{Soft limit}
On the right is
shown the interference between a diagram with emission from parton $I$
and one with emission 
from parton $K$. The resulting term has propagator singularities when
both partons $I$ and $K$ go on shell, which can happen simultaneously 
if parton $j$ is soft. 
\index{Eikonal}
This generates the $2s_{ik}/(s_{ij}s_{jk})$ soft 
singularity, also called the soft eikonal factor or the dipole factor.

We now understand the fundamental origin of the IR singularities, why
they are universal, and why amplitudes factorise in the soft and
collinear limits --- 
the singularities are simply 
generated by intermediate parton propagators going on shell, which is
independent of the nature of the hard process, and hence can be
factorised from it. 

\index{Antennae}
Thus, for each pair of (massless) color-connected partons $I$ and $K$ in
$F$, the squared amplitude for $F+1$ gluon, 
$|{\cal M}_{F+1}|^2$, will include a factor
\begin{equation}
\left\vert{\cal M}_{F+1}\right\vert^2 \ = \ 
  \underbrace{g_s^2\,N_C\, \left(\frac{2s_{ik}}{s_{ij}s_{jk}}~+~\mbox{collinear
  terms}\right)}_{\mbox{Antenna Function}} \, \left\vert{\cal M}_F\right\vert^2~,\label{eq:eikonal2}
\end{equation} 
where $g_s^2 = 4\pi\alpha_s$ is the strong coupling,
$i$ and $k$ represent partons $I$ and $K$ after the branching (i.e.,
they include possible recoil effects) and $s_{ij}$ is
the invariant between parton $i$ and the emitted parton, $j$. 

\begin{figure}[t]
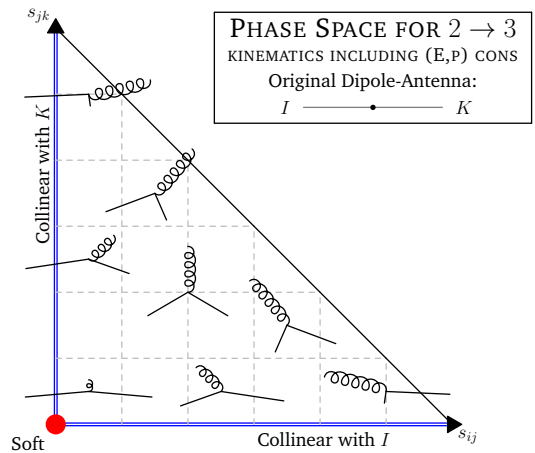

\centering
\vspace*{5mm}
\include{fmfphase}
\caption{Illustration of the branching phase space for $q\bar{q}\to
  qg\bar{q}$, with the original dipole-antenna oriented
  horizontally, the two parents
  sharing the transverse component of recoil, and the azimuthal angle
  $\phi$ (representing rotations of the emitted parton 
  around the dipole axis) chosen such
  that the gluon is radiated upwards. From~\cite{Giele:2011cb}.
\label{fig:branchingps}}
\end{figure}

The branching phase space of a color dipole (i.e., a pair of partons
connected by a color-index contraction) is illustrated
in \figRef{fig:branchingps}. 
Expressed in the branching invariants, $s_{ij}$ and $s_{jk}$, the
phase space has a characteristic triangular shape, imposed by the
relation $s = s_{ij} + s_{jk} + s_{ik}$ (assuming massless
partons). Sketchings of the post-branching parton momenta have been
inserted in various places in the figure, for illustration. The soft
singularity is located at the origin of the plot and the collinear
regions lie along the axes. 

The collinear terms for a $q\bar{q}\to qg\bar{q}$ ``antenna'' are
unambiguous and are  
given in \secRef{sec:subtraction}. Since gluons are in the adjoint
representation, they carry both a color and an anticolor index 
(one corresponding to the rows and the other to the
columns of the Gell-Mann matrices), and there is therefore some
ambiguity concerning how to partition collinear radiation among the two
antennae they participate in. This is discussed in more detail
in \cite{Giele:2011cb}. Differences are subleading, however, and for
our purposes here we shall consider gluon antenna ends as radiating
just like quark ones. The difference between quark and gluon
radiation then arise mainly because gluons participate in two
antennae, while quarks only participate in one. This is related to the
difference between the color factors, $C_A \sim 2C_F$. 

The problem that plagued the fixed-order truncations
in \secRef{sec:pQCD} is clearly visible in \eqRef{eq:eikonal2}: 
if we integrate over the
entire phase space including the region $s_{ij}\to 0$, $s_{jk}\to 0$,
we end up with a double pole. If we instead regulate the divergence by 
cutting off the integration at some minimal \emph{perturbative cutoff
  scale} $\mu_\mrm{IR}^2$, we end up with a logarithm squared of that
scale. This is
a typical example of ``large logarithms'' being generated by the
presence of scale hierarchies. Also note that 
the precise definition of $\mu_\mrm{IR}$ is not
  unique. Any scale choice that properly isolates the singularities
  from the rest of phase space will do, with some typical choices being, for
  example, invariant-mass and/or transverse-momentum scales.

Before we continue, it is worth noting that  \eqRef{eq:eikonal2} is often
rewritten in other forms to emphasise specific aspects of it. 
One such rewriting is thus to reformulate the invariants
$s_{ij}$ appearing in it in terms of energies and angles, 
\begin{equation}
s_{ij} = 2 E_iE_j\left(1-\cos\theta_{ij}\right)~. 
\end{equation}
Rewritten in this way, the differentials can be partial-fractioned,
\begin{equation}
\frac{\dd{s_{ij}}}{s_{ij}}\frac{\dd{s_{jk}}}{s_{jk}} \propto
\frac{\dd{E_j}}{E_j}\frac{\dd{\theta_{ij}}}{\theta_{ij}} + 
\frac{\dd{E_j}}{E_j}\frac{\dd{\theta_{jk}}}{\theta_{jk}} ~. 
\end{equation}
This kind of rewriting 
enables an intuitively appealing categorisation of the singularities 
as related to vanishing energies and angles, explaining why they are 
called \emph{soft} and \emph{collinear}, respectively. Arguments based
on this rewriting have led to important insights in QCD.
For instance, within the framework of conventional parton showers,
it was shown in a sequence of publications (see 
\cite{Marchesini:1983bm,Marchesini:1987cf} and references therein) that 
the destructive interference effects between two 
colour-connected partons (\emph{coherence})\index{Angular
ordering}\index{Coherence} 
can be described, on average\footnote{Averaged over azimuthal emission angles.}, by using the opening angle of the emissions as the shower ordering variable.
One should still keep in mind, however,  
that Lorentz non-invariant formulations come with similar caveats and
warnings as do  gauge non-invariant formulations of quantum field
theory: while they can be practical to work with at 
intermediate stages of a calculation, one should be careful with
any physical conclusions that rely explicitly on them.

We shall therefore here restrict ourselves to a 
Lorentz-invariant formalism based directly on
\eqRef{eq:eikonal2}, pioneered by the dipole formulation of QCD
cascades~\cite{Gustafson:1987rq}. 
The collinear limit is then replaced by a more
general \emph{single-pole} limit in which a single 
parton-parton invariant vanishes (as, \emph{for instance}, when 
a pair of partons become collinear),
while the soft limit is replaced by one in which two (or more) 
invariants involving the
same parton vanish simultaneously (as, for instance by that parton
becoming soft in a frame defined by two or more hard partons). This
avoids frame-dependent ambiguities from entering into the language,
at the price of a slight reinterpretation of what
is meant by collinear and soft. 

\begin{figure}[t]
\centering
\small
\begin{tabular}{cccc}
Dipole $p_\perp$ & Dipole Mass & Angle & Pythia $p_\perp$\\
\includegraphics[scale=0.375]{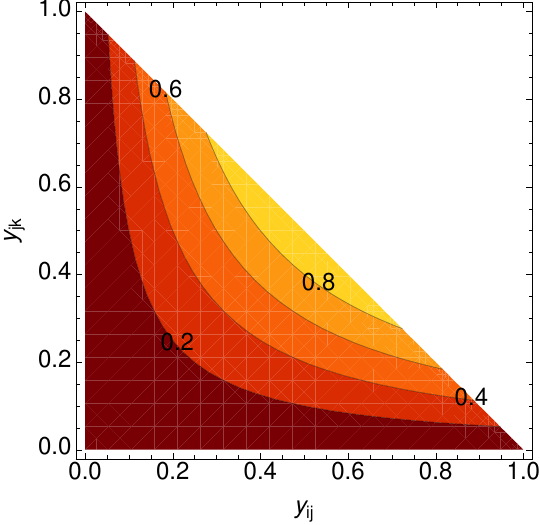} &
\includegraphics[scale=0.375]{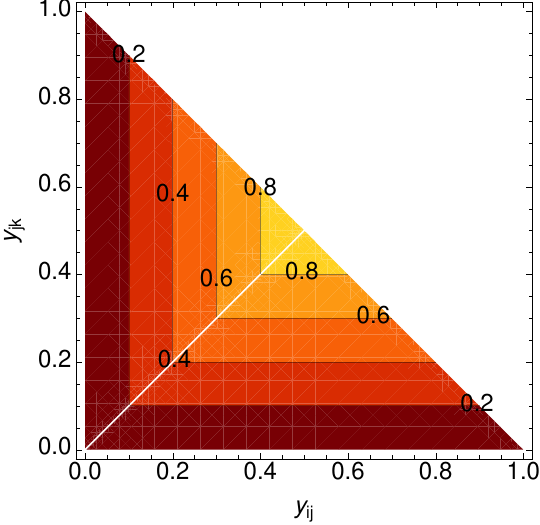} & 
\includegraphics[scale=0.3225]{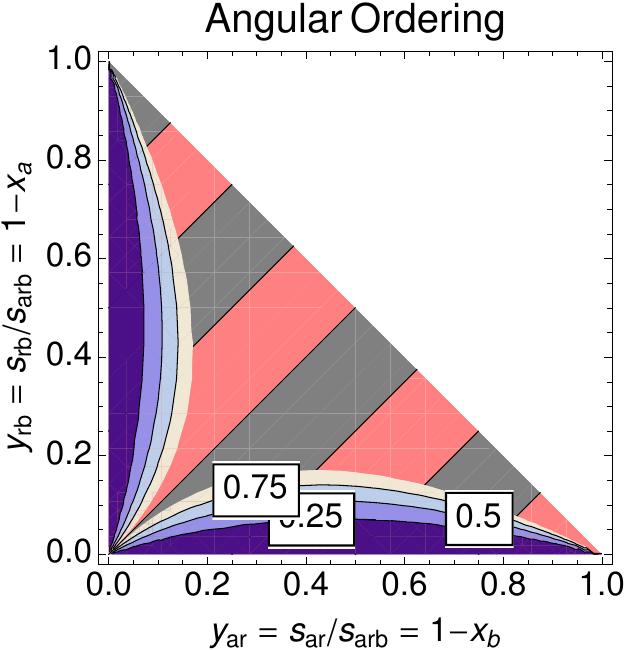} &
\includegraphics[scale=0.3225]{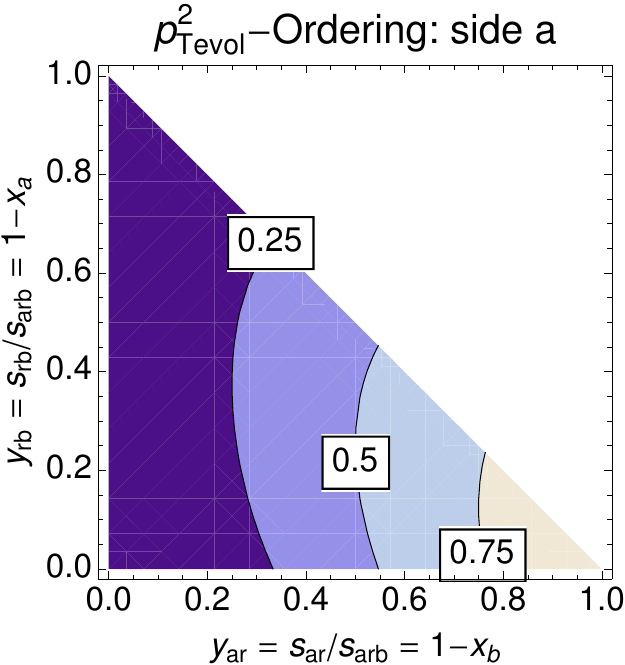} 
\end{tabular}
\caption{A selection of parton-shower evolution variables, represented
as contours over the dipole phase space. Note: the right-most variable
corresponds to evolution of only one of the parents, the one with no
collinear singularity along the bottom of the plot. \label{fig:qe}}
\end{figure}

In the generator landscape, \emph{angular ordering} is used by the 
\Hw~\cite{Marchesini:1987cf}
and \textsc{Herwig$++$}~\cite{Gieseke:2003rz} programs,  and an \emph{angular
veto} is imposed on the virtuality-ordered evolution
in \tsc{Pythia~6}~\cite{Bengtsson:1986et}.
Variants of the dipole/antenna approach is used 
by the \Ar~\cite{Lonnblad:1992tz}, \Sh~\cite{Nagy:2005aa,Schumann:2007mg},
and \Vc~\cite{Fischer:2016vfv} programs,
while the $p_\perp$-ordered showers in \textsc{Pythia~6} and \tsc{8} 
represent a
hybrid, combining collinear splitting kernels with dipole
kinematics \cite{Sjostrand:2004ef}. Phase-space 
contours of equal value of some of
these choices are illustrated in \figRef{fig:qe}. 
During the shower
evolution, each model effectively ``sweeps'' over 
phase space in the order implied
by these contours. E.g., a $p_\perp$-ordered dipole shower (leftmost
plot in \figRef{fig:qe}) will treat a hard-collinear branching as
occurring ``earlier'' than a soft one, while a mass-ordered dipole
shower (second plot) will tend to do the opposite. This affects
the tower of virtual corrections generated by each shower model 
via the so-called Sudakov factor, discussed below. Experimental tests
of the subleading aspects of shower models can therefore quite
important, see e.g.~\cite{Fischer:2014bja} for a recent example.

Independently of rewritings and philosophy, 
the real power of \eqRef{eq:eikonal2} lies in the fact
that it is \emph{universal}. Thus, for
\emph{any} process $F$, we can apply \eqRef{eq:eikonal2} in order to
get an approximation for $\dd{\sigma_{F+1}}$. We may then, for instance,  
take our newly obtained expression for $F+1$ as our arbitrary process
and crank \eqRef{eq:eikonal2} again, to obtain an approximation for
$\dd{\sigma_{F+2}}$, and so forth. What we have here is therefore a
very simple recursion relation that can be used to generate approximations to
leading-order cross sections with arbitrary numbers of additional
legs. The quality 
of this approximation is governed by how many terms besides the leading
one shown in \eqRef{eq:eikonal} are included in the game. Including
all possible terms, the most general form for the cross section at 
$F+n\,$ jets, restricted to the phase-space region above some 
infrared cutoff scale $\mu_{\mrm{IR}}$, has the following algebraic structure,
\index{Transcendentality}
\begin{equation}
\sigma_{F+n}^{(0)} = \alpha_s^n \left( 
   \ln^{2n} 
 + \ln^{2n-1} 
 + \ln^{2n-2} 
 + \ldots 
 + \ln{}  
 + {\cal F} \right)  \label{eq:transcend}
\end{equation}
where we use the notation $\ln^\lambda$ without an argument  to denote 
generic functions of \emph{transcendentality} $\lambda$ (the
logarithmic function 
to the power $\lambda$ being a ``typical'' example of a function 
with transcendentality $\lambda$ appearing in cross section expressions, but
also dilogarithms and higher logarithmic functions\footnote{Note: 
due to the theorems 
that allow us, for instance, to rewrite dilogarithms in different
ways with logarithmic and lower ``spillover'' terms, the coefficients at each 
$\lambda$ are only well-defined up to reparametrisation ambiguities
involving the  terms with transcendentality greater than $\lambda$.} of
transcendentality $>1$ should be implicitly understood to belong to
our notation $\ln^\lambda$). The last term, $\cal F$, represents a rational
function of transcendentality 0. We shall also use the nomenclature 
\emph{singular} and \emph{finite} for  the $\ln^\lambda$ and 
${\cal F}$ terms, respectively,  a terminology which reflects their respective
behavior in the limit $\mu_{\mrm{IR}}\to 0$. 

The simplest approximation one can build on \eqRef{eq:transcend}, dropping
all but the leading $\ln^{2n}$ term in the parenthesis, 
is thus the \emph{leading-transcendentality} approximation. This
approximation is better known as 
the DLA (double logarithmic approximation), since it generates the
correct coefficient for terms which have two powers of logarithms for
each power of $\alpha_s$, while terms of lower transcendentalities are not
guaranteed to have the correct coefficients. In so-called LL
(leading-logarithmic) parton shower algorithms, one generally expects
to reproduce the correct coefficients for the $\ln^{2n}$ and
$\ln^{2n-1}$ terms. In addition, several formally subleading
improvements are normally also introduced in such algorithms 
(such as explicit momentum
conservation, gluon polarisation and other 
\index{Spin correlations}%
spin-correlation effects
\cite{Collins:1987cp,Knowles:1988vs,Richardson:2001df},
higher-order coherence effects~\cite{Marchesini:1983bm}, 
renormalisation scale choices
\cite{Catani:1990rr}, finite-width effects \cite{Gigg:2008yc}, etc), 
as a means to improve the agreement
with some of the more subleading coefficients as well, if not in every
phase-space point then at least on average. 
\index{Monte Carlo!Tuning}%
Though LL showers do not magically acquire NLL
(next-to-leading-log) precision from such procedures, one therefore
still expects a significantly better average performance from them 
than from corresponding ``strict'' LL analytical resummations. A side
effect of this is that it is often possible to ``tune'' shower
algorithms to give 
better-than-nominal agreement with experimental distributions, by
adjusting the parameters controlling the treatment of subleading
effects. One should remember, however, that there is a limit to how
much can be accomplished in this way --- at some point, agreement with
one process will only come at the price of disagreement with another,
and at this point further tuning would be meaningless.  

\begin{figure}
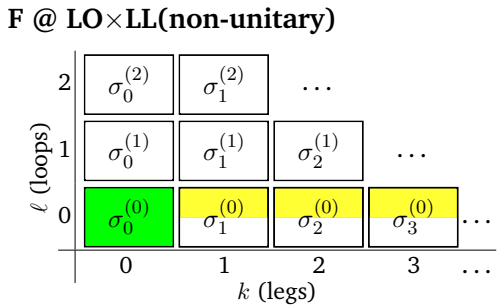

\centering
\scalebox{0.9}{
\begin{tabular}{l}
\large\bf F @ LO$\times$LL(non-unitary) \\[2mm]
\begin{loopsnlegs}[c]{p{0.25cm}|ccccc}
 \small 2&~\wbox{\pqcd[2]{0}} & \wbox{\pqcd[2]{1}} & \ldots &
\\[2mm]
 \small 1&~\wbox{\pqcd[1]{0}} & \wbox{\pqcd[1]{1}}  
   & \wbox{\pqcd[1]{2}} & \ldots \\[2mm]
 \small 0&~\gbox{\pqcd[0]{0}} & \ywbox{\pqcd[0]{1}} 
   & \ywbox{\pqcd[0]{2}} &\ywbox{\pqcd[0]{3}} & \ldots \\
\hline
& \small 0 & \small 1 & \small 2 & \small 3 & \ldots
 \end{loopsnlegs}
\end{tabular}}
\caption{Coefficients of the perturbative series covered by LO + LL 
  approximations to higher-multiplicity tree-level matrix elements.
Green (darker) shading represents the full perturbative
  coefficient at the respective $k$ and $\ell$. Yellow (lighter)
  shading represents an LL approximation to it. Half-shaded boxes
  indicate phase spaces in which we are prohibited from integrating
  over the IR singular region, as discussed in
  \secsRef{sec:fixed-order} and \ref{sec:matching}.
\label{fig:LL}}
\end{figure}
Applying such an iterative process on a Born-level cross section, one
obtains the description of the full perturbative series illustrated in
\figRef{fig:LL}. The yellow (lighter) shades
used here for $k\ge 
1$ indicate that the coefficient obtained is not the exact one, but rather an 
approximation to it that only gets its leading singularities
right. However, since this is still only an approximation to
infinite-order \emph{tree-level} cross sections 
(we have not yet included any virtual corrections), 
we cannot yet integrate this approximation over all of
phase space, as illustrated by the yellow boxes being only
half filled in \figRef{fig:LL}; otherwise, the summed total cross section
would still be infinite. This particular approximation would therefore
still appear to be  
very useless indeed --- on one hand, 
it is only guaranteed to get the singular terms right, but on the
other, it does not actually allow us to  integrate over the singular
region. In order to obtain a truly \emph{all-orders} calculation, the
constraint of unitarity must also be explicitly imposed, which
furnishes an approximation to all-orders loop corrections as well.
Let us therefore emphasise that \figRef{fig:LL} is included for
pedagogical purposes only; all resummation calculations, whether
analytical or parton-shower based, include virtual corrections as well
and consequently yield finite total cross sections, as will now be described.

\subsubsection{Step Two: Infinite Loops}

\index{KLN theorem}%
\index{Unitarity}%
Order-by-order unitarity, such as used in the KLN theorem, implies
that the singularities caused by integration over unresolved radiation
in the tree-level matrix elements must be canceled, order by order, by
equal but opposite-sign singularities in the virtual corrections at
the same order. That is, from \eqRef{eq:eikonal2}, we immediately 
know that the 1-loop correction to $\dd{\sigma_F}$ \emph{must} contain a
term, 
\begin{equation}
2\mrm{Re}[{\cal M}_F^{(0)}{\cal M}_F^{(1)*}] \ \supset \ 
- g_s^2 \, N_C \, \left\vert {\cal M}_F^{(0)} \right\vert^2
\, 
\int 
\frac{\dd{s_{ij}}\dd{s_{jk}}}{16\pi^2 s_{ijk}} \left( 
\frac{2s_{ik}}{s_{ij}s_{jk}}~+~\mbox{less singular
  terms}\right)~,\label{eq:eikonalv}
\end{equation}
that cancels the divergence coming from \eqRef{eq:eikonal2}
itself. Further, since this is universally true, we may apply
\eqRef{eq:eikonalv} again to get an approximation to the 
corrections generated by \eqRef{eq:eikonal2} at the next order and so
on. By adding such terms explicitly, order by order, we may now
bootstrap our way around the entire perturbative series, using
\eqRef{eq:eikonal2} to move horizontally and \eqRef{eq:eikonalv} to
move along diagonals of constant $n=k+\ell$. 
\index{Unitarity}
\begin{figure}
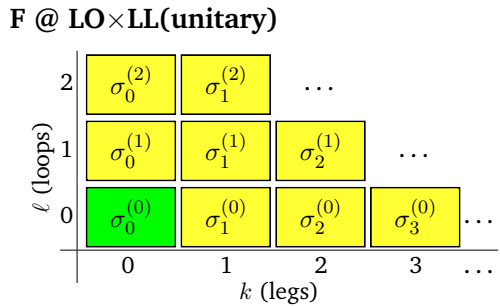

\centering
\scalebox{0.9}{
\begin{tabular}{l}
\large\bf F @ LO$\times$LL(unitary)\\[2mm]
\begin{loopsnlegs}[c]{p{0.25cm}|ccccc}
 \small 2&~\ybox{\pqcd[2]{0}} & \ybox{\pqcd[2]{1}} & \ldots & 
\\[2mm]
 \small 1&~\ybox{\pqcd[1]{0}} & \ybox{\pqcd[1]{1}}  
   & \ybox{\pqcd[1]{2}} & \ldots \\[2mm]
 \small 0&~\gbox{\pqcd[0]{0}} & \ybox{\pqcd[0]{1}} 
   & \ybox{\pqcd[0]{2}} &\ybox{\pqcd[0]{3}} & \ldots \\
\hline
& \small 0 & \small 1 & \small 2 & \small 3 & \ldots
 \end{loopsnlegs}
\end{tabular}}
\caption{Coefficients of the perturbative series covered by LO + LL 
  calculations, imposing unitarity order by order for each $n =
  k+\ell$. Green (darker) shading represents the full perturbative
  coefficient at the respective $k$ and $\ell$. Yellow (lighter)
  shading represents an LL approximation to it. 
\label{fig:LLu}}
\end{figure}
Since real-virtual cancellations are now explicitly restored, we may
finally extend the integrations over all of phase space, resulting in
the picture shown in \figRef{fig:LLu}. 

The picture shown in \figRef{fig:LLu}, not the one in \figRef{fig:LL}, 
corresponds to what is actually done in
\emph{resummation} calculations, both of the analytic and
parton-shower types\footnote{In the way these calculations are 
formulated in practice, they
in fact rely on one additional property, called exponentiation, that allows us
to move along straight vertical lines in the loops-and-legs diagrams. However, 
since the two different directions furnished by \eqsRef{eq:eikonal2} and
\eqref{eq:eikonalv} are already sufficient to move freely in the
full 2D coefficient space, we shall use exponentiation
without extensively justifying it here.}.
\index{Unitarity}
Physically, there is a significant and intuitive meaning to the imposition of
unitarity, as follows. 

\index{Jets}
\index{Inclusive cross sections}
\index{Exclusive cross sections}
\index{Event evolution|see{Evolution}}
\index{Evolution}
Take a jet algorithm, with some measure of jet resolution, $Q$, and
apply it to an arbitrary sample of events, say dijets. At a very crude
resolution scale, corresponding to a high value for $Q$, 
you find that everything is clustered back to a dijet configuration,
and the 2-jet cross section is equal to the total inclusive cross
section,
\begin{equation}
\sigma_\mrm{tot} = \sigma_{F;\mrm{incl}} ~.
\end{equation}
At finer resolutions, decreasing $Q$, you see that 
some events that were previously classified as 2-jet events contain
additional, lower-scale jets, that you can now resolve, and hence
those events now migrate to the 3-jet bin, while the total inclusive
cross section of course remains unchanged,
\index{Inclusive cross sections}
\index{Exclusive cross sections}
\index{Evolution}
\begin{equation}
\sigma_\mrm{tot} = \sigma_{F;\mrm{incl}} = \sigma_{F;\mrm{excl}}(Q)
+ \sigma_{F+1;\mrm{incl}}(Q)~, \label{eq:incexc}
\end{equation}
where ``incl'' and ``excl'' stands for inclusive and exclusive cross
sections\footnote{$F$ \emph{inclusive} $=$ $F$ plus
  anything. $F$ \emph{exclusive} $=$ $F$ and only
  $F$. Thus, $\sigma_{F;\mathrm{incl}}=\sum_{\mrm{k}=
    0}^{\infty}\sigma_{F+k;\mrm{excl}}$},  
respectively, 
and the $Q$-dependence in the two terms on the 
right-hand side must cancel so that the total inclusive cross
section is independent of $Q$. Later, some 3-jet events now migrate 
further, to 4 and higher jets, 
while still more 2-jet events migrate \emph{into} the 3-jet
bin, etc. For arbitrary $n$ and $Q$, we have
\index{Inclusive cross sections}
\index{Exclusive cross sections}
\index{Evolution}
\begin{equation}
\sigma_{F+n;\mrm{incl}}(Q) = \sigma_{F;\mrm{incl}}
- \sum_{m=0}^{n-1} \sigma_{F+m;\mrm{excl}}(Q)~. 
\end{equation}
This equation expresses the trivial fact that the cross section for
$n$ or more jets can be computed as the total inclusive cross section for $F$
minus a sum over the cross sections for $F$ + exactly $m$ jets including
all $m<n$. On the theoretical side, it is these negative terms which must
be included in the calculation, for each order $n=k+\ell$, to restore
unitarity. Physically, they 
express that, at a given scale $Q$, each event will be classified
as having \emph{either} 0, 1, 2, or whatever jets. Or, equivalently,
for each event we gain in the 3-jet bin as $Q$ is lowered, we must loose one
event in the 2-jet one; the negative contribution to the 2-jet bin
is exactly minus the integral of the positive contribution to the
3-jet one, and so on. 
\index{Detailed balance}
\index{Evolution}
We may perceive of this \emph{detailed balance}
 as an \emph{evolution} of the event structure with $Q$, for each 
event, which is effectively what is done in parton-shower
algorithms, to which we shall return in \SecRef{sec:Markov}.

\subsection{Perturbation Theory with Markov Chains \label{sec:Markov}}
Consider again the Born-level cross section for an arbitrary hard process,
$F$, differentially in an arbitrary infrared-safe observable $\cal O$,
as obtained from \eqRef{eq:fixed-order}:
\begin{equation}
\left.   \frac{\dd{\sigma^{(0)}_F}}{\dd{\cal O}}\right\vert_{\mbox{\textcolor{black}{Born}}}
 = \int \dPS{F} \ |{\cal M}_F^{(0)}|^2 \ \delta({\cal
   O}-{\cal O}(\PS{F}))~,
\label{eq:starting}
\end{equation}
where the integration runs over the full final-state on-shell phase space of
$F$ (this expression and those below would also apply to hadron collisions
were we to include integrations over the parton distribution functions
in the initial state), and the $\delta$ function projects out a
1-dimensional slice defined by $\cal O$ evaluated on the 
set of final-state momenta which we denote $\PS{F}$.
 
\index{Evolution}
\index{Parton showers}
To make the connection to parton showers, 
we insert an operator, ${\cal S}$, that acts on the Born-level
final state \emph{before} the observable is evaluated, i.e., 
\begin{equation}
\left.   \frac{\dd{\sigma_F}}{\dd{\cal
    O}}\right\vert_{\mbox{\textcolor{black}{${\cal S}$}}}
 = \int \dd{\Phi_F} \ |{\cal M}_F^{(0)}|^2 \ {\cal S}(\PS{F},{\cal O})~.
\end{equation}
Formally, this operator --- the evolution operator --- will be
responsible for generating all (real and virtual) 
higher-order corrections to the Born-level expression.
The measurement $\delta$ function appearing explicitly in
\eqRef{eq:starting} is now implicit in ${\cal S}$.

\index{Evolution}
\index{Factorisation!Factorisation scale}
\index{Parton showers}
\index{Parton showers!Evolution variable}
Algorithmically, parton showers cast $\cal S$ as an iterative Markov
(i.e., history-independent) chain, 
with an evolution parameter, $Q_E$, that formally 
represents the factorisation scale of the event, below which all
structure is summed over inclusively. Depending on the particular
choice of shower algorithm, $Q_E$ may be defined as a parton
virtuality (virtuality-order showers), as a transverse-momentum scale
\index{pT-ordering@$p_\perp$-ordering}
\index{Transverse-momentum-ordering|see{$p_\perp$-ordering}}
($p_\perp$-ordered showers), or as a combination of
\index{Angular ordering}
energies times angles (angular ordering). Regardless of the specific
form of $Q_E$, 
the evolution parameter will go towards zero as the Markov chain
develops, and the event structure
will become more and more exclusively resolved. 
\index{Parton showers!Infrared cutoff|see{Hadronisation scale}}
\index{Hadronisation scale}
A transition from a
perturbative evolution to a non-perturbative one can also be
inserted, when the evolution reaches an appropriate scale, typically 
around 1~GeV. This scale, called the \emph{hadronisation scale}, thus 
represents the lowest perturbative scale that can appear in the
calculations, with all perturbative corrections below it
summed over inclusively.

Working out the precise form that $\cal S$ must have in order to give the
correct expansions discussed in \secRef{sec:parton-showers} takes a
bit of algebra, and is beyond the scope we aim to cover in these
lectures. Heuristically, the procedure is as follows. 
We noted that the singularity structure of QCD is universal
and that at least its first few terms are known to us. We also saw
that we could iterate 
that singularity structure, using universality and unitarity, 
thereby bootstrapping our way around the entire perturbative
series. This was illustrated by 
\figRef{fig:LLu} in \secRef{sec:parton-showers}. 

\index{Evolution}
\index{Markov chains}
Skipping intermediate steps, the form of the all-orders pure-shower
Markov chain, for the evolution of an event between two scales $Q_{1}
> Q_E > Q_{2}$, is,  
\begin{equation}\begin{array}{rcl}
\displaystyle \hspace*{-1mm} {\cal S}(\PS{F},Q_{1},Q_{2},\obs) \hspace*{-1mm}
&  \hspace*{-1mm}= \hspace*{-1mm} &\displaystyle \hspace*{-1mm}
  \underbrace{
\Delta(\PS{F},Q_{1},Q_{2})\ 
\delta\left(\obs-\obs(\PS{F})\right) 
}_{\mbox{$F+0$ exclusive above $Q_{2}$}}\\[9mm]
& & \hspace*{-1.2cm}+ \displaystyle 
\underbrace{
\sum_r \int_{Q_{E2}}^{Q_{E1}}
\frac{\dPS[r]{F+1}}{\dPS{F}} 
\ S_r(\PS{F+1}) \ \Delta(\PS{F},Q_{1},Q_{F+1})
 \ {\cal S}(\PS{F+1},Q_{F+1},Q_{2},\obs)}_{\mbox{$F+1$ inclusive above $Q_{2}$}}
~,\label{eq:markov} \hspace*{-1mm}
\end{array}
\end{equation}
\index{Sudakov factor}
\index{Parton showers!Sudakov factor}
with the so-called \emph{Sudakov factor},
\begin{equation} 
\Delta(\PS{F},Q_{1},Q_{2}) = \exp\left[-\sum_r\int_{Q_{2}}^{Q_{1}} 
\frac{\dPS[r]{F+1}}{\dPS{F}} S_r(\PS{F+1}) \right]~, \label{eq:sudakov}
\end{equation}
defining the probability that there is \emph{no evolution} (i.e., no
emissions) between the scales $Q_{1}$ and $Q_{2}$, according to the
\emph{radiation functions} $S_r$ to which we shall return below. 
\index{Markov chains}
The term on the first line of
\eqRef{eq:markov} thus represents all events that \emph{did not}
evolve as the resolution scale was lowered from $Q_{1}$ to $Q_{2}$,
while the second line contains a sum and phase-space integral over
those events that \emph{did} evolve --- including the insertion of ${\cal
  S}(\PS{F+1})$ representing the possible further evolution of the
event and completing the iterative definition of the Markov chain. 

The factor $\dPS[r]{F+1}/\dPS{F}$
defines the chosen phase space factorisation. Our favorite is the
so-called dipole-antenna factorisation, whose principal virtue is that
it is the simplest Lorentz invariant 
factorisation which is simultaneously exact over
all of phase space while only involving on-shell momenta. For
completeness, its form is
\begin{equation}  
\frac{\dPS[r]{F+1}}{\dPS{F}} = \frac{\dPS[r]{3}}{\dPS{2}} =
\dd{s_{a1}} \dd{s_{1b}} \frac{\dd{\phi}}{2\pi} \frac{1}{16\pi^2
  s_r}~, \label{eq:phasespace} 
\end{equation}
which involves just one color-anticolor pair for each $r$, with
invariant mass squared $s_r = (p_a+p_1+p_b)^2$. 
Other choices, such as purely collinear ones (only exact in the
collinear limit \emph{or} involving explicitly off-shell momenta), 
more global ones involving all partons in the event (more complicated,
in our opinion), or less global ones with a single parton playing the
dominant role as emitter, are also possible, again depending 
on the specific algorithm considered.

The radiation functions $S_r$ obviously play a crucial role in these
equations, driving the emission probabilities. For example, if 
$S_r \to 0$, then  $\Delta \to \exp(0) = 1$ and all events stay in the
top line of \eqRef{eq:markov}. Thus, in regions of phase  
space where $S_r$ is small, there is little or no evolution. 
Conversely, for $S_r\to \infty$, we have $\Delta\to 0$, implying that \emph{all}
events evolve. 
One possible choice for the radiation functions $S_r$ was
implicit in \eqRef{eq:eikonal2}, in which we took them to include only
the leading 
(double) singularities, with $r$ representing color-anticolor
pairs. In general, the shower
may exponentiate the entire set of universal singular terms, or only
a subset of them (for example, the terms leading in the number of colors
$N_C$), which is why we here let the explicit form of $S_r$ be
unspecified. 
\index{DGLAP kernels}%
Suffice it to say that 
in traditional parton showers, $S_r$ would simply be the DGLAP
splitting kernels (see, e.g., \cite{Dissertori:2003pj}), 
\index{Antennae}%
while they would be so-called dipole or antenna radiation
functions in the various dipole-based approaches to QCD 
(see, e.g.,
\cite{Azimov:1986sf,Gustafson:1987rq,Catani:1996vz,GehrmannDeRidder:2005cm,Giele:2007di,Schumann:2007mg,Giele:2011cb,LopezVillarejo:2011ap}).   

The procedure for how to technically ``construct'' a shower algorithm
of this kind, using random numbers to generate scales distributed according to
\eqRef{eq:markov}, is described more fully in \cite{Giele:2011cb},
using a notation that closely parallels the one used here. The
procedure is also described at a more technical level in the review
\cite{Buckley:2011ms}, though using a slightly different
notation. Various aspects of the Sudakov veto algorithm are discussed
in \cite{Platzer:2011dq,Lonnblad:2012hz,Mrenna:2016sih}.
\index{Monte Carlo!Pedagogical introduction}%
Finally, pedagogical introduction to Monte Carlo
methods in general can be found in \cite{James:1980yn,Weinzierl:2000wd}. 

\subsection{Decays of Unstable Particles \label{sec:decays}}
\index{Resonance decays}%
\index{Monte Carlo!Decay distributions}%
In most BSM processes and some SM ones, an important aspect of the
event simulation is how decays of short-lived particles, such as top
quarks, EW and Higgs bosons, and new BSM resonances, are
handled. We here briefly summarise the spectrum of possibilities, but
emphasise that there is no universal standard. Users are advised to check
 whether the treatment of a given code is adequate for the physics
study at hand. 

The appearance of an unstable resonance as a physical particle
at some intermediate stage of the event generation implies that its
production and decay processes are treated as being factorised. 
\index{Narrow width approximation}%
This is called the \emph{narrow width approximation} and is 
valid up to corrections of order $\Gamma/m_0$, with $\Gamma$
the width and $m_0$ the pole mass of the particle. 
States whose widths are a substantial fraction
of their mass should not be treated in this way, but 
rather as intrinsically virtual internal propagator lines. 

For states treated as physical particles, two aspects are relevant:
the mass distribution of the decaying particle itself and the
distributions of its decay products. For the mass distribution, 
the simplest is to use a 
$\delta$ function at $m_0$. The next level up, typically
used in general-purpose Monte Carlos, 
\index{Breit-Wigner distribution}%
is to use a Breit-Wigner distribution (relativistic or
non-relativistic), which formally resums  higher-order virtual corrections to
the mass distribution. Note, however, that this 
still only generates an improved picture for \emph{moderate} fluctuations
away from $m_0$. Similarly to above, particles 
that are significantly off-shell (in units of $\Gamma$) should not be
treated as resonant, but rather as internal off-shell propagator
lines. In most Monte Carlo codes, some further refinements are also 
included, for instance
by letting $\Gamma$ be a function of $m$ (``running widths'') and by
limiting the magnitude of the allowed fluctuations away from
$m_0$. See also \cite{Seymour:1995qg} for an elaborate discussion of the
Higgs boson lineshape. 

For the distributions of the decay products, the simplest treatment is
again to assign them their respective $m_0$ values, with a uniform
(i.e., isotropic, or ``flat'')
phase-space distribution. A more sophisticated treatment distributes
the decay products according to the differential decay matrix
elements, capturing at least the internal dynamics and helicity
structure of the decay process, including EPR-like correlations. 
\index{Spin correlations}%
Further refinements include polarisations of the external 
states~\cite{Collins:1987cp,Knowles:1988vs,Richardson:2001df} (see
also \cite{Stelzer:1995gc,Parke:1996pr,Smillie:2005ar} 
for phenomenological studies)
 and assigning the decay products their own
Breit-Wigner distributions, the latter of which 
opens the possibility to  include also intrinsically off-shell decay
channels, like $H\to W W^*$. Please refer to the physics manual of the
code you are using and/or make simple cross checks like plotting the
distribution of phase-space invariants it produces.

\index{Resonance decays}%
During  subsequent showering of the decay products, 
most parton-shower models will preserve the total invariant 
mass of each resonance-decay system separately, 
so as not to skew the original resonance shape.


\clearpage
\index{Matching}%
\index{LO}%
\section{Matching at LO and NLO\label{sec:matching}}
The essential problem that leads to matrix-element/parton-shower
matching can be illustrated in a very simple way.
\begin{figure}
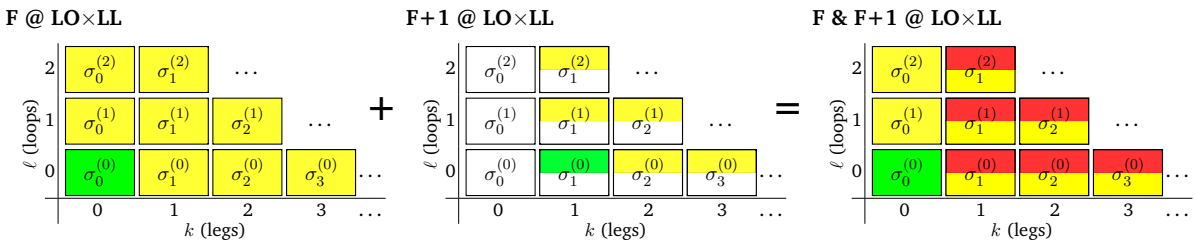

\begin{center}%
\scalebox{0.70}{\begin{tabular}{l}
\large\bf F @ LO$\times$LL \\[2mm]
\begin{loopsnlegs}[c]{p{0.25cm}|ccccc}
 \small 2&~\ybox{\pqcd[2]{0}} & \ybox{\pqcd[2]{1}} & \ldots &
\\[2mm]
 \small 1&~\ybox{\pqcd[1]{0}} & \ybox{\pqcd[1]{1}}  
   & \ybox{\pqcd[1]{2}} & \ldots \\[2mm]
 \small 0&~\gbox{\pqcd[0]{0}} & \ybox{\pqcd[0]{1}} 
   & \ybox{\pqcd[0]{2}} &\ybox{\pqcd[0]{3}} & \ldots \\
\hline
& \small 0 & \small 1 & \small 2 & \small 3 & \ldots
 \end{loopsnlegs}
\end{tabular}\hspace*{-7mm}\raisebox{0.2cm}{\huge\bf +}\hspace*{-1.5mm}
\begin{tabular}{l}
\large\bf F+1 @ LO$\times$LL\\[2mm]
\begin{loopsnlegs}[c]{p{0.25cm}|ccccc}
 \small 2&~\wbox{\pqcd[2]{0}} & \ywbox{\pqcd[2]{1}} & \ldots & 
\\[2mm]
 \small 1&~\wbox{\pqcd[1]{0}} & \ywbox{\pqcd[1]{1}}  
   & \ywbox{\pqcd[1]{2}} & \ldots \\[2mm]
 \small 0&~\wbox{\pqcd[0]{0}} & \gwbox{\pqcd[0]{1}} 
   & \ywbox{\pqcd[0]{2}} &\ywbox{\pqcd[0]{3}} & \ldots \\
\hline
& \small 0 & \small 1 & \small 2 & \small 3 & \ldots
 \end{loopsnlegs}
\end{tabular}
\hspace*{-7mm}\raisebox{0.2cm}{\huge\bf =}\hspace*{-1.5mm}
\begin{tabular}{l}
\large\bf F \& F+1 @ LO$\times$LL\\[2mm]
\begin{loopsnlegs}[c]{p{0.25cm}|ccccc}
 \small 2&~\ybox{\pqcd[2]{0}} & \rybox{\pqcd[2]{1}} & \ldots & 
\\[2mm]
 \small 1&~\ybox{\pqcd[1]{0}} & \rybox{\pqcd[1]{1}}  
   & \rybox{\pqcd[1]{2}} & \ldots \\[2mm]
 \small 0&~\gbox{\pqcd[0]{0}} & \rybox{\pqcd[0]{1}} 
   & \rybox{\pqcd[0]{2}} &\rybox{\pqcd[0]{3}} & \ldots \\
\hline
& \small 0 & \small 1 & \small 2 & \small 3 & \ldots
 \end{loopsnlegs}
\end{tabular}}
\caption{The double-counting problem caused by naively
  adding cross sections involving matrix elements with different
  numbers of legs.\label{fig:doublecounting}}
\end{center}
\end{figure}
 Assume we have computed the LO cross section for some process, $F$,
 and that we have added an LL shower to it, as in the left-hand pane
 of \figRef{fig:doublecounting}. We know that this only gives us
 an LL description of $F+1$. We now wish to improve this from LL to LO  
 by adding the actual LO matrix element for $F+1$. Since we also want to
 be able to hadronize these events, etc, we again add an LL shower off
 them. However, since the matrix element for $F+1$ is divergent, we must
 restrict it to cover only the phase-space region with at least one
 hard resolved jet, illustrated by the half-shaded boxes in the middle
 pane of \figRef{fig:doublecounting}. 

\index{Matching}%
\index{Double counting}%
\index{Inclusive cross sections}%
\index{Exclusive cross sections}%
Adding these two samples,
 however, we end up counting the LL terms of the inclusive cross
 section for $F+1$ twice, since we are now getting them once from the shower
 off $F$ and once from the matrix element for $F+1$, illustrated by
 the dark shaded (red) areas of the right-hand pane of
 \figRef{fig:doublecounting}. This \emph{double-counting} problem
 would grow worse if we attempted to add more matrix elements, with
 more legs. The cause is very simple. Each such calculation
 corresponds to an \emph{inclusive} cross section, and hence naive
 addition would give
\begin{equation}
\sigma_{\mrm{tot}} =\sigma_{0;\mathrm{incl}} +
  \sigma_{1;\mathrm{incl}} =  \sigma_{0;\mathrm{excl}} +  2\,\sigma_{1;\mathrm{incl}}~.
\end{equation}
Recall the definition of inclusive and exclusive cross
 sections, \eqRef{eq:incexc}: $F$ \emph{inclusive} $=$ $F$ plus
  anything. $F$ \emph{exclusive} $=$ $F$ and only
  $F$. Thus, $\sigma_{F;\mathrm{incl}}=\sum_{\mrm{k}=
    0}^{\infty}\sigma_{F+k;\mrm{excl}}$.

 Instead, we must \emph{match} the coefficients calculated
 by the two parts of the full calculation --- showers and matrix
 elements --- more systematically, for each order in perturbation
 theory, so that the nesting of inclusive and exclusive cross sections
 is respected without overcounting.

Given a parton shower and a matrix-element generator, there are
fundamentally three different ways in which we can consider matching
the two \cite{Giele:2011cb}: slicing, subtraction, and unitarity. The
following subsections will briefly introduce each of these.

\index{Matching}%
\index{Matching!Slicing}%
\index{HERWIG}%
\subsection{Slicing}
\begin{figure}
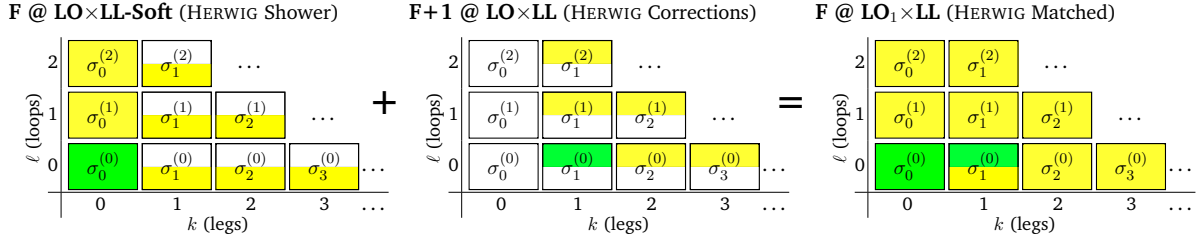

\begin{center}%
\scalebox{0.70}{\begin{tabular}{l}
{\large\bf F @ LO$\times$LL-Soft} (\Hw\ Shower)\\[2mm]
\begin{loopsnlegs}[c]{p{0.25cm}|ccccc}
 \small 2&~\ybox{\pqcd[2]{0}} & \wybox{\pqcd[2]{1}} & \ldots &
\\[2mm]
 \small 1&~\ybox{\pqcd[1]{0}} & \wybox{\pqcd[1]{1}}  
   & \wybox{\pqcd[1]{2}} & \ldots \\[2mm]
 \small 0&~\gbox{\pqcd[0]{0}} & \wybox{\pqcd[0]{1}} 
   & \wybox{\pqcd[0]{2}} &\wybox{\pqcd[0]{3}} & \ldots \\
\hline
& \small 0 & \small 1 & \small 2 & \small 3 & \ldots
 \end{loopsnlegs}
\end{tabular}\hspace*{-7mm}\raisebox{0.2cm}{\huge\bf +}\hspace*{-1.5mm}
\begin{tabular}{l}
{\large\bf F+1 @ LO$\times$LL} (\Hw\ Corrections)\\[2mm]
\begin{loopsnlegs}[c]{p{0.25cm}|ccccc}
 \small 2&~\wbox{\pqcd[2]{0}} & \ywbox{\pqcd[2]{1}} & \ldots & 
\\[2mm]
 \small 1&~\wbox{\pqcd[1]{0}} & \ywbox{\pqcd[1]{1}}  
   & \ywbox{\pqcd[1]{2}} & \ldots \\[2mm]
 \small 0&~\wbox{\pqcd[0]{0}} & \gwbox{\pqcd[0]{1}} 
   & \ywbox{\pqcd[0]{2}} &\ywbox{\pqcd[0]{3}} & \ldots \\
\hline
& \small 0 & \small 1 & \small 2 & \small 3 & \ldots
 \end{loopsnlegs}
\end{tabular}
\hspace*{-7mm}\raisebox{0.2cm}{\huge\bf =}\hspace*{-1.5mm}
\begin{tabular}{l}
{\large\bf F @ LO$_1\times$LL} (\Hw\ Matched)\\[2mm]
\begin{loopsnlegs}[c]{p{0.25cm}|ccccc}
 \small 2&~\ybox{\pqcd[2]{0}} & \ybox{\pqcd[2]{1}} & \ldots & 
\\[2mm]
 \small 1&~\ybox{\pqcd[1]{0}} & \ybox{\pqcd[1]{1}}  
   & \ybox{\pqcd[1]{2}} & \ldots \\[2mm]
 \small 0&~\gbox{\pqcd[0]{0}} & \gybox{\pqcd[0]{1}} 
   & \ybox{\pqcd[0]{2}} &\ybox{\pqcd[0]{3}} & \ldots \\
\hline
& \small 0 & \small 1 & \small 2 & \small 3 & \ldots
 \end{loopsnlegs}
\end{tabular}}
\index{HERWIG}%
\caption{\Hw's original matching
 scheme~\cite{Seymour:1994we,Seymour:1994df}, in which the dead zone
 of the \Hw\ shower was used as an 
  effective ``matching scale'' for one emission beyond a basic hard
 process. 
\label{fig:herwig}}
\end{center}
\end{figure}

 The most commonly encountered matching type is
  currently based on separating (slicing)
  phase space into two regions, one of which is supposed to be
  mainly described by hard matrix elements and the other of which is
  supposed to be described by the shower. This type of 
  approach was first 
  used in \Hw~\cite{Corcella:2000bw}, to
  include matrix-element corrections for one emission beyond the 
  basic hard process \cite{Seymour:1994we,Seymour:1994df}.
  This is illustrated in \figRef{fig:herwig}.
\index{Sudakov factor}
\index{CKKW|see{Matching}}%
\index{MLM|see{Matching}}%
\index{Matching!CKKW}%
\index{Matching!L-CKKW}%
\index{Matching!MLM}%
  The method has since been generalized by several
  independent groups to include
  arbitrary numbers of additional legs, the most well-known of these
  being the CKKW~\cite{Catani:2001cc},
  CKKW-L~\cite{Lonnblad:2001iq,Lavesson:2005xu}, and
  MLM~\cite{Mangano:2006rw,Mrenna:2003if} approaches. 

Effectively,  the shower approximation is set to zero
  above some scale, either due to the presence of explicit dead zones
  in the shower, as in \Hw, or by vetoing any emissions above a certain
  \emph{matching scale}, as in the (L)-CKKW 
  and MLM approaches. The empty part of phase space can then be filled
  by separate events generated according to 
  higher-multiplicity tree-level matrix elements (MEs). In the
  (L)-CKKW and MLM schemes, this process can be iterated to include
  arbitrary numbers of additional hard legs (the practical limit being
  around 3 or 4, due to computational complexity). 

\index{Matching}%
In order to match smoothly with the shower calculation, the
  higher-multiplicity matrix elements 
  must be associated with Sudakov form factors (representing the
  virtual corrections that would have been generated if a shower had 
  produced the same phase-space configuration), and their $\alpha_s$
  factors must be chosen so that, at least at the matching scale, 
  they become identical to the choices made on the shower
  side~\cite{Cooper:2011gk}.  
  The CKKW and MLM approaches do this by 
  constructing ``fake parton-shower histories'' for the
  higher-multiplicity matrix elements. By applying a sequential jet
  clustering algorithm, a tree-like branching structure can be
  created that at least has the same dominant structure as
  that of a parton shower. Given the fake shower tree, $\alpha_s$
  factors can be chosen for each vertex with argument
  $\alpha_s(p_\perp)$ and Sudakov factors can be
  computed for each internal line in the tree. In the CKKW method,
  these Sudakov factors are estimated analytically, while the MLM and CKKW-L
  methods compute them numerically, from the actual shower
  evolution.
  
  Thus, the matched result is identical
  to the matrix element (ME) in the region above the matching scale, 
  modulo higher-order (Sudakov and $\alpha_s$) corrections. We may
  sketch this as  
\begin{equation}
\mbox{Matched (above matching scale)} =
\color{gray}\overbrace{\color{black}\mbox{Exact}}^{\small\mbox{ME}}~
\ {\color{black}\times} \ \overbrace{\color{black}(1 +
\mathcal{O}(\alpha_s))}^{\small\mbox{corrections}} \label{eq:scalebased1}~,
\end{equation}
where the ``shower-corrections'' include the approximate Sudakov factors
and $\alpha_s$ 
reweighting factors applied to the matrix elements in order to obtain
a smooth transition to the shower-dominated region.

\begin{figure}
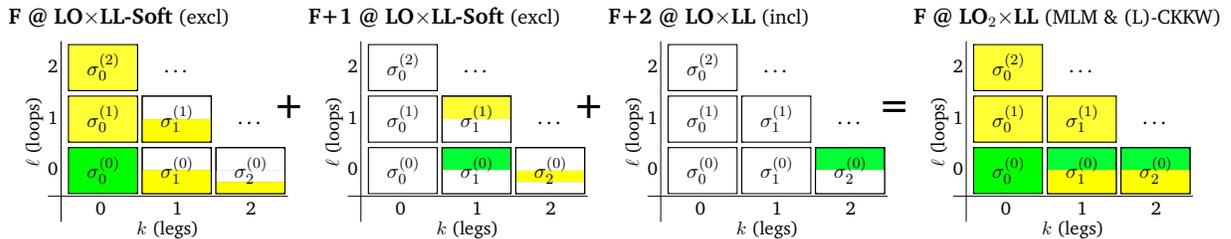

\begin{center}%
\scalebox{0.70}{\begin{tabular}{l}
{\large\bf F @ LO$\times$LL-Soft} (excl)\\[2mm]
\begin{loopsnlegs}[c]{p{0.25cm}|ccccc}
 \small 2&~\ybox{\pqcd[2]{0}} &  \ldots &
\\[2mm]
 \small 1&~\ybox{\pqcd[1]{0}} & \wybox{\pqcd[1]{1}}  
   & \ldots \\[2mm]
 \small 0&~\gbox{\pqcd[0]{0}} & \wybox{\pqcd[0]{1}} 
   & \wwybox{\pqcd[0]{2}} & \\
\hline
& \small 0 & \small 1 & \small 2 &
 \end{loopsnlegs}
\end{tabular}\hspace*{-6mm}\raisebox{0.2cm}{\huge\bf +}\hspace*{-2.5mm}
\begin{tabular}{l}
{\large\bf F+1 @ LO$\times$LL-Soft} (excl)\\[2mm]
\begin{loopsnlegs}[c]{p{0.25cm}|ccccc}
 \small 2&~\wbox{\pqcd[2]{0}} & \ldots & 
\\[2mm]
 \small 1&~\wbox{\pqcd[1]{0}} & \ywbox{\pqcd[1]{1}}  
   &  \ldots \\[2mm]
 \small 0&~\wbox{\pqcd[0]{0}} & \gwbox{\pqcd[0]{1}} 
   & \wywbox{\pqcd[0]{2}} &  \\
\hline
& \small 0 & \small 1 & \small 2 & 
 \end{loopsnlegs}
\end{tabular}\hspace*{-6mm}\raisebox{0.2cm}{\huge\bf +}\hspace*{-2.5mm}
\begin{tabular}{l}
{\large\bf F+2 @ LO$\times$LL} (incl)\\[2mm]
\begin{loopsnlegs}[c]{p{0.25cm}|ccccc}
 \small 2&~\wbox{\pqcd[2]{0}} & \ldots & 
\\[2mm]
 \small 1&~\wbox{\pqcd[1]{0}} & \wbox{\pqcd[1]{1}}  
   &\ldots \\[2mm]
 \small 0&~\wbox{\pqcd[0]{0}} & \wbox{\pqcd[0]{1}} 
   & \gwbox{\pqcd[0]{2}} &  \\
\hline
& \small 0 & \small 1 & \small 2 & 
 \end{loopsnlegs}
\end{tabular}
\hspace*{-6mm}\raisebox{0.2cm}{\huge\bf =}\hspace*{-2.5mm}
\begin{tabular}{l}
{\large\bf F @ LO$_2\times$LL} (MLM \& (L)-CKKW)\\[2mm]
\begin{loopsnlegs}[c]{p{0.25cm}|ccccc}
 \small 2&~\ybox{\pqcd[2]{0}} &  \ldots & 
\\[2mm]
 \small 1&~\ybox{\pqcd[1]{0}} & \ybox{\pqcd[1]{1}}  
   & \ldots \\[2mm]
 \small 0&~\gbox{\pqcd[0]{0}} & \gybox{\pqcd[0]{1}} 
   & \gybox{\pqcd[0]{2}} & \\
\hline
& \small 0 & \small 1 & \small 2 &
 \end{loopsnlegs}
\end{tabular}}
\caption{Slicing, with up
  to two additional emissions beyond the basic process. The showers off
  $F$ and $F+1$ are set to zero above a specific ``matching
  scale''. (The number of coefficients 
  shown was reduced a bit in these plots to make them fit in one row.)
\label{fig:slicing}}
\end{center}
\end{figure}

Below the matching scale, the
small difference between the matrix elements and the shower
approximation can be dropped (since their leading singularities are
identical and this region by definition includes no hard jets), 
yielding the pure shower answer in that region,
\begin{eqnarray}
\mbox{Matched (below matching scale)} & = &
\color{gray}\overbrace{\color{black}\mbox{Approximate}}^{\mbox{shower}}
\ {\color{black}+} \ \overbrace{\color{black}(\mbox{Exact} -
\mbox{Approximate})}^{\mbox{correction}}\nonumber\\ 
& = & \mbox{Approximate} \ + \ \mbox{non-singular} \nonumber \\
& \to & \mbox{Approximate}~.
\end{eqnarray}
This type of strategy is illustrated in \figRef{fig:slicing}.

\index{Matching}%
As emphasized above, 
since this strategy is discontinuous across phase space, a main point
here is to ensure that the behavior across the matching scale be as
\index{CKKW}smooth as possible. CKKW showed \cite{Catani:2001cc} that it is
possible to remove any  dependence on the matching scale through  NLL
precision by careful choices of all ingredients in the matching;
technical details of the implementation 
(affecting the 
$\mathcal{O}(\alpha_s)$ terms in eq.~(\ref{eq:scalebased1}))
are important, and the dependence on the unphysical matching scale
may be larger than NLL unless the implementation
matches the theoretical algorithm
precisely~\cite{Lonnblad:2001iq,Lavesson:2005xu,Lavesson:2008ah}. 
Furthermore, since the Sudakov
\index{Matching!L-CKKW}\index{Matching!MLM}factors are generally computed using showers
(MLM, L-CKKW) or a  
\index{Matching!CKKW}shower-like formalism (CKKW), while the real corrections
are computed 
using matrix elements, care must be taken not to (re-)introduce
differences that could break the detailed real-virtual balance that
ensures unitarity among the singular parts, see
e.g.,~\cite{Cooper:2011gk}. 

\index{Matching}%
\index{QCD!Scale invariance}%
\index{Matching scale|see{Matching}}%
\index{Matching!Matching scale}%
It is advisable not to choose the matching scale too low. This is
again  essentially due to the approximate scale invariance of QCD
imploring us to 
write the matching scale as a ratio, rather than as an absolute
number. 
If one uses a
very low matching 
scale, the higher-multiplicity matrix elements will already be quite
singular, leading to very large LO cross sections before matching. 
After matching, these large cross sections are tamed by
the Sudakov factors produced by the matching scheme, and hence the
final cross sections may still look reasonable. But 
the higher-multiplicity matrix elements in
general contain subleading singularity structures, beyond those
accounted for by the shower, and hence the delicate
line of detailed balance that ensures unitarity has most
assuredly been overstepped. We therefore recommend not to take the 
matching scale lower than about an order of magnitude below the
 characteristic scale of the hard process. 

One should also be aware that all strategies of this type are 
quite computing intensive. This is basically due to the fact that 
a separate phase-space generator is required for each of the
$n$-parton correction terms, with each such sample a priori consisting
of weighted events such that a separate unweighting step (often with
quite low efficiency) is needed before an
unweighted sample can be produced. 

\index{Matching}%
\index{Matching!Subtraction}%
\index{Subtraction!In the context of matching}%
\subsection{Subtraction}
\begin{figure}
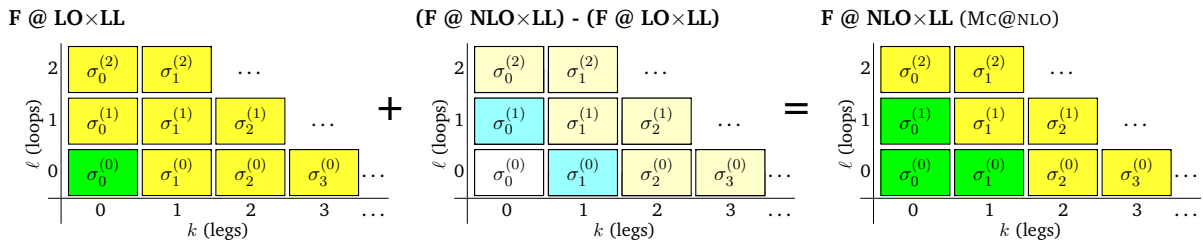

\begin{center}%
\scalebox{0.70}{\begin{tabular}{l}
{\large\bf F @ LO$\times$LL}\\[2mm]
\begin{loopsnlegs}[c]{p{0.25cm}|ccccc}
 \small 2&~\ybox{\pqcd[2]{0}} & \ybox{\pqcd[2]{1}} & \ldots &
\\[2mm]
 \small 1&~\ybox{\pqcd[1]{0}} & \ybox{\pqcd[1]{1}}  
   & \ybox{\pqcd[1]{2}} & \ldots \\[2mm]
 \small 0&~\gbox{\pqcd[0]{0}} & \ybox{\pqcd[0]{1}} 
   & \ybox{\pqcd[0]{2}} &\ybox{\pqcd[0]{3}} & \ldots \\
\hline
& \small 0 & \small 1 & \small 2 & \small 3 & \ldots
 \end{loopsnlegs}
\end{tabular}
\hspace*{-7mm}\raisebox{0.2cm}{\huge\bf +}\hspace*{-1.5mm}
\begin{tabular}{l}
{\large\bf (F @ NLO$\times$LL) - (F @ LO$\times$LL)}\\[2mm]
\begin{loopsnlegs}[c]{p{0.25cm}|ccccc}
 \small 2&~\eggbox{\pqcd[2]{0}} & \eggbox{\pqcd[2]{1}} & \ldots &
\\[2mm]
 \small 1&~\cyanbox{\pqcd[1]{0}} & \eggbox{\pqcd[1]{1}}  
   & \eggbox{\pqcd[1]{2}} & \ldots \\[2mm]
 \small 0&~\wbox{\pqcd[0]{0}} & \cyanbox{\pqcd[0]{1}} 
   & \eggbox{\pqcd[0]{2}} &\eggbox{\pqcd[0]{3}} & \ldots \\
\hline
& \small 0 & \small 1 & \small 2 & \small 3 & \ldots
 \end{loopsnlegs}
\end{tabular}
\hspace*{-7mm}\raisebox{0.2cm}{\huge\bf =}\hspace*{-1.5mm}
\begin{tabular}{l}
{\large\bf F @ NLO$\times$LL} (\Fw)\\[2mm]
\begin{loopsnlegs}[c]{p{0.25cm}|ccccc}
 \small 2&~\ybox{\pqcd[2]{0}} & \ybox{\pqcd[2]{1}} & \ldots &
\\[2mm]
 \small 1&~\gbox{\pqcd[1]{0}} & \ybox{\pqcd[1]{1}}  
   & \ybox{\pqcd[1]{2}} & \ldots \\[2mm]
 \small 0&~\gbox{\pqcd[0]{0}} & \gbox{\pqcd[0]{1}} 
   & \ybox{\pqcd[0]{2}} &\ybox{\pqcd[0]{3}} & \ldots \\
\hline
& \small 0 & \small 1 & \small 2 & \small 3 & \ldots
 \end{loopsnlegs}
\end{tabular}}
\caption{\Fw. In the middle
  pane, cyan boxes
  denote non-singular correction terms, while the egg-colored ones 
  denote showers off such corrections, which cannot lead to
  double-counting at the LL level.
\label{fig:fw}}
\end{center}
\end{figure}

Another way of matching two calculations is by subtracting one
  from the other and correcting by the difference, schematically
\begin{equation}
\mbox{Matched} =
\color{gray}\overbrace{\color{black}\mbox{Approximate}}^{\mbox{shower}}
\color{black} \ + \ \color{gray}\overbrace{\color{black}
(\mbox{Exact}-\mbox{Approximate})}^{\mbox{correction}}~. \label{eq:additive}
\end{equation}
\index{NLO}%
\index{Matching!MCatNLO}
This looks very much like the structure of a subtraction-based NLO
fixed-order calculation, \secRef{sec:subtraction}, in which the shower
approximation here plays the role of subtraction terms, and indeed
this is what is used  in 
strategies like \Fw{}
\cite{Frixione:2002ik,Frixione:2003ei,Frixione:2008ym}, illustrated in
\figRef{fig:fw}.
In this type of approach, however, negative-weight events will
generally occur, for instance in  
phase-space points where the approximation is larger than the exact
answer. 

Negative weights are not in principle an insurmountable
problem. Histograms can be filled with each event counted 
according to its weight, as usual. However, negative weights do affect
efficiency. Imagine a worst-case scenario in which 1000
positive-weight events have been generated, along with 999 negative-weight
ones (assuming each event weight has the same absolute value, the
closest one can get to an unweighted sample in the presence of
negative weights). 
The statistical precision of the MC answer would be equivalent
to one event, for 2000 generated, i.e., a big loss in convergence
rate. 

\begin{figure}
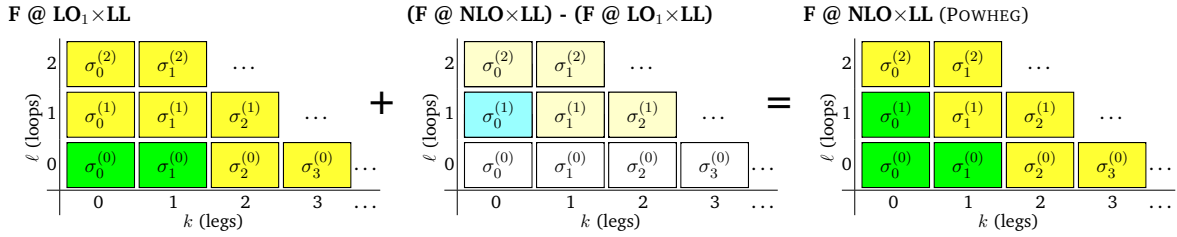

\begin{center}%
\scalebox{0.685}{\begin{tabular}{l}
{\large\bf F @ LO$_1\times$LL}\\[2mm]
\begin{loopsnlegs}[c]{p{0.25cm}|ccccc}
 \small 2&~\ybox{\pqcd[2]{0}} & \ybox{\pqcd[2]{1}} & \ldots &
\\[2mm]
 \small 1&~\ybox{\pqcd[1]{0}} & \ybox{\pqcd[1]{1}}  
   & \ybox{\pqcd[1]{2}} & \ldots \\[2mm]
 \small 0&~\gbox{\pqcd[0]{0}} & \gbox{\pqcd[0]{1}} 
   & \ybox{\pqcd[0]{2}} &\ybox{\pqcd[0]{3}} & \ldots \\
\hline
& \small 0 & \small 1 & \small 2 & \small 3 & \ldots
 \end{loopsnlegs}
\end{tabular}
\hspace*{-7mm}\raisebox{0.2cm}{\huge\bf +}\hspace*{-1.5mm}
\begin{tabular}{l}
{\large\bf (F @ NLO$\times$LL) - (F @ LO$_1\times$LL)}\\[2mm]
\begin{loopsnlegs}[c]{p{0.25cm}|ccccc}
 \small 2&~\eggbox{\pqcd[2]{0}} & \eggbox{\pqcd[2]{1}} & \ldots &
\\[2mm]
 \small 1&~\cyanbox{\pqcd[1]{0}} & \eggbox{\pqcd[1]{1}}  
   & \eggbox{\pqcd[1]{2}} & \ldots \\[2mm]
 \small 0&~\wbox{\pqcd[0]{0}} & \wbox{\pqcd[0]{1}} 
   & \wbox{\pqcd[0]{2}} &\wbox{\pqcd[0]{3}} & \ldots \\
\hline
& \small 0 & \small 1 & \small 2 & \small 3 & \ldots
 \end{loopsnlegs}
\end{tabular}
\hspace*{-7mm}\raisebox{0.2cm}{\huge\bf =}\hspace*{-1.5mm}
\begin{tabular}{l}
{\large\bf F @ NLO$\times$LL} (\Pw)\\[2mm]
\begin{loopsnlegs}[c]{p{0.25cm}|ccccc}
 \small 2&~\ybox{\pqcd[2]{0}} & \ybox{\pqcd[2]{1}} & \ldots &
\\[2mm]
 \small 1&~\gbox{\pqcd[1]{0}} & \ybox{\pqcd[1]{1}}  
   & \ybox{\pqcd[1]{2}} & \ldots \\[2mm]
 \small 0&~\gbox{\pqcd[0]{0}} & \gbox{\pqcd[0]{1}} 
   & \ybox{\pqcd[0]{2}} &\ybox{\pqcd[0]{3}} & \ldots \\
\hline
& \small 0 & \small 1 & \small 2 & \small 3 & \ldots
 \end{loopsnlegs}
\end{tabular}}
\caption{\Pw. In the middle
  pane, cyan boxes
  denote non-singular correction terms, while the egg-colored ones 
  denote showers off such corrections, which cannot lead to
  double-counting at the LL level.
\label{fig:powheg}}
\end{center}
\end{figure}

\index{Matching!Subtraction}%
\index{NLO}%
\index{MCatNLO|see{Matching}}%
\index{Matching!MCatNLO}%
In practice, generators like MC@NLO ``only'' produce around 10\% or
less events with negative weights, so the convergence rate should not
be severely affected for ordinary applications. Nevertheless,  
the problem of negative weights 
motivated the development of the so-called \Pw\ approach 
\index{POWHEG|see{Matching}}\index{Matching!POWHEG}\cite{Frixione:2007vw}, illustrated in \figRef{fig:powheg}, 
which is
constructed specifically to prevent negative-weight events from
occurring and simultaneously to be more independent of which
parton-shower algorithm it is used with. In the \Pw\ method, 
one effectively modifies the real-emission probability for the first
emission
to agree with
the $F+1$ matrix element (this is covered under unitarity, below). One
is then left with a purely virtual correction, which will typically be
positive, at least for processes for which the NLO cross section is
larger than the LO one. 

\begin{figure}
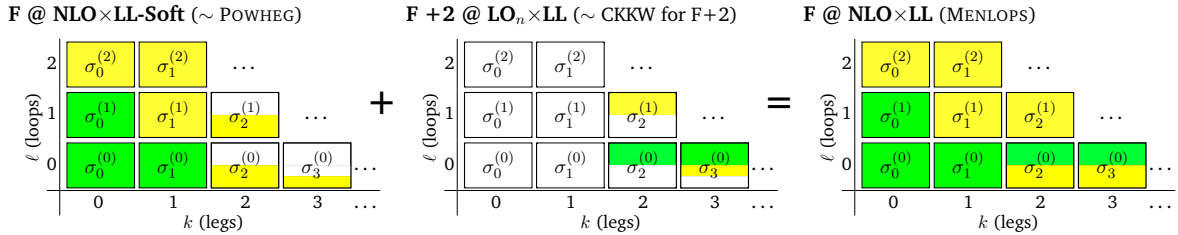

\begin{center}%
\scalebox{0.685}{\begin{tabular}{l}
{\large\bf F @ NLO$\times$LL-Soft} ($\sim$ \Pw)\\[2mm]
\begin{loopsnlegs}[c]{p{0.25cm}|ccccc}
 \small 2&~\ybox{\pqcd[2]{0}} & \ybox{\pqcd[2]{1}} & \ldots &
\\[2mm]
 \small 1&~\gbox{\pqcd[1]{0}} & \ybox{\pqcd[1]{1}}  
   & \wybox{\pqcd[1]{2}} & \ldots \\[2mm]
 \small 0&~\gbox{\pqcd[0]{0}} & \gbox{\pqcd[0]{1}} 
   & \wybox{\pqcd[0]{2}} &\wwybox{\pqcd[0]{3}} & \ldots \\
\hline
& \small 0 & \small 1 & \small 2 & \small 3 & \ldots
 \end{loopsnlegs}
\end{tabular}
\hspace*{-7mm}\raisebox{0.2cm}{\huge\bf +}\hspace*{-1.5mm}
\begin{tabular}{l}
{\large\bf F +2 @ LO$_n\times$LL} ($\sim$ CKKW for F+2)\\[2mm]
\begin{loopsnlegs}[c]{p{0.25cm}|ccccc}
 \small 2&~\wbox{\pqcd[2]{0}} & \wbox{\pqcd[2]{1}} & \ldots &
\\[2mm]
 \small 1&~\wbox{\pqcd[1]{0}} & \wbox{\pqcd[1]{1}}  
   & \ywbox{\pqcd[1]{2}} & \ldots \\[2mm]
 \small 0&~\wbox{\pqcd[0]{0}} & \wbox{\pqcd[0]{1}} 
   & \gwbox{\pqcd[0]{2}} &\gywbox{\pqcd[0]{3}} & \ldots \\
\hline
& \small 0 & \small 1 & \small 2 & \small 3 & \ldots
 \end{loopsnlegs}
\end{tabular}
\hspace*{-7mm}\raisebox{0.2cm}{\huge\bf =}\hspace*{-1.5mm}
\begin{tabular}{l}
{\large\bf F @ NLO$\times$LL} (\textsc{Menlops})\\[2mm]
\begin{loopsnlegs}[c]{p{0.25cm}|ccccc}
 \small 2&~\ybox{\pqcd[2]{0}} & \ybox{\pqcd[2]{1}} & \ldots &
\\[2mm]
 \small 1&~\gbox{\pqcd[1]{0}} & \ybox{\pqcd[1]{1}}  
   & \ybox{\pqcd[1]{2}} & \ldots \\[2mm]
 \small 0&~\gbox{\pqcd[0]{0}} & \gbox{\pqcd[0]{1}} 
   & \gybox{\pqcd[0]{2}} &\gybox{\pqcd[0]{3}} & \ldots \\
\hline
& \small 0 & \small 1 & \small 2 & \small 3 & \ldots
 \end{loopsnlegs}
\end{tabular}}
\caption{\textsc{Menlops}. Note that each of the \Pw\ and CKKW samples
 are composed 
  of separate sub-samples, as illustrated in \figsRef{fig:slicing} and
  \ref{fig:powheg}. 
\label{fig:menlops}}
\end{center}
\end{figure}

\index{NLO}%
\index{Matching}%
The advantage of these methods
is obviously that NLO corrections to the Born level can be
systematically incorporated. However, a systematic way of
extending this strategy beyond the first additional emission is not
available, save for combining them with a slicing-based strategy
\index{MENLOPS|see{Matching}}\index{Matching!MENLOPS}for the additional legs, as in \textsc{Menlops}
\cite{Hamilton:2010wh}, illustrated in \figRef{fig:menlops}. 
These issues are, however, no more severe than
in ordinary fixed-order NLO 
approaches, and hence they are not viewed as disadvantages if the
point of reference is an NLO computation. 

\index{Matching!Unitarity}%
\index{Matching}%
\subsection{Unitarity}
The oldest, and in my view most attractive, 
approach \cite{Bengtsson:1986et,Bengtsson:1986hr}
  consists of working out the 
 shower approximation to a given fixed order, and correcting
 the shower splitting functions at that order by a multiplicative
 factor given by the ratio of the matrix element
 to the shower approximation, phase-space point by phase-space
 point. We may sketch this as 
\begin{equation}
\mbox{Matched} =
\color{gray}\overbrace{\color{black}\mbox{Approximate}}^{\mbox{shower}}
\ {\color{black}\times} \
\overbrace{\color{black}\frac{\mbox{Exact}}{\mbox{Approximate}}}^{\mbox{correction}}~. \label{eq:multiplicative}
\end{equation}
When these correction factors are inserted back into the
shower evolution, they guarantee that the shower evolution off $n-1$
partons correctly reproduces the $n$-parton matrix elements, 
without the need to generate a separate $n$-parton sample. 
That is, the shower approximation is essentially used as a 
pre-weighted (stratified) all-orders phase-space generator, on which a
more exact answer can subsequently be
imprinted order by order in perturbation theory. Since the shower is 
already optimized for exactly the kind of singular structures that
occur in QCD, very fast computational speeds can therefore be obtained
with this method~\cite{LopezVillarejo:2011ap}.

\index{PYTHIA}%
\index{Matching}%
In the original
approach \cite{Bengtsson:1986et,Bengtsson:1986hr}, used by \Py
\cite{Sjostrand:2006za,Sjostrand:2014zea}, this was only 
worked out for one additional emission beyond the basic hard
process. 
\index{POWHEG}%
In \Pw~\cite{Frixione:2007vw,Alioli:2010xd}, 
it was extended to include also virtual corrections to the Born-level
matrix element. 
\index{VINCIA}%
\index{VINCIA!Matching}%
\index{Matching!VINCIA}%
Finally, in \Vc, it has been 
extended to include arbitrary numbers of emissions at tree
level~\cite{Giele:2011cb,LopezVillarejo:2011ap} and one emission at
loop level~\cite{Hartgring:2013jma}, though that method has so far
only been applied to final-state showers.

\begin{figure}[t]
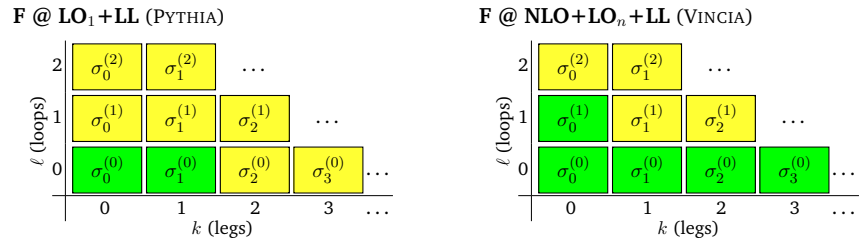

\begin{center}%
\scalebox{0.70}{
\begin{tabular}{l}
{\large\bf F @ LO$_{1}$+LL} (\Py)\\[2mm]
\begin{loopsnlegs}[c]{p{0.25cm}|ccccc}
 \small 2&~\ybox{\pqcd[2]{0}} & \ybox{\pqcd[2]{1}} & \ldots &
\\[2mm]
 \small 1&~\ybox{\pqcd[1]{0}} & \ybox{\pqcd[1]{1}}  
   & \ybox{\pqcd[1]{2}} & \ldots \\[2mm]
 \small 0&~\gbox{\pqcd[0]{0}} & \gbox{\pqcd[0]{1}} 
   & \ybox{\pqcd[0]{2}} &\ybox{\pqcd[0]{3}} & \ldots \\
\hline
& \small 0 & \small 1 & \small 2 & \small 3 & \ldots
 \end{loopsnlegs}
\end{tabular}\hspace*{1cm}
\begin{tabular}{l}
{\large\bf F @ NLO+LO$_{n}$+LL} (\Vc)\\[2mm]
\begin{loopsnlegs}[c]{p{0.25cm}|ccccc}
 \small 2&~\ybox{\pqcd[2]{0}} & \ybox{\pqcd[2]{1}} & \ldots & 
\\[2mm]
 \small 1&~\gbox{\pqcd[1]{0}} & \ybox{\pqcd[1]{1}}  
   & \ybox{\pqcd[1]{2}} & \ldots \\[2mm]
 \small 0&~\gbox{\pqcd[0]{0}} & \gbox{\pqcd[0]{1}} 
   & \gbox{\pqcd[0]{2}} &\gbox{\pqcd[0]{3}} & \ldots \\
\hline
& \small 0 & \small 1 & \small 2 & \small 3 & \ldots
 \end{loopsnlegs}
\end{tabular}}
\caption{\Py~(left) and \Vc~(right). Unitarity-based. Only one event
 sample is produced by 
  each of these methods, and hence no sub-components are shown. 
\label{fig:match-unitary}}
\end{center}
\end{figure}
An illustration of the perturbative coefficients that can be 
included in each of  these approaches is given in
\figRef{fig:match-unitary}, as usual with green (darker shaded) boxes
representing exact coefficients and yellow (light shaded) boxes
representing logarithmic approximations. 

\index{Matching}%
\index{Sudakov factor}%
Finally, two more
properties unique to this method deserve mention. Firstly, 
since the corrections modify the actual shower evolution kernels, the
corrections are automatically \emph{resummed} in the Sudakov
exponential, which should improve the logarithmic precision once
$k\ge2$ is included, 
and secondly, since the shower is \emph{unitary},  an initially unweighted sample
of $(n-1)$-parton configurations remains unweighted, with 
no need for a separate event-unweighting or event-rejection
step.

\clearpage
\section{Hadronisation and Soft Hadron-Hadron Physics \label{sec:soft}}
\index{Hadronisation}\index{Underlying event}%
We here give a very brief overview of the main aspects of soft
QCD that are relevant for hadron-hadron collisions, such as
hadronisation, minimum-bias and soft-inclusive physics, and
the so-called underlying event. This will be kept at a
pedestrian level and is largely based on the reviews in
\cite{pdg2012,Buckley:2011ms,Skands:2011pf}. 

\index{Hadronisation}%
\index{Primary hadrons}%
\index{Hadronisation scale}%
\index{Monte Carlo!Event generators}%
In the context of event generators,  \emph{hadronisation} 
denotes the process by which a set of coloured partons (\emph{after}
showering) is 
transformed into a set of colour-singlet \emph{primary} hadrons, which
may then subsequently decay further.  
This non-perturbative transition takes place at the \emph{hadronisation
scale} $Q_{\mathrm{had}}$, which by construction 
is identical to the infrared cutoff of the parton
shower.  
In the absence of a first-principles solution to the relevant
dynamics, event generators use QCD-inspired phenomenological models to
describe this transition. 

\index{Confinement}%
The problem can
be stated as follows: given a set of partons resolved at a scale of
$Q_\mathrm{had}\sim$ 1 GeV, we need a ``mapping'' 
from this set onto a set of on-shell colour-singlet (i.e., confined)
hadronic states. MC models do this in three steps:
\index{String model}%
\begin{enumerate}
\item Map the partonic system  onto a continuum of high-mass 
hadronic states (called  ``strings'' or ``clusters'').
\item Iteratively map strings/clusters onto discrete set of primary
  hadrons (via string breaks / cluster splittings / cluster decays).
\item Sequential decays into secondaries ($\rho \to
  \pi\pi$, $\Lambda\to n \pi$, $\pi^0 \to \gamma\gamma$, ...).  
\end{enumerate}
The physics governing this mapping is non-perturbative. However, 
we do have some 
knowledge of the properties that such a solution must have. 
For instance, Poincar\'e invariance, unitarity, and 
causality are all concepts that apply beyond 
perturbation theory. 
\index{Lattice QCD}
In addition, lattice QCD provides us a means
of making explicit quantitative studies in a genuinely non-perturbative
setting (albeit only of certain questions). 

\begin{figure}[t]
\centering\includegraphics*[scale=0.86]{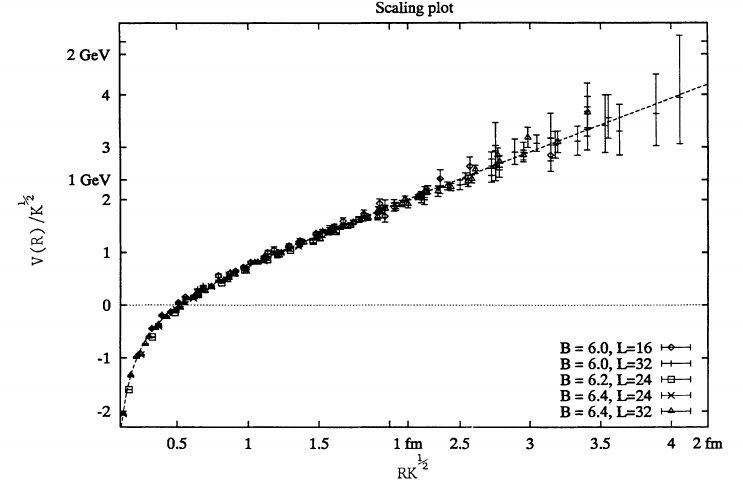}
\caption{Static quark-antiquark potential, as a function of separation 
  distance, in quenched lattice QCD, from 
\cite{Bali:1992ab}\label{fig:linear}. Note that the axes are
scaled by units of the string tension $\sqrt{\kappa}\sim 420$
MeV. Additional labels corresponding to 1 GeV and 2 GeV are also
provided, on the $y$ axis, and to 1 fm and 2 fm, on the $x$ axis.
A constant term, $V_0$, has been subtracted from all the results. 
The dashed line corresponds to $V(R) = R - \pi/(12R)$.}
\end{figure}

\index{Confinement}%
\index{Linear confinement|see{Confinement}}%
An important result in ``quenched'' lattice
QCD\footnote{Quenched QCD implies no ``dynamical'' quarks, i.e., no
  $g\to q\bar{q}$ splittings  allowed.} is that the potential
of the colour-dipole field between a charge and an anticharge 
appears to grow linearly with the separation of the charges, at
distances greater than about 0.5 femtometers; this behavior is
illustrated by the plot shown in 
\figRef{fig:linear}, from~\cite{Bali:1992ab}. (Note that the axes are
scaled by units of the string tension $\sqrt{\kappa}\sim 420$
MeV. Additional labels corresponding to 1 GeV and 2 GeV are also
provided, on the $y$ axis, and to 1 fm and 2 fm, on the $x$ axis.)  
\index{String model}%
This is known as ``linear confinement'', and it forms the starting
point for the \emph{string model of hadronisation}, discussed below in
\secRef{sec:stringModel}. 
\index{Preconfinement}%
\index{Cluster model}%
Alternatively, a property of 
perturbative QCD called ``preconfinement''~\cite{Amati:1979fg} 
is the basis of the 
\emph{cluster model of hadronisation}, described in
\cite{Buckley:2011ms,pdg2012}. 

\index{PYTHIA}%
\index{HERWIG}%
\index{SHERPA}%
In the generator landscape,
\Py~\cite{Sjostrand:2006za,Sjostrand:2014zea}, \Qgsjet~\cite{Ostapchenko:2007qb}, and \Sibyll~\cite{Riehn:2015oba}
use string fragmentation models (as do \Ar~\cite{Lonnblad:1992tz}, \Dpmjet~\cite{Bopp:2005cr},
\Phojet~\cite{Bopp:1998rc}, and \Vc~\cite{Fischer:2016vfv} via interfaces to \Py), while
\Hw~\cite{Bellm:2015jjp} and \Sh~\cite{Gleisberg:2008ta} use cluster fragmentation. \Epos~\cite{Pierog:2013ria} uses a combination of
string hadronisation and a hydrodynamics-inspired statistical
hadronisation model.

\index{Parton level}%
\index{Hadronisation scale}%
It should be emphasised that the so-called \emph{parton level}
that can be obtained by switching off hadronisation in
an MC generator, is not a universal concept, 
since each model defines the hadronisation scale 
differently. E.g., the hadronisation scale can be defined by a cutoff
in invariant mass, transverse momentum, or some other quantity, 
with different tunes using different values for the cutoff. The former
is equivalent to using different effective factorisation schemes, and
the latter corresponds to different factorisation scales, for the soft
non-perturbative component of the modelling. 
Comparisons to
distributions at this level (i.e., with hadronisation switched off) 
may therefore be used to provide an idea of the 
overall impact of hadronisation corrections within a given model, 
but should be avoided in the context of physical observables.
\index{Fragmentation functions}%
Note also that the corresponding MC \emph{fragmentation functions} 
are intrinsically defined at the hadronisation scale. They can
therefore not be compared directly to those that are 
used in fixed-order / analytical-resummation contexts, which are
typically defined at a factorisation scale of order the scale of the
hard process.

We use the term ``soft hadron-hadron physics'' to 
comprise all scattering processes for which a hard, perturbative scale
is not required to be present\footnote{Note, however,
that while a hard scale is not \emph{required} to be present, 
it is not explicitly required to be absent either. 
Thus, both diffractive, minimum-bias, pile-up and
underlying-event processes will have tails towards high-$p_\perp$ 
 physics as well. For example, even $t\bar{t}$ pair production
   can be viewed as a tail of minimum-bias interactions, and there is
   a tail of diffractive processes in which hard dijets can be
   produced diffractively (see, e.g.,~\cite{Navin:2010kk}).}.
This includes elastic, diffractive, minimum-bias, and pile-up
processes, as well as the
physics contributing to the so-called underlying event. We 
give a brief introduction to such processes
in \secRef{sec:soft-processes}. 

We round off with a discussion of the data constraints that enter in
the tuning of Monte Carlo models in \secRef{sec:tuning}, and give an
outline of a procedure that could be followed in a realistic set-up.

\subsection{The String Model of Hadronisation\label{sec:stringModel}}
\index{String model}%
Starting from early concepts developed by Artru and Mennessier 
\cite{Artru:1974hr}, several hadronisation models based on strings
were proposed in the late 1970'ies and early 80'ies. 
Of these, the most widely used today is
the so-called Lund model, implemented in the \Py~code. 
\index{PYTHIA}%
We shall therefore concentrate on that particular model here, though
many of the overall concepts would be shared by any string-inspired
method. (More extended discussions can be found in Andersson's
book~\cite{Andersson:1998tv} and in an older comprehensive Physics Reports
review~\cite{Andersson:1983ia}.)

\index{Quarks!As string endpoints}%
Consider the production of a $q\bar{q}$ pair from vacuum, for instance
in the process $e^+e^-\to \gamma^*/Z\to q\bar{q} \to
\mbox{hadrons}$. As the quarks 
move apart, linear confinement implies that a potential 
\index{Confinement}%
\begin{equation}
V(R) = \kappa\, R \label{eq:string}
\end{equation}
is asymptotically reached for large distances, $R$. At short 
distances, there is a Coulomb term proportional to $1/R$ as well,
cf.~\figRef{fig:linear}, but
this is neglected in the Lund model. Such a potential describes a 
string with tension (energy per unit length) $\kappa$, with the
value~\cite{Bali:1992ab} 
\begin{equation}
\kappa~\sim~(420\,\mbox{MeV})^2~\sim~0.18\,\mbox{GeV}^2~\sim~0.9\,\mbox{GeV/fm}~, 
\end{equation}
which, for comparison with the world of macroscopic objects, would be 
sufficient to lift a 16-ton truck~\cite{travis}. 

\begin{figure}[t]
\centering
\scalebox{0.9}{
\begin{tabular}{c}
\rotatebox{90}{\small $1\,$fm}
\end{tabular}\hspace*{-7mm}
\begin{tabular}{c}
\includegraphics*[scale=1.0]{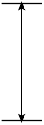}
\end{tabular}
\begin{tabular}{c}
\includegraphics*[scale=0.75]{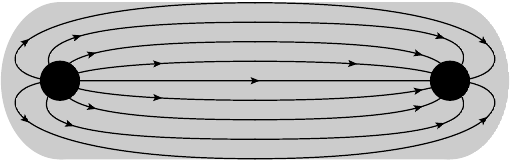}
\end{tabular}}
\caption{Illustration of the transition between a Coulomb potential
  at short distances to the string-like one of \eqRef{eq:string} at
  large $q\bar{q}$ separations.\label{fig:string}}
\end{figure}

The string can be thought of as 
parameterizing the position of the axis of 
a cylindrically symmetric flux tube, illustrated in 
\figRef{fig:string}. Such simple $q-\bar{q}$ strings form the starting
point for the string model. 
\index{Gluons!As kinks on strings}%
More complicated multi-parton topologies
are treated by  
representing gluons as transverse ``kinks'', e.g., 
$q-g-\bar{q}$. The space-time evolution is then 
slightly more involved~\cite{Andersson:1998tv}, and modifications
to the fragmentation 
model to handle stepping across gluon corners have to be included,
but the main point is that there are no separate free parameters for
gluon jets. Differences with respect to 
quark  fragmentation arise simply because quarks are only
connected to a single string piece, while gluons have one on either
side, increasing the energy loss per unit (invariant) time 
from a gluon to the string by a
factor of 2 relative to quarks, which can be compared to the ratio of
colour Casimirs $C_A/C_F = 2.25$. Another appealing feature
of the model is that low-energy gluons are absorbed smoothly into the
string, without leading to modifications. This improves the stability
of the model with respect to variations of the infrared behavior of
the parton shower.

As the partonic string endpoints move apart, their kinetic energy is gradually
converted to potential energy, stored in the growing string spanned
between them. In the ``quenched'' approximation, in which $g\to
q\bar{q}$ splittings are not allowed, this process would continue
until the endpoint quarks have lost \emph{all} their
momentum, at which point they would reverse direction and be
accelerated by the now shrinking string. 

\begin{figure}[t]
\centering
\scalebox{0.9}{
\begin{tabular}{ccc}
\begin{tabular}{c}
\includegraphics*[scale=0.75]{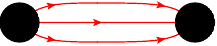} \ \
\includegraphics*[scale=0.75]{strBrk3.pdf}\\[1mm]
\includegraphics*[scale=0.75]{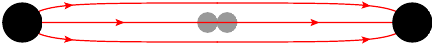}\\[1mm]
\includegraphics*[scale=0.75]{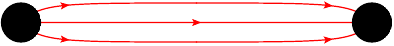}
\end{tabular}\hspace*{1.2cm}
& 
\begin{tabular}{c}
\rotatebox{90}{\small time}
\end{tabular}\hspace*{-2.2cm}
\begin{tabular}{r}
\includegraphics*[scale=1.4]{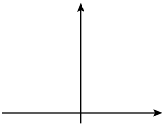}\\[-3mm]
{\small space}
\end{tabular}\hspace*{-0.5cm}&
\raisebox{0.9cm}{\begin{tabular}{c}\includegraphics*[scale=0.6]{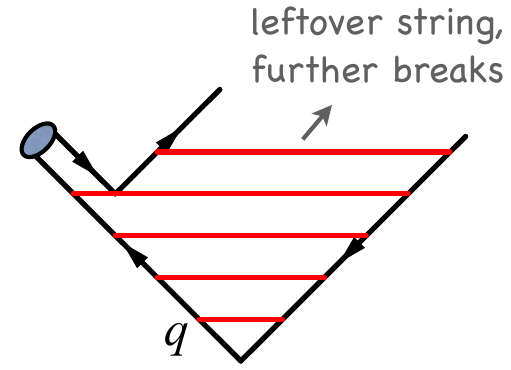}
\end{tabular}}\\
\hspace*{-1cm} a) & & b)
\end{tabular}
}
\caption{{\sl a)} Illustration of string breaking by quark pair creation in the
  string field. {\sl b)} Illustration of the algorithmic choice to
  process the fragmentation from the outside-in, splitting off a
  single on-shell hadron in each step.\label{fig:stringbreak}}
\end{figure}

In the real world,
 quark-antiquark fluctuations inside the string field
can make the transition to become real particles by absorbing 
energy from the string, thereby screening the original endpoint 
charges from each other and breaking the string into two separate
colour-singlet pieces, $(q\bar{q}) \to
(q\bar{q}')+(q'\bar{q})$, illustrated in \figRef{fig:stringbreak}\!a.
This process then continues until only ordinary hadrons remain. (We
will give more details on the individual string breaks below.)

\index{Causality}%
Since the string breaks are causally disconnected
(as can easily be realised from space-time diagrams like the one in
\figRef{fig:stringbreak}, see also \cite{Andersson:1998tv}), they do
not have to be considered in any 
specific time-ordered sequence. In the Lund model, the string breaks are
instead generated starting 
with the leading (``outermost'') hadrons, 
containing the endpoint quarks, and iterating inwards
towards the center of the string, alternating randomly between
fragmentation off the left- and right-hand sides, respectively,
\figRef{fig:stringbreak}\!\!b.  
One can thereby split off a single well-defined hadron in each step,
with a mass that, for unstable hadrons, is selected according to a
Breit-Wigner distribution.  

\index{String breaks}%
\index{Tunneling}%
The details of the individual string breaks 
are not known from first principles. The
Lund model invokes the idea of quantum mechanical tunneling, with a Gaussian suppression of the transverse momenta and masses imparted to
the produced quarks\footnote{The full form of the series essentially accounts for the probability that no string-decay event already happened and incorporates the exclusion principle that no more than one event can occur for each given quantum state of the produced pair, while the limiting Gaussian form is the na\"\i ve probability per phase-space volume; see~\cite{Casher:1978wy}.}, 
\begin{eqnarray}
\mathrm{Prob}(m_q^2,\pt[q]^2)&\propto&-\ln\left[1-\exp\left(\frac{-\pi
    (m_q^2 + \pt[q]^2)}{\kappa}\right) \right]  ~= ~
\sum_{n=1}^{\infty} \frac{1}{n} \exp\left(\frac{-n \pi (
    m_q^2 + \pt[q]^2 ) }{\kappa}\right) \nonumber
\\[2mm]
& \!\!\stackrel{m_q^2 \gg \kappa/\pi}{\to}\!\! & \exp\left(\frac{-\pi
    m_q^2}{\kappa}\right)\exp\left( \frac{-\pi \pt[q]^2}{\kappa}\right)
~, \label{eq:tunneling}
\end{eqnarray}
where $m_q$ is the mass of the produced quark (the relevant range of values for the \emph{constituent} up, down, and strange quark masses is typically taken to be $\sim$ 300 -- 500 MeV) and $\pt$ is the transverse momentum imparted to it by the breakup process (with the $\bar{q}$ having the opposite
\pt).

\index{Schwinger Mechanism}%
This form of the suppression factor was originally derived in 1951 by Schwinger in QED, for creation of electron-positron pairs in a strong constant electric field~\cite{Schwinger:1951nm}\footnote{Note that extremely strong electric fields are required, so the effect has not been observed in the lab yet, although attempts are being made at large laser facilities, see e.g.~\cite{Dunne:2008kc}.}. Its generalisation to QCD was first treated using WKB tunneling approximations~\cite{Casher:1978wy,Glendenning:1983qq}, and there is also a more recent claim of an exact path-integral solution~\cite{Nayak:2005pf}, in all cases finding the same basic form as in the QED case\footnote{We note that the form in~\cite{Nayak:2005pf} also contains a term proportional to $|d_{abc}E^aE^bE^c|^2$ with $d_{abc}$ the symmetric structure constants of QCD and $E_a$ the chromoelectric field with index $a\in[1,8]$.}. Since the Schwinger tunneling probability is derived for an infinitely extended constant field, while QCD flux tubes in nature have a finite size, modifications to the above baseline picture may be considered. E.g., a flux tube with radius $r_s$ could imply a cutoff for momenta $p_{\perp \mathrm{max}} \sim 2 r_s \kappa/\pi$~\cite{Casher:1978wy}. It is not straightforward to define the relevant radius unambiguously, however, and we note that, within the effective description in an MC program, tails to higher $p_\perp$ values may be justified anyway due to unresolved perturbative effects below the shower cutoff. A high-$p_\perp$ cutoff is therefore typically not imposed in the MC implementation.
See~\cite{Fischer:2016zzs} for a phenomenological study that explores some alternative assumptions (e.g., a thermal suppression) in an event-generator context.  

\index{Flavour!In string breaks}%
Due to the factorisation of the \pt and $m$ dependence
implied by \eqRef{eq:tunneling}, the \pt spectrum of produced
quarks in this model is independent of the quark flavour, with a
universal average value of 
\begin{eqnarray}
\sigma^2 & \equiv & \left<\pt[q]^2\right> =  \left< p_x^2 + p_y^2\right> \\
& = & \frac{\pi}{\kappa}\int_{-\infty}^{\infty} \!\mathrm{d}p_x \int_{-\infty}^{\infty}
      \!\mathrm{d}p_y \ (p_x^2 + p_y^2) \ \exp\left(\frac{-\pi(p_x^2 +
      p_y^2)}{\kappa}\right) \\ 
& = & \frac{\pi}{\kappa} \int_0^\infty \! \mathrm{d}p_\perp^2 \
      p_\perp^2 \ 
\exp\left(-\pi p_\perp^2/ \kappa\right) \\ 
& = & \kappa/\pi \sim
(240\,\mbox{MeV})^2~. 
\end{eqnarray}
Bear in mind that ``transverse'' is here defined
with respect to the string axis.  Thus, the
\pt in a frame where the string is moving is modified by a Lorentz
boost factor. Also bear in mind that $\sigma^2$ is here 
a purely non-perturbative parameter. 
\index{Hadronisation scale}%
In a Monte Carlo model
with a fixed shower cutoff $Q_\mrm{had}$, 
the effective amount of ``non-perturbative'' \pt 
may be larger than this,  due to effects of additional unresolved
soft-gluon radiation below $Q_\mrm{had}$ (for example, the ``Monash
Tune'' of \Py~8~\cite{Skands:2014pea} has a final-state shower cutoff
at $p_{\perp\mathrm{min}} = 500\, \mathrm{MeV}$ and uses a value of
$\sigma = 335\,\mbox{MeV}$). 
\index{String model}%
\index{Monte Carlo!Tuning}%
In principle, the magnitude of this additional component 
should scale with the cutoff, but in practice it is up to
the user to enforce this by retuning (see \secRef{sec:tuning}) 
the effective $\sigma$ parameter 
when changing the hadronisation scale. Since hadrons
receive $\pt$ contributions from two breakups, one on either side, 
their average transverse momentum squared will be twice as large,
\begin{equation}
\left<\pt[h]^2\right> = 2\sigma^2~. 
\end{equation}
Finally, we note that the assumption $m_q^2 \gg \kappa/\pi$ which is used to justify neglecting the subleading terms \eqRef{eq:tunneling} may be open to question and could lead to deviations of order 10\% from the universal Gaussian spectrum, for realistic constituent masses. 

\index{Flavour!In string breaks}%
\index{Mesons}%
The mass suppression implied by \eqRef{eq:tunneling} is however less 
straightforward to interpret. Integrated over $p_\perp$, the Schwinger prediction is that the differential string decay probability (per unit space-time volume) via pair production of quarks with constituent mass $m_q$,  is proportional to \index{Schwinger Mechanism}  
\begin{equation}
P(m_q^2) \propto \sum_{n=1}^\infty \frac{1}{n^2} \exp\left( -\frac{n \pi m_q^2}{\kappa} \right)\stackrel{m_q^2\gg \kappa/\pi}{\to} \exp\left( -\frac{\pi m_q^2}{\kappa} \right).
\end{equation}
However, since quark masses
are notoriously difficult to define for light quarks, the value of the
strangeness suppression must effectively be extracted from experimental
measurements, e.g., of the $K/\pi$ ratio, with a
resulting suppression of roughly $s/u \sim s/d \sim$ 0.2 -- 0.3.
Inserting even comparatively low values for the charm quark mass in
\eqRef{eq:tunneling}, however, one obtains a relative suppression of
charm of the order of $10^{-11}$. Heavy quarks can therefore safely be
considered to be produced only in the perturbative stages 
and not by the soft fragmentation.

\index{Baryons}%
\index{Diquarks}%
Baryon production can be incorporated in the same basic picture
\cite{Andersson:1981ce},  
by  allowing  string breaks to occur also by the production of
pairs of so-called \emph{diquarks}, 
loosely bound states of two quarks in an overall $\bar{3}$
representation (e.g., ``red + blue $\sim$ antigreen'', cf.~the
rules for colour combinations in \secRef{sec:L}). Again,
the relative  
rate of diquark-to-quark production is not known a priori and must be
extracted from experimental measurements, e.g., of the $p/\pi$ ratio. 
More advanced
scenarios for baryon production have also been proposed, in particular
the so-called popcorn model \cite{Andersson:1984af,Eden:1996xi}, which
is normally used in addition 
to the diquark picture and then acts to decrease the correlations
among neighboring baryon-antibaryon pairs by allowing mesons to be formed
inbetween them. 
\index{PYTHIA}%
\index{String junctions}%
Within the \Py\ framework, 
a fragmentation model including explicit \emph{string
  junctions}~\cite{Sjostrand:2002ip} has so far only been applied to
baryon-number-violating new-physics processes  and to the 
description of beam remnants (and then acts to increase baryon
stopping  \cite{Sjostrand:2004pf}). 

This brings us to the next step of the algorithm: 
assignment of the produced quarks within hadron
multiplets. 
\index{Flavour!In string breaks}%
\index{Mesons}%
Using a nonrelativistic classification of spin states, 
the fragmenting $q$ ($\bar{q}$) may combine with the $\bar{q}'$ 
($q'$) from a newly created breakup to produce either a vector or
a pseudoscalar meson, or, if diquarks are involved, either a spin-1/2
or spin-3/2 baryon. Unfortunately, the string model is  entirely
unpredictive in this respect, and this is therefore the sector that
contains the largest amount of free parameters. 
 From spin counting alone, one would expect the ratio $V/S$ of
vectors to pseudoscalars to be 3, but this is modified by the $V$--$S$
mass splittings, which implies a phase-space suppression of vector production,
with corresponding suppression parameters to be extracted from data. 

Especially for the light flavours, the substantial difference
in phase space caused by the $V$--$S$ mass splittings 
implies a rather large suppression of vector production. Thus,
for $D^*/D$, the effective  
ratio is already reduced to about $\sim$ 1.0~--~2.0, while for $K^*/K$
and $\rho/\pi$,  extracted values range from 0.3~--~1.0. (Recall, as 
always, that these are production ratios of \emph{primary hadrons},
hence feed-down from secondary decays of heavier hadrons 
complicates the extraction of these parameters from
experimental data, in particular for the lighter hadron species.)

\index{Mesons}%
\index{PYTHIA}%
The production of higher meson resonances is assumed to be low in a string
framework\footnote{The four $L = 1$ multiplets are implemented in \Py, 
but are disabled by default, largely because several states are poorly
known and thus may result in a worse overall description when
included.}. 
For diquarks, separate parameters control the relative
rates of spin-1 diquarks vs.\ spin-0 ones and, likewise, have to
extracted from data, with resulting values of order $(qq)_1/(qq)_0
\sim$ 0.075~--~0.15. 

\index{Fragmentation functions}%
\index{Lund symmetric fragmentation function}%
\begin{figure}[tp]
\centering
\begin{tabular}{ccc}
\underline{The $a$ parameter} & & \underline{The $b$ parameter}\\[2mm]
 \hspace*{1.8cm}$\color{red}a=0.9$ \hspace*{4mm}
$\color{blue}a=0.1$
 & & \hspace*{1cm}$\color{red}b=0.5$ \hspace*{7mm} $\color{blue}b=2.0$ 
\\
\includegraphics*[scale=1.1]{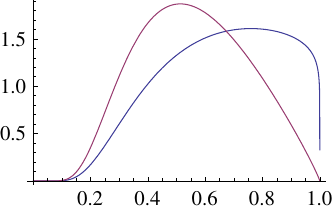} & \hspace*{1cm} &
\includegraphics*[scale=1.1]{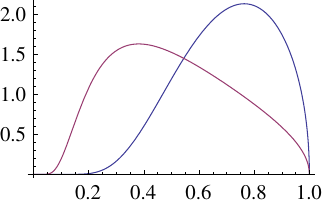}\\
$b=1$, $m_\perp=1$ GeV & &
$a=0.5$, $m_\perp=1$ GeV 
\end{tabular}
\caption{Normalised Lund symmetric fragmentation
function, for fixed $m_\perp = 1$ GeV. 
{\sl Left:} variation of the $a$ parameter, from 0.1 (blue)
to 0.9 (red), with fixed $b = 1$ GeV$^{-2}$. {\sl Right:}
variation of the $b$ parameter, from 0.5 (red) to 2 (blue) GeV$^{-2}$, 
with fixed $a=0.5$.\label{fig:ffab}}
\end{figure}
With $\pt^2$ and $m^2$ now fixed, the final step is to select the
fraction, $z$, of the fragmenting endpoint quark's longitudinal
momentum that is carried by the created hadron. In this respect, the
string picture is substantially more predictive than for the flavour
selection. Firstly, the requirement  that the
fragmentation be independent of the sequence in which breakups are
considered (causality) imposes a ``left-right symmetry'' on the
possible form of the fragmentation function, $f(z)$, 
with the solution~\cite{Andersson:1983jt}
\begin{equation}
f(z) \propto \frac{1}{z} (1-z)^a \exp\left(-\frac{b\,(m_h^2 +
  \pt[h]^2)}{z}\right)~, \label{eq:lund-symm}
\end{equation}
which is known as the \emph{Lund symmetric fragmentation function}
(normalised to unit integral). 
The $a$ and $b$ parameters, illustrated in \figRef{fig:ffab}, 
are the only free
parameters of the fragmentation function, though $a$ may in principle
be flavour-dependent. Note that the explicit mass dependence in $f(z)$ 
implies a harder fragmentation function for heavier hadrons (in the
rest frame of the string). 

\index{Lund symmetric fragmentation function!Bowler modification}%
For massive endpoints (e.g., $c$ and $b$ quarks), which do not
move along straight lightcone sections, the exponential suppression
with string area leads to modifications of the form
\cite{Bowler:1981sb}, 
\begin{equation}
f(z) \to f(z)/z^{b\,m_Q^2}~, 
\end{equation}
with $m_Q$  the heavy-quark mass. Strictly speaking, this is 
the only fragmentation function that is consistent with causality in
the string model, though a few alternative forms are typically
provided as well. 

As a by-product, the probability distribution in invariant time $\tau$
of $q'\bar{q}'$ breakup vertices, or equivalently $\Gamma = (\kappa
\tau)^2$, is also obtained, with $\mathrm{d}P/\mathrm{d}\Gamma \propto
\Gamma^a \exp(-b\Gamma)$ 
\index{Area law}%
implying an area law for the colour
flux~\cite{Wilson:1974sk}, and
the average breakup time lying along a 
hyperbola of constant invariant time $\tau_0 \sim
10^{-23}\mathrm{s}$~\cite{Andersson:1998tv}. 

We may also ask, e.g., how many units of rapidity does the particle
production from a string span? Measuring $p_z$ along the string
direction and defining rapidity by 
\begin{equation}
y = \frac12\ln\left(\frac{E+p_z}{E-p_z}\right)~,
\end{equation}
the absolute highest rapidity that can be reached, by a pion traveling
exactly along the string direction and taking all of the endpoint
quark's energy, is $y_\mrm{max} = \ln(2E_q/m_\pi)$. I.e., 
the rapidity region covered by a fragmenting string scales
logarithmically with the energy, and since the density of hadrons
produced per unit rapidity is roughly constant (modulo endpoint
effects), the average number of hadrons produced by string fragmentation
likewise scales logarithmically with energy.

\begin{figure}[tp]
\centering
\includegraphics*[scale=0.65]{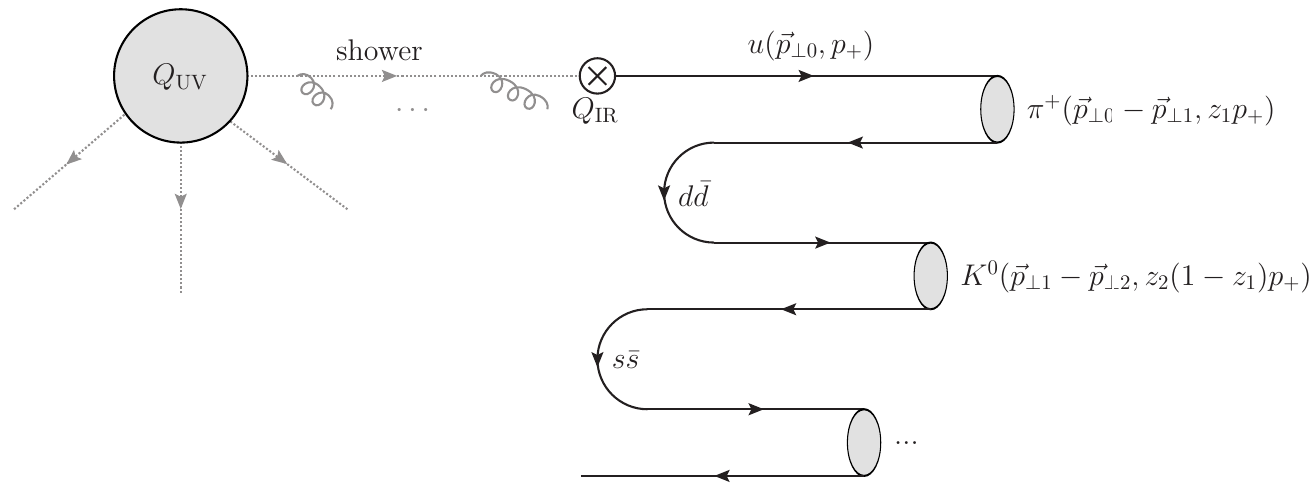}
\caption{Iterative selection of flavours and
  momenta in the Lund string-fragmentation model. \label{fig:iterative}}
\end{figure}

The iterative selection of flavours, \pt,
and $z$ values is illustrated in \figRef{fig:iterative}.
A parton produced in a hard process at some high scale
$Q_{\mathrm{UV}}$ emerges from 
  the parton shower, at the hadronisation scale $Q_{\mathrm{IR}}$, 
  with 3-momentum $\vec{p}=(\vec{p}_{\perp 0},p_+)$, where the ``+'' on the third
  component denotes ``light-cone'' momentum, $p_\pm = E\pm p_z$. 
  Next, an adjacent $d\bar{d}$ pair from the
  vacuum is created, with relative transverse momenta $\pm\pt[1]$. 
  The fragmenting quark combines with the $\bar{d}$ from the breakup
  to form a $\pi^+$, which carries off a fraction $z_1$ of the total
  lightcone momentum $p_+$. The next hadron carries off a fraction
  $z_2$ of the remaining momentum, etc. 

\subsection{Soft Hadron-Hadron Processes \label{sec:soft-processes}}
\index{Minimum-bias events}%
\index{Diffractive scattering}%
\index{Underlying event}%
\index{Elastic scattering}%
\index{Inelastic scattering}%
The total hadron-hadron ($hh$) cross section is around 100 mb at LHC
energies~\cite{Antchev:2013paa}, growing slowly with the CM energy, 
$\sigma_\mrm{tot}(s) \propto s^{0.096}$~\cite{Donnachie:2013xia}. 
There are essentially four types of physics processes,
which together make up $\sigma_\mrm{tot}$:
\begin{enumerate}
\item Elastic scattering:  $hh\to hh$;
\item Single diffractive dissociation: $hh\to h+{\rm gap}+X$, with
``gap'' denoting an empty rapidity region, and $X$ anything that is
not the original beam particle;
\item Double diffractive dissociation: $hh\to
X+{\rm gap}+X$ (both hadrons ``blow up'');
\item Inelastic non-diffractive scattering: everything else.
\end{enumerate}
In principle, higher ``multi-gap'' diffractive components may be
defined as well, the most important one being central diffraction: $hh\to
h+{\rm gap}+X+{\rm gap}+h$, see the discussion of diffraction 
in \secRef{sec:diffraction} below. 

Some important differences exist between theoretical and experimental 
terminology\cite{Khoze:2010by}.  In the experimental setting, 
diffraction is defined by an observable rapidity gap, with 
$|\Delta y|_\mrm{gap} \gsim 3$ typically giving clean diffractive
samples. 
In the MC context, however, 
each diffractive process type produces a
whole spectrum of gaps, with small ones suppressed but not
excluded. Likewise, events of non-diffractive origin may produce accidental
rapidity gaps, now with large ones suppressed (exponentially) 
but not excluded, and in
the transition region there could even be quantum mechanical
interference between the two. Due to this unphysical 
model dependence of theoretical definitions of diffraction, 
we strongly advise to phrase measurements first and foremost in terms
of physical observables, and only seek to connect with theory
models as a second, separate, step. 

The distinction between elastic and
inelastic events \emph{is}, however, unambiguous (modulo $pp \to
pp\gamma$ events); the final state either contains just two 
protons, or not.
The total hadron-hadron cross section can therefore be written as
a sum of these two physically distinguishable components, 
\begin{equation}
\sigma_{\mathrm{tot}}(s) = 
\sigma_{\mathrm{el}}(s) +
\sigma_{\mathrm{inel}}(s)~, 
\end{equation}
where $s=(p_A+p_B)^2$ is the beam-beam center-of-mass energy squared. 

Another potentially confusing term is ``minimum bias'' (MB). This 
originates from the experimental requirement of a minimal energy or 
number of hits
in a given (experiment-dependent) instrumented region near the beam,
used to determine whether  
there was any non-trivial activity in the event, or not. This
represents the smallest possible 
``trigger bias'' that the corresponding experiment is capable of. 
There is thus no universal definition of ``min-bias''; each
experiment has its own. We give a brief discussion of minimum bias 
in \secRef{sec:MB} below.

Finally, in events containing a hard parton-parton interaction,
the \emph{underlying event} (UE) can be roughly conceived of as
the \emph{difference} between QCD with and without including the
remnants of the original beam hadrons.  Without such ``beam
remnants'', only the hard interaction itself, and its  
associated parton showers and hadronisation, would contribute to the
observed particle production. In reality, after the partons that
participate in the hard interaction have been taken out, 
the remnants still contain whatever is left of the incoming beam hadrons,
including also a partonic substructure, which leads to the possibility
of  \emph{multiple parton interactions}
(MPI). We discuss MPI-based models of the UE in \secRef{sec:mpi}
below. 
Other useful reviews of MPI-based MC models can be found
in \cite{Buckley:2011ms,Skands:2011pf}. Analytical models are mostly
formulated only for double parton scattering, see
e.g., \cite{Blok:2010ge,Blok:2011bu,Gaunt:2012wv,Manohar:2012jr}.  

\subsubsection{Diffractive Scattering \label{sec:diffraction}}
\index{Diffractive scattering}%
As mentioned above, if the beam particles $A$ and/or $B$ are not
elementary, the inelastic final states may be 
 divided into ``diffractive'' and ``non-diffractive''
topologies. This is a qualitative classification, usually based on
whether the final state looks like
the decay of an excitation of the beam particles 
(diffractive\footnote{An example of a process that would be labeled as
  diffractive would be if one the protons is excited to a
$\Delta^+$ which then decays back to $p^++\pi^0$, without anything else
  happening in the event.  In
  general, a whole tower of possible diffractive excitations are
  available, which in the continuum limit can be described by a mass
  spectrum falling roughly as $\dd{M^2}/M^2$.}), 
or not (non-diffractive), or upon the presence of a
large rapidity gap somewhere in the final state which would separate
such excitations. 

Given that an event has been labeled as diffractive, either within the
context of a theoretical model, or by a final-state observable, 
we may distinguish
between three different classes of diffractive topologies, which it is
possible to distinguish between physically, at least in principle. 
In double-diffractive (DD) events, both of the beam particles are
diffractively excited and hence none of them survive the collision
intact. In single-diffractive (SD) events, only one of the beam
particles gets excited and the other survives intact. The last 
diffractive topology is  central diffraction (CD),  in which 
both of the beam particles survive intact, leaving an excited system
in the central region between them. (This latter topology includes
``central exclusive production'' where a single particle is produced
in the central region.)  
That is,  
\begin{equation}
\sigma_{\mathrm{inel}}(s) = 
\sigma_{\mathrm{SD} }(s)
+
\sigma_{\mathrm{DD}}(s) +
\sigma_{\mathrm{CD}}(s) + 
\sigma_{\mathrm{ND}}(s) ~, \label{eq:diff}
\end{equation}
where ``ND'' (non-diffractive, here understood not to include elastic
scattering) contains no gaps in the event
consistent with the chosen definition of diffraction. Further, 
each of the diffractively excited systems in the events labeled SD,
DD, and CD, respectively, may in principle consist of
several subsystems with gaps between them. Eq.~(\ref{eq:diff}) may 
thus be defined to be exact, within a specific definition of
diffraction, even in the presence of multi-gap events. 
Note, however, that different
theoretical models almost always use different (model-dependent) definitions of
diffraction, and therefore the individual components in one model are
in general not directly comparable to those of another. It is
therefore important that data be presented at the level of physical
observables if unambiguous conclusions are to be drawn from them.

\subsubsection{Minimum Bias \label{sec:MB}}
\index{Minimum-bias events}%
  In principle, \emph{everything} that produces a hit in a given
  experiment's minimum-bias trigger, is a subset of
minimum-bias (MB). In particular, since there is no explicit veto on
  hard activity, it is useful to keep in mind that 
minimum bias includes a diverse mixture of both soft and hard
  processes, though the  fraction that is made up of hard
  high-$p_\perp$ processes is only a small tail compared to the total
  minimum-bias cross section\footnote{
Furthermore, since only a tiny fraction of the total
minimum-bias rate can normally be stored, the minimum-bias sample
would give quite poor statistics if used for hard physics studies. Instead,
separate dedicated 
hard-process triggers are typically included in addition to the
minimum-bias one, in order to ensure maximal statistics also for hard
physics processes.}.

In theoretical contexts, 
the term ``minimum-bias'' is often used with a slightly different
meaning; to denote specific (classes of) inclusive soft-QCD
subprocesses in a given model. 
Since these two usages are not exactly identical, in these lectures 
we have chosen to reserve the term ``minimum bias''  to pertain strictly to
definitions of experimental measurements, and instead use  
the term ``soft inclusive'' physics as a generic descriptor for the
class of processes which generally dominate the various experimental
``minimum-bias'' measurements in theoretical models. This parallels
the terminology used in the review \cite{Buckley:2011ms}, from which
most of the discussion here has been adapted. 
See \eqRef{eq:diff} above for a compact overview of the types of
physical processes that contribute to minimum-bias data samples.
For a more detailed description of Monte Carlo models of this physics, 
see \cite{Buckley:2011ms}.

\subsection{Underlying Event and Multiple Parton Interactions \label{sec:mpi}}

In this subsection, we focus on the physics of 
multiple parton interactions (MPI) as a theoretical basis for
understanding both inelastic, non-diffractive 
processes (minimum-bias), as well as the so-called underlying event
(a.k.a.\ the jet pedestal effect). 
Keep in mind, however, that especially at low
multiplicities, and when rapidity gaps are present, the contributions from
diffractive processes should not be ignored.

Due to the simple fact that hadrons are composite, multi-parton
interactions (several distinct parton-parton interactions in one and
the same hadron-hadron collision) will always be there --- but how
many, and how much additional energy and tracks do they deposit in a
given measurement region? The first detailed Monte Carlo model for 
perturbative MPI was proposed by Sj\"ostrand in 
\cite{Sjostrand:1987su}, and with some variation 
this still forms the basis for most modern
implementations~\cite{Buckley:2011ms}.

\begin{figure}[t]
\centering
\includegraphics[scale=0.42]{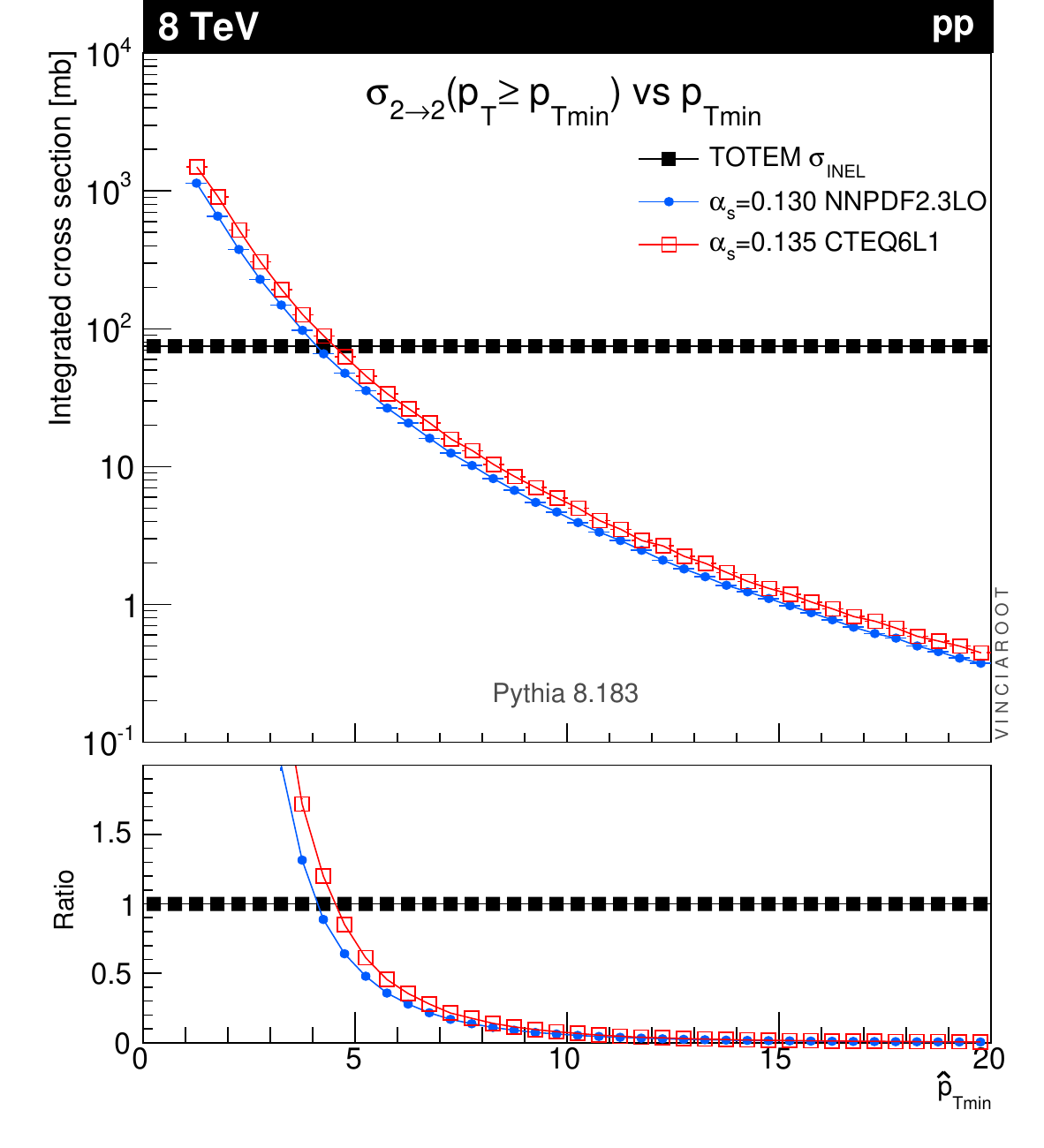}
\caption{Proton-proton collisions at 8 TeV. LO QCD parton-parton
  cross section (integrated above 
  $p_{T\mrm{min}}$, for two different $\alpha_s$ and PDF choices)
  compared to the total inelastic hadron-hadron 
  cross section. Towards the right of the plot, we see, as
  expected, that hard dijet events is only a tiny fraction of the
  total cross section. The fact that the curves cross at a scale of
  order 5 GeV is interpreted to mean that this is a characteristic
  scale relevant for MPI. \cite{Skands:2014pea}.
\label{fig:sigmaQCD}
}
\end{figure}

The first crucial observation is that the $t$-channel propagators
appearing in perturbative QCD $2\to2$ scattering
almost go on shell at low $\pt{}$, causing the differential cross 
sections to become very large, behaving roughly as
\begin{equation}
d \sigma_{2\to 2} \propto \ 
\frac{d{t} }{{t}^2} \ \sim \ 
 \frac{{d}{\pt^2}}{\pt^4}~. \label{eq:dpt4}
\end{equation}
At LHC energies, this \emph{parton-parton} cross section
becomes larger than the total \emph{hadron-hadron} cross section at
$p_\perp$ scales of order $4-5~\mbox{GeV}$. This is illustrated in
\figRef{fig:sigmaQCD}, in which we compare the integrated QCD
parton-parton cross 
section (with two different $\alpha_s$ and PDF choices) to the total
inelastic cross section measured by TOTEM~\cite{Antchev:2013paa}, for
$pp$ collisions at $\sqrt{s} =8~\mrm{TeV}$.
In the context
of MPI models, this is interpreted straightforwardly 
to mean that \emph{each} hadron-hadron collision
contains \emph{several} few-GeV parton-parton collisions. 

In the limit that all the partonic interactions are independent and equivalent,
one would simply have a Poisson distribution in the number of MPI, with average
\begin{equation}
 \langle n \rangle(\ptmin{}) = \frac{\sigma_{2\to2}(\ptmin{})}
{\sigma_{\rm tot}} ~,
\end{equation}
with $\ptmin{}$ a lower cutoff scale which we shall return to below,
and $\sigma_\mrm{tot}$ a measure of the inelastic
hadron-hadron cross section. 
This simple reinterpretation in fact expresses unitarity;
instead of the total interaction cross
section diverging as $\ptmin{} \to 0$ (which would violate
unitarity), we have restated the problem so
that it is now the {\it number of MPI per collision} that
diverges, with the total cross section remaining finite. 

Two important ingredients remain to fully
regulate the remaining divergence. Firstly, the interactions cannot use up more
momentum than is available in the parent hadron. This 
 suppresses the large-$n$ tail of the estimate
above.
\index{PYTHIA}%
In \Py-based models, the
MPI are ordered in $\pt{}$~\cite{Sjostrand:1987su,Sjostrand:2004ef},
and the parton 
densities for each successive 
interaction are explicitly constructed so
that the sum of $x$ fractions can never be greater than unity. In
the \Hw\ models~\cite{Butterworth:1996zw,Bahr:2009ek}, instead the
uncorrelated estimate of  
$\langle n \rangle$ above is used as an initial guess, but
the generation of actual MPI is stopped once the
energy-momentum conservation limit is reached.

The second ingredient invoked to suppress the number of interactions,
at low $\pt{}$ and $x$, is colour screening; if 
the wavelength $\sim$ $1/\pt{}$ of an exchanged coloured parton 
becomes larger than a typical colour-anticolour separation distance, 
it will only see an {\it average} colour charge that vanishes in the limit
$\pt{} \to 0$, hence leading to suppressed interactions. 
This provides an infrared cutoff for
MPI similar to that provided by the hadronisation
scale for parton showers. 
A first estimate of the colour-screening cutoff would be 
the proton size, $
\ptmin{} \approx \hbar/r_{p} \approx
0.3~\mbox{GeV} \approx \Lambda_{\rm QCD}$, 
but empirically this appears to be far too low. In current models, one
replaces the proton radius $r_{p}$ in the above formula by a ``typical
colour screening distance,'' i.e.,\ an average size of a region within
which the net compensation of a given colour  
charge occurs. This number is not known from first 
principles, though it may be related to
saturation~\cite{Ryskin:2011qe}. In current MPI models, it is 
perceived of simply as an effective cutoff parameter, to be determined
from data. 

Note that  the partonic cross sections depend upon the PDF set used,
and therefore the optimal value to use for the 
cutoff will also depend on this
choice~\cite{Schulz:2011qy}. Note also that 
the cutoff does not have to be energy-independent. 
Higher energies imply that parton densities can be probed at smaller
$x$ values, where the number of partons rapidly increases. Partons
then become closer packed and the colour-screening distance $d$
decreases. The uncertainty on the scaling of the cutoff is a major concern when
extrapolating between different collider 
energies~\cite{Skands:2010ak,Schulz:2011qy,Skands:2013asa}. 

We now turn to the origin of the observational fact that hard jets appear to sit on top of a ``pedestal'' of underlying activity, which on average appears to be distributed evenly at all rapidities (i.e., also far from the jet cores). 
\begin{quote}
  ``Outside the [jet], a constant ET plateau is observed, whose height is independent of the jet ET. Its value is substantially higher than the one observed for minimum bias events.''~(\textbf{UA1} Collaboration (1983)~\cite{Arnison:1983gw})
\end{quote}
That is, the so-called ``underlying event'' (UE) is 
substantially more active, with larger fluctuations, than the average min-bias
event. In the MPI context, this is interpreted as a consequence of
impact-parameter-dependence: in peripheral
collisions, only a small fraction of events contain any
high-$\pt{}$ activity, whereas central collisions are more
likely to contain at least one hard scattering; a high-$\pt{}$ triggered
sample will therefore be biased towards small impact parameters, $b$,
with a larger number of MPI (and associated increased activity).

\begin{figure}[t]
  \centering
  \includegraphics[scale=0.375]{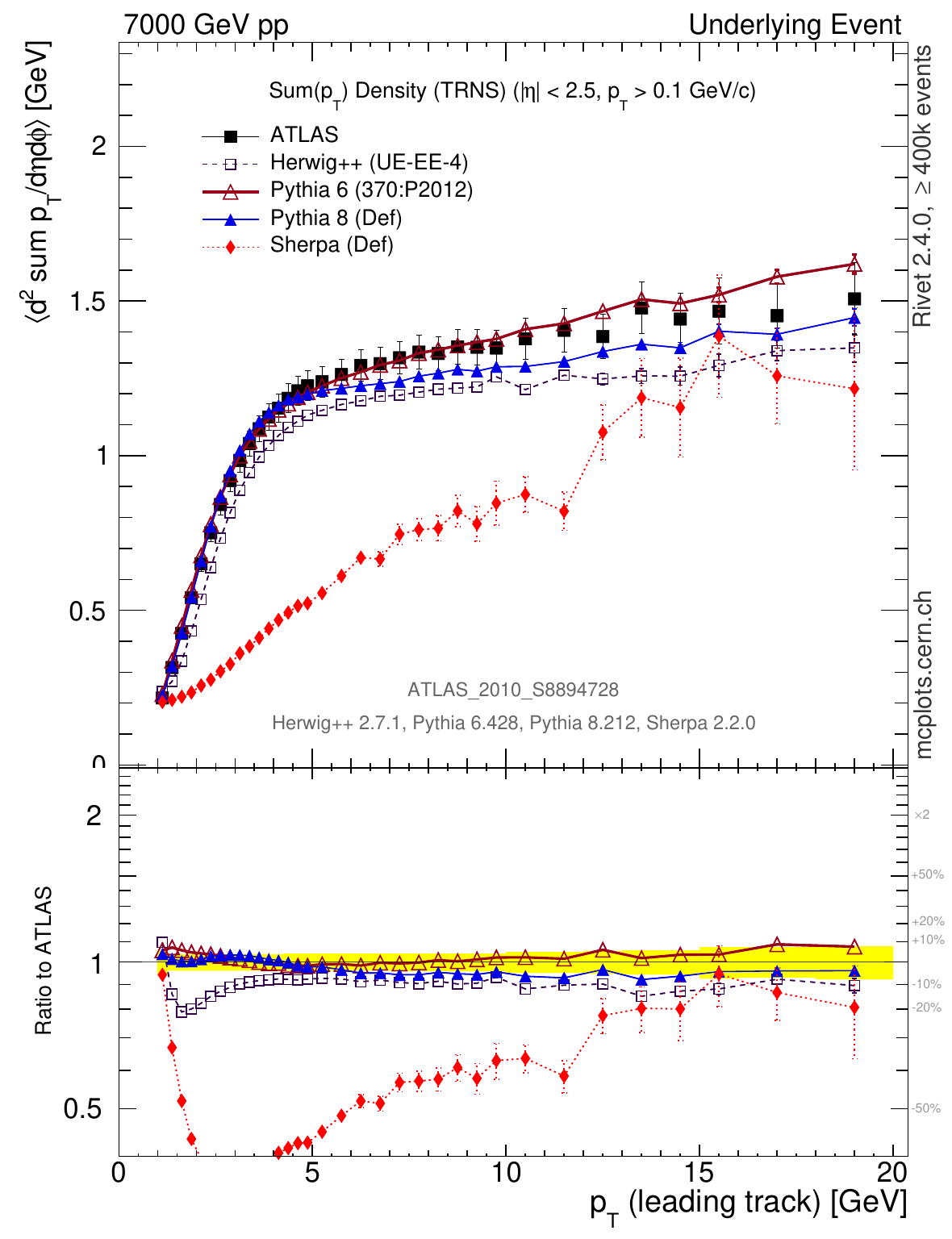}~
  \includegraphics[scale=0.38]{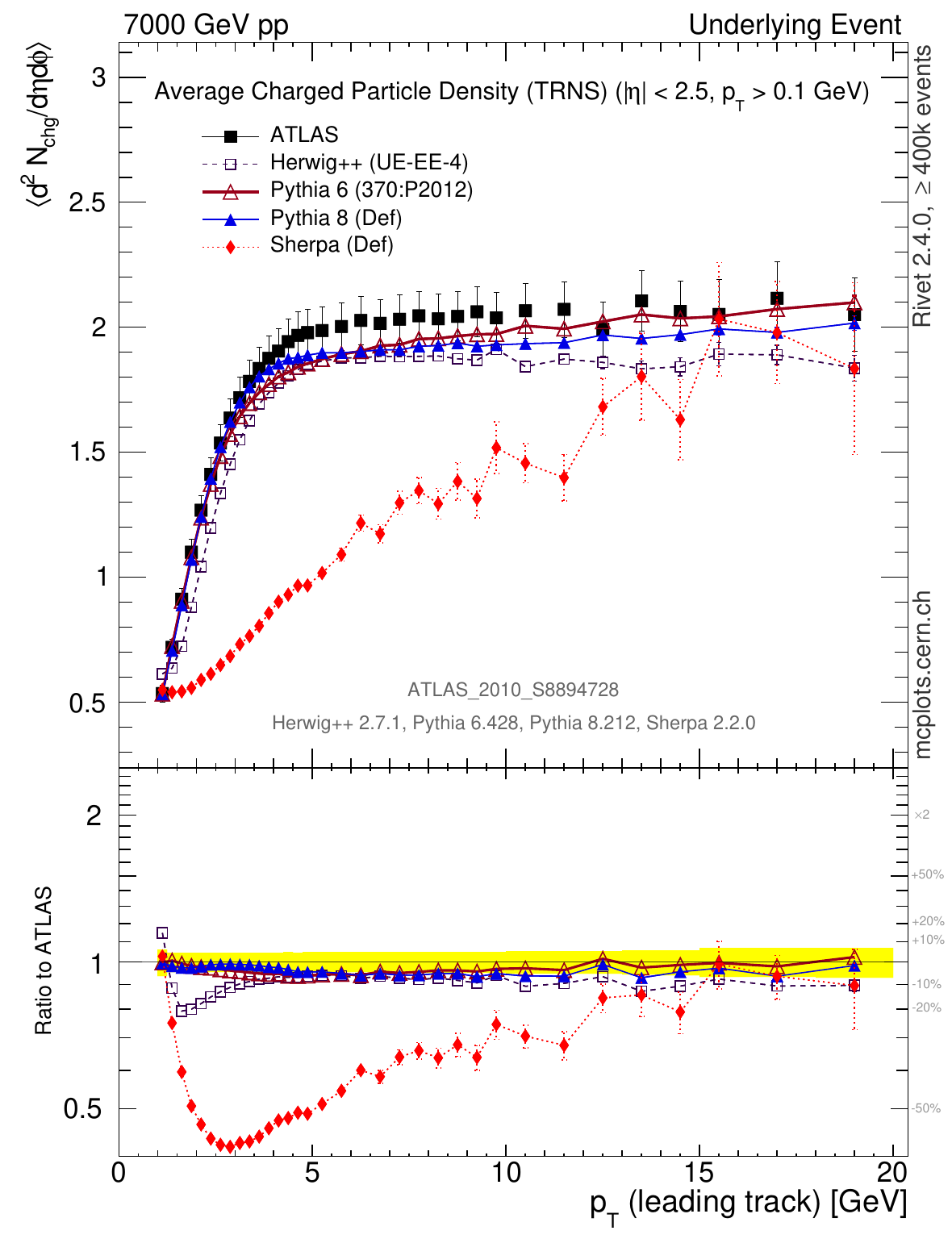}
  \caption{The rise of the Underlying Event, as a function of leading-track $p_\perp$, as measured in the ``Transverse Region'' defined by azimuthal angles $60^\circ< \Delta\phi <120^\circ$ from the leading track, averaged over the available rapidity region. {\sl Left:}~average summed $p_\perp$ (per unit $\Delta\eta\Delta\phi$) for all tracks with $p_\perp>0.1$\,GeV. {\sl Right:}~average number of charged tracks (per unit $\Delta\eta\Delta\phi$) with $p_\perp>0.1$\,GeV. 
    Plots from  \href{http://mcplots.cern.ch}{mcplots.cern.ch}~\cite{Karneyeu:2013aha}.\label{fig:UElevel}}
\end{figure}
The rise of the pedestal level from low to high trigger-object\footnote{Different measurements and detectors employ different types of trigger objects, such as the leading track, leading track-jet, leading calorimeter jet, etc. In principle, the more inclusive (IR safe) the trigger is defined, the cleaner the measurement of the pedestal rise will be.} $p_\perp$ is illustrated in \figRef{fig:UElevel}. The leveling off of the distributions above leading-track $p_\perp$ values $\sim$ 5 GeV can be interpreted as an effect of ``maximum bias''; when the trigger $p_\perp$ is high enough, the selected events are essentially already maximally biased to small impact parameters, and from that point onwards the UE essentially becomes ``independent of the jet ET''~\cite{Arnison:1983gw} (modulo spillover of bremsstrahlung radiation from the hard-jet regions, which will still increase with the trigger-jet $p_\perp$; this component can be partially suppressed e.g.\ by vetoing events with bremsstrahlung; by considering only the least active of the two UE regions; or by systematically classifying all of the activity in the event as either ``jet-like'' or ``plateu-like''~\cite{Cacciari:2009dp}).
The ability of a theory model to describe the rise of the pedestal from the min-bias level to the UE plateau therefore depends upon its modeling of the $b$-dependence, and correspondingly the impact-parameter distribution (or, equivalently, the assumed mass distributions of the proton in $b$-space)
constitutes another main tuning parameter.   
\index{Monte Carlo!Tuning}%
A detailed discussion of impact-parameter dependent models goes beyond
the scope of these lectures;  
see \cite{Sjostrand:1987su,Sjostrand:2004pf,oai:arXiv.org:1101.5953}.  

For hard processes at the LHC at 13 TeV, the transverse energy, $E_T$, in the
UE is expected to be about $3.3~\mbox{GeV}$ per unit $\Delta
R=\sqrt{\Delta\phi^2+\Delta\eta^2}$ 
area~\cite{Skands:2013asa} (for a reference case of 100-GeV dijets, 
including both charged and neutral particles, with no cut on $p_\perp$), 
though with large event-to-event fluctuations of order $\pm
2~\mbox{GeV}$~\cite{Aad:2010fh}. Thus, for example, the total $E_T$
originating from the UE,  
in a cone with radius $0.4$ can be estimated to
be $E_{T\mrm{UE}}(R=0.4)\sim 1.6\pm 1~\mbox{GeV}$, 
while the $E_T$ in a cone with radius $1.0$ would be
$E_{T\mrm{UE}}(R=1.0)\sim 10\pm 6~\mbox{GeV}$. Note that the magnetic
field in realistic detectors will deflect some fraction of the soft charged
component of the underlying event into helix trajectories that will
hence not contribute to the energy deposition in the
calorimeters. 

\subsection{Tuning \label{sec:tuning}} 
\index{Tuning|see{Monte Carlo}}%
\index{Monte Carlo!Tuning}%
\index{Monte Carlo!Event generators}%
\index{Monte Carlo!Uncertainties}%

A main virtue of general-purpose Monte Carlo event generators
is their ability to
provide a complete and fully differential picture of collider final
states, down to the level of individual particles. 
As has been emphasised in these lectures, 
the achievable accuracy depends both on the
inclusiveness of the chosen observable and on the  
sophistication of the simulation itself. An important driver for the
latter is obviously the development of improved theoretical models,
e.g., by including matching to higher-order matrix elements, more
accurate resummations, or better non-perturbative models,  as
discussed in the previous sections; but it  
also depends crucially on the available constraints on the remaining
free parameters of the model. Using existing data (or more precise
calculations) to constrain these is referred to as generator
tuning. 

Keep in mind that generators attempt to deliver a \emph{global}
description of the data; a tune is no good if it fits one distribution
perfectly, but not any others. It is therefore crucial to study the
simultaneous degree of agreement or disagreement over many, mutually
complementary, distributions. 
A useful online resource for making such comparisons can be found at
the \href{http://mcplots.cern.ch}{MCPLOTS} web
site\cite{Karneyeu:2013aha} (which relies on computing power donated
by volunteers,
via the \href{http://lhcathome.web.cern.ch/test4theory}{LHC@home} project~\cite{Lombra?aGonz?lez:2012zz}). The
analyses come from the comprehensive \textsc{Rivet} analysis
toolkit~\cite{Buckley:2010ar}, which can also be run stand-alone to
make your own MC tests and comparisons.

\index{Monte Carlo!Event generators}%
Although MC models may appear to have a
bewildering number of independently adjustable parameters, 
it is worth noting that most  of these only control
relatively small (exclusive) details of the event generation. The
majority of the 
(inclusive) physics is determined by only a few, very important ones, 
such as the value of the strong coupling, in the perturbative
domain, and the form of the fragmentation function for massless
partons, in the non-perturbative one. 

Armed with a good understanding of the underlying model, an expert would
therefore normally take a highly factorised approach to constraining
the parameters, 
first constraining the perturbative ones (using IR safe observables
and/or more precise theory calculations) and thereafter the
non-perturbative ones, each ordered in a measure of their relative
significance to the overall modeling. This  allows one 
to concentrate on just a few parameters and a few carefully chosen 
distributions at a time, reducing the full parameter space to manageable-sized
chunks. Still, each step will often involve more than one single
parameter, and non-factorisable 
correlations may still necessitate additional iterations from
the beginning before a fully satisfactory set of parameters is obtained. 

Recent years have seen the emergence
of automated tools that attempt to reduce the amount of both computer
and manpower required for this task, for instance 
by making full generator runs only for a
limited set of parameter points, and then interpolating between
these  to obtain approximations to what the true generator result
would have been for any intermediate parameter point, as, e.g., in 
\textsc{Professor}~\cite{Buckley:2009bj}. 
Automating the human expert input is more difficult. 
Currently, this is addressed by a combination of input solicited from
the generator authors (e.g., which parameters and ranges to consider,
which observables constitute a complete set, etc)
and the elaborate construction of non-trivial weighting
functions that determine how much weight is assigned to each 
individual bin in each distribution. The field is still
burgeoning, and future sophistications are to be
expected. 
Nevertheless, at this point the overall quality of the tunes
obtained with automated methods appear to at least be competitive with
the manual ones. 

\index{Monte Carlo!Event generators}%
However, though we have very good LHC tunes for essentially all the
general-purpose generators by now, there are two important aspects
which have so far been neglected, and which it is becoming
increasingly urgent to address. The first is that a central tune is
not really worth much, unless you know what the uncertainty on it is. 
A few individual proposals for systematic tuning variations have
been made~\cite{Skands:2010ak,Richardson:2012bn}, but so far there is
no general approach for 
establishing MC uncertainties by tune variations. 
(Note: in 2016 all of the general-purpose MC collaborations
published strategies for automated evaluations of
perturbative shower uncertainties, which we highly recommend the
community to start using, see~\cite{Badger:2016bpw,Mrenna:2016sih,Bellm:2016voq}.)
The second issue is
that virtually all generator tuning is done at the ``pure'' LL shower
level,  and not much is known about what happens
to the tuning when matrix-element matching is subsequently
included\footnote{One aspect, consistent $\alpha_s$ variations, is
  discussed in ~\cite{Cooper:2011gk}.}.

Finally, rather than performing one global tune to all the
data, as is usually done,  a more systematic check on the validity of
the underlying physics model could be obtained by instead performing
several independent 
optimisations of the model parameters for a range of different 
phase-space windows and/or collider environments.
In regions in which consistent parameter sets are obtained (with
reasonable $\Delta\chi^2$ values), 
the underlying 
model can be considered as interpolating well, i.e., it is universal. 
If not, a breakdown in the ability of the model to span different
physical regimes has been identified, and can be addressed, with the
nature of the deviations giving clues as to the nature of the breakdown. 
With the advent of automated tools,  
such systematic studies are now becoming feasible, with a first
example given in \cite{Schulz:2011qy}. 

\index{Monte Carlo!Event generators}%
\index{Monte Carlo!Tuning}%
We round off by giving a sketch of a reasonably complete tuning
procedure, without going into details about the parameters that
control each of these sectors in individual Monte Carlo models 
(a recent detailed discussion in the context of PYTHIA 8 can be found
in~\cite{Skands:2014pea}): 

{\bf 1) Keep in mind} that inabilities of models to
 describe data is a vital part of the feedback cycle between
 theory and experiment. Also keep in mind that
 perturbation theory at (N)LO+LL is doing \emph{very well} if it gets
 within 10\% of a given IR safe measurement. An agreement of 5\% should be
 considered the absolute sanity limit, beyond which it does not make
 any sense to tune further. For some quantities, e.g., ones for which
 the underlying modeling is \emph{known} to be poor, an order-of-magnitude
  agreement or worse may have to be accepted. 
 
\index{String model}%
\textbf{2) Final-State Radiation and Hadronisation:} 
 mainly using LEP and other $e^+e^-$ collider data. On the IR safe
 side, there are event shapes and jet observables. On the IR sensitive
 side, multiplicities and particle spectra. Pay attention to 
 the high-$z$ tail of the fragmentation,
 where a single hadron carries a large fraction of an entire jet's
 momentum (most likely to give ``fakes''). 
Depending on the focus of the tuning, attention should also
 be paid to identified-particle rates and ratios (perhaps with a nod
 to hadron-collider measurements), and to fragmentation
 in events containing heavy quarks and/or gluon jets. 
 Usually, more weight is given to
 those particles that are most copiously produced. The scaling
 properties of IR safe vs.\ IR sensitive contributions can be
 tested by comparing data at several different $e^+e^-$ collider
 energies.  
 
\textbf{3) Initial-State Radiation, and 
  ``Primordial\footnote{Primordial $k_T$: an
  additional soft \pt\ component that is injected on top of the
  \pt\ generated by the initial-state shower itself, see
  \cite[Section 7.1]{Buckley:2011ms}.} $k_T$'':} the main constraining
  distribution is the dilepton \pt\ distribution in Drell-Yan events in
  hadron-hadron collisions. Ideally, one would like to use 
  several different $Q^2$ values, and/or complementary processes,
  like $p_\perp(\mrm{dijet})$ or $p_\perp(t\bar{t})$. 
For any observables containing explicit
  jets, be aware that the UE can produce small 
  horisontal shifts in jet \pt\ distributions, which may in turn 
  result in larger-than-expected 
  vertical changes if the distributions are falling sharply. Also note
  that the ISR evolution is sensitive to the choice of PDFs.

\textbf{4) Initial-Final Connections:} (radiation from
  colour lines connecting the initial and final states): 
\index{DIS}%
DIS and jet broadening in hadron collisions. This is one of the
  most poorly controlled parts of most MC models, highly sensitive to
  the treatment of coherence (see e.g.~\cite{Ritzmann:2012ca} for
  illustrations).  
  Keep in mind that it
  is \emph{not} directly constrained by pure final-state observables,
  such as LEP fragmentation, nor by pure initial-state observables,
  such as the Drell-Yan \pt\ spectrum, which is why we list it as a
  separate item here. The modeling of this
  aspect can have important effects on specific observables, a recent
  example being the $t\bar{t}$ forward-backward asymmetry at the
  Tevatron~\cite{Skands:2012mm}.

\textbf{5) Underlying Event:} Good constraints on the overall level of the
  underlying event can be obtained by counting the summed transverse
  energy (more IR safe) and/or particle multiplicities and average
  transverse momenta (more IR sensitive) in regions \emph{transverse}
  to a hard trigger jet (more IR safe) or particle (more IR
  sensitive), see e.g.~\cite{Field:2011iq}. 
  Constraints on the \emph{fluctuations} of the underlying
  event are also important, and can be obtained, e.g., by comparing to
  measurements of the RMS of such distributions~\cite{Aad:2010fh} or
  by plotting salient quantities along an axis from low to high UE
  activity~\cite{Martin:2016igp}. Again, note
  that the UE modelling can be very sensitive to the choice of
  PDFs~\cite{Schulz:2011qy,Skands:2014pea}. Finally, the modeling of
  the UE should be \emph{universal}, hence the same UE model should
  --- ideally --- be able to describe UE distributions not only in
  inclusive-jet events, but also the UE in processes like
  Drell-Yan~\cite{Aaltonen:2010rm,Chatrchyan:2012tb,Aad:2014jgf,Alioli:2016wqt}
  or  $t\bar{t}$ production~\cite{CMS:2013mfa,CMS:2015usp}.  

\index{String model}%
\index{Colour connections}%
\textbf{6) Colour (Re-)Connections and other Final-State Interactions:}  
 By Final-State Interactions,
  we intend a broad spectrum of possible collective effects that may
  be included to a greater or lesser extent in various models. These
  effects include: Bose-Einstein correlations (see,
  e.g., \cite{Lonnblad:1997kk}), 
rescattering (see, e.g., \cite{Corke:2009tk}), 
colour reconnections / string
  interactions  (see,
  e.g., \cite{Rathsman:1998tp,Skands:2007zg,Bopp:2011fw,Gieseke:2012ft,Christiansen:2015yqa}),  
  hydrodynamics (see, e.g., \cite{Werner:2011zzc}), 
etc. 
  As a rule, these effects
  are soft and/or non-perturbative and hence should not modify hard IR safe
  observables appreciably. They can, 
  however, have \emph{drastic} effects on IR sensitive ones, such as
  particle multiplicities, momentum distributions, and correlations, 
  wherefore useful constraints are typically furnished by 
  measurements of spectra and correlations as functions of
  quantities believed to 
  serve as indicators of the strength of these phenomena 
  (such as event multiplicity~\cite{Petrov:2012hca,Bierlich:2015rha,Adam:2016emw} or underlying-event
  activity~\cite{Martin:2016igp}), and/or by collective-flow-type 
  measurements. 
  Finally, if the model includes a universal description of
  underlying-event and soft-inclusive QCD, as many MPI-based models do, then
  minimum-bias data can also be used as a control sample, though one must
  then be careful either to address diffractive contributions properly
  or to include only gap-suppressed data samples. A complete MB and UE
  model should also be able to describe the rise of the pedestal from
  MB to UE, e.g., in the transverse UE observables (see above).

\textbf{7) Beam Remnants:} Constraints on beam-remnant fragmentation
(see, e.g., \cite{Sjostrand:2004pf}) are most directly obtained in the forward
  region~\cite{Chatrchyan:2011wm,Chatrchyan:2011wb,Aspell:2012ux,Chatrchyan:2013gfi,Aaij:2014pza,Chatrchyan:2014qka,Antchev:2014lez}. Especially
  for soft-inclusive triggers, cuts designed to isolate diffractive
  topologies are then required to distinguish effectively between the
  fragmentation of a (diffractive) colour-singlet excitation of the
  beam particle and that of a (non-diffractive) colour-charged
  remnant having exchanged some (non-zero) net amount of
  colour charge with the other colliding beam hadron. To lowest
  order, the remnant should be in a triplet (octet) colour state after
  having exchanged a quark (gluon) with the other beam particle, but
  taking MPI into account much larger amounts of colour may be
  transferred, with correspondingly larger possible colour
  representations of the remnant (see, e.g.,~\cite{Christiansen:2015yqa}). 
The amount of baryon transport from the remnant
  to a given rapidity region~\cite{Erhan:1979ba,Aaij:2011va,Abazov:2015sri} can
  also be used to probe how much the colour 
  structure of the remnant was effectively disturbed, with more baryon
  transport indicating a larger amount of ``beam baryon blowup''.
Ideally one would also correlate the forward activity with a (central) measure of
how much total colour charge is scattered out of the remnant(s), such as
an observable sensitive to the number of MPI.

\section*{Acknowledgments}
Thanks to C.~Brust, M.~Cacciari, Y.~Dokshitzer, L.~Hartgring, A.~Kronfeld, 
A.~Larkoski, J.~Lopez-Villarejo,
C.~Murphy, P.~Nason, P.~Pigard, S.~Prestel, G.~Salam, and T.~Sj\"ostrand
whose valuable comments and sharing of insight contributed to these lectures. 
In addition, I have used material from my 2010 ESHEP
lectures~\cite{Skands:2011pf} and 2014 AEPSHEP lectures, and from 
the ESHEP lectures by Mangano \cite{Ellis:2009zzb}, by
Salam \cite{Salam:2010zt,Grojean:2010ab}, by Sj\"ostrand \cite{Sjostrand:2006su}, 
and by Stirling \cite{Ellis:2008zzf}, as well as the recent review on
Monte Carlo event generators by the MCnet collaboration \cite{Buckley:2011ms}. 

\addcontentsline{toc}{section}{References}
\bibliographystyle{utphys}
\bibliography{tasi-skands}
\section*{}
\clearpage
\addcontentsline{toc}{section}{Index}
\printindex

\end{document}